\documentclass[a4paper,11pt]{article}
\pdfoutput=1 

\usepackage{jheppub} 
\usepackage[T1]{fontenc} 

\usepackage{amssymb}
\usepackage{enumerate} 

\title{\boldmath Production and decay of the Higgs boson in association with top quarks}

\author{Daniel Stremmer}
\author{and Malgorzata  Worek}
\affiliation{ Institute for Theoretical Particle Physics
and Cosmology, RWTH Aachen University, \\D-52056 Aachen, Germany}

\emailAdd{daniel.stremmer@rwth-aachen.de}
\emailAdd{worek@physik.rwth-aachen.de}

\abstract{
We report on the calculation of the next-lo-leading order QCD corrections to Higgs boson production and decay in association with top quarks. We consider leptonic decays of top quarks leading to the hadronic process $pp\to e^+\nu_e\, \mu^-\bar{\nu}_\mu \, b\bar{b}\,H\, (H\to X)$ at the LHC with $\sqrt{s}=13$ TeV. All resonant as well as  non-resonant Feynman diagrams, interferences and off-shell effects are included for the top quark and $W$ gauge boson. Decays of the Higgs boson, on the other hand,  are included in the narrow-width approximation. Specifically, we consider Higgs boson decays into $b\bar{b}$, $\tau^+\tau^-$, $\gamma\gamma$ and $e^+e^-e^+e^-$. Numerical results are given at the integrated and differential fiducial level for various factorisation and renormalisation scale choices and different PDF sets. We study the main theoretical uncertainties that are associated with neglected higher order terms in the perturbative expansion and with different parametrisations of the PDFs. Furthermore, we examine the size of the off-shell effects by an explicit comparison to the calculation in the full narrow-width approximation. Finally, the impact of the contributions induced by the bottom-quark parton density is investigated.
}

\dedicated{\rm P3H-21-064,  TTK-21-38}


\begin{document} 
\maketitle
\flushbottom

%
\section{Introduction}
\label{sec:introduction}
%

After the discovery of the Higgs boson in $2012$ by the ATLAS and CMS collaborations \cite{ATLAS:2012yve,CMS:2012jmk}, one of the main goals of the Large Hadron Collider (LHC) is the precise determination of its properties and its couplings to other Standard Model (SM) particles. The Higgs boson couples to fermions via the Yukawa interaction, which is proportional to the mass of fermions and therefore of particular interest for the top quark, the heaviest fermion in the SM. Within the SM, the top quark Yukawa coupling $(Y_t)$ is close to unity. Its  measurement can not only serve as an important test of the SM but also can be used to constrain various models beyond the SM (BSM). Indeed, in many new physics scenarios $Y_t$ is expected to differ from the SM value. Higgs production at the LHC is dominated by gluon fusion, $gg\to H$. Since this process is loop induced, it only provides an indirect way to access $Y_t$, and it can be affected by heavy BSM particles. On the other hand, Higgs production in association with top quarks, $pp\to t\bar{t}H$, gives a window to directly probe this coupling already at tree level. This production mechanism contributes only about $1\%$ \cite{deFlorian:2016spz} to the total Higgs boson production cross section and was observed in $2018$ by the ATLAS and CMS collaborations \cite{ATLAS:2018mme,CMS:2018iho}. 

On the other hand, the Higgs boson can only be observed in the detector by its decay products. The first observation of $t\bar{t}H$ production in a single decay channel was reported recently by the ATLAS and CMS collaborations \cite{CMS:2020htp,ATLAS:2020ior} through Higgs decay into two photons $H\to\gamma\gamma$. Even though this decay channel is loop induced and has a very small branching ratio of about $0.2\%$ \cite{deFlorian:2016spz}, it is  experimentally very well accessible due to the small background and good mass resolution for $M_{\gamma\gamma}$. Measurements of the Higgs boson production rate in association with top quarks in final states with electrons, muons and tau leptons have been carried out very recently by CMS \cite{CMS:2020mpn}. This analysis aimed at events that contain $H\to W^+W^-$, $H\to \tau^+ \tau^-$ or $H\to ZZ$ decays with the corresponding branching ratios of about $21\%$, $6\%$ and $3\%$ respectively. Further searches are reported by ATLAS and CMS for $t\bar{t}H$ with a Higgs boson decay into a bottom quark pair $H\to b\bar{b}$ \cite{ATLAS:2017fak,CMS:2018hnq}, which has the largest branching ratio of about $58\%$ \cite{deFlorian:2016spz}. From the experimental point of view, however, this decay channel is the most challenging one, because of the complicated Higgs boson reconstruction from the so-called combinatorial background caused by a multitude of $b$-jets coming both from the two top quarks and the Higgs boson. Furthermore, this decay channel suffers from large reducible $t\bar{t}jj$ background  and irreducible QCD and  $Z$-peak backgrounds, $t\bar{t}b\bar{b}$  and $t\bar{t}Z(Z\to b\bar{b})$ respectively.

For $t\bar{t}H$ production in the case of stable top quarks,  next-to-leading order (NLO) QCD corrections are known for many years \cite{Beenakker:2001rj,Beenakker:2002nc,Reina:2001sf,Dawson:2002tg,Dawson:2003zu} and are combined in the literature with NLO electroweak  corrections \cite{Frixione:2014qaa,Zhang:2014gcy,Frixione:2015zaa,Frederix:2018nkq}. A further step towards a more precise modelling of the $t\bar{t}H$ process has been achieved by including soft gluon resummation effects with (next-to-)next-to-leading-logarithmic ((N)NLL) accuracy \cite{Kulesza:2015vda, Broggio:2015lya,Broggio:2016lfj,Kulesza:2017ukk,Broggio:2019ewu, Kulesza:2020nfh}. To include decays of the top quarks various approaches have been used. On the one hand, NLO QCD calculations have been matched to parton-shower programs (NLO+PS) \cite{Frederix:2011zi,Garzelli:2011vp,Hartanto:2015uka,Maltoni:2015ena}. On the other,   the narrow-width approximation (NWA) has  been employed  \cite{Zhang:2014gcy}. Furthermore, full off-shell effects of top quarks in the di-lepton decay channel have already been included with NLO QCD corrections in Ref. \cite{Denner:2015yca} and with additional electroweak corrections in Ref. \cite{Denner:2016wet}.

In this article, we report on the computation of the NLO QCD corrections to off-shell $t\bar{t}H$ production in the di-lepton top quark decay channel with Higgs boson decays in the NWA. In a first step, we compute independently the NLO QCD corrections to $pp\to e^+\nu_e\, \mu^-\bar{\nu}_\mu \, b\bar{b}\,H +X$ at order ${\cal O}(\alpha_s^3\alpha^5)$ at the LHC with $\sqrt{s}=13$ TeV and perform a comparison with the results presented in Ref. \cite{Denner:2015yca} at the integrated and differential fiducial level. In a next step, we extend the analysis of $t\bar{t}H$ production and investigate the dependence of our results upon variation of renormalisation and factorisation scales and parton distribution functions in the quest for an accurate estimate of the theoretical uncertainties. Additionally, we explore a few possibilities for a dynamical scale choice with the goal of stabilizing the perturbative convergence of the differential cross sections in the high $p_T$ tails.  Afterwards, we study the off-shell effects of the top quarks at the integrated and differential level by a comparison with the calculations in the  NWA and NWA with leading order (LO) top quark decays (NWA${}_{\rm LOdecay}$). Having  NWA${}_{\rm LOdecay}$ results at our disposal, we can additionally estimate the size of the NLO QCD corrections to top-quark decays. We close the discussion on $t\bar{t}H$ production  by investigating the $b$-initiated contributions that are often neglected in similar calculations with off-shell top quarks because of their assumed numerical insignificance. At last, we include decays of  the Higgs boson in the NWA while we keep all off-shell effects of the top quark and $W$ gauge boson. Even though the non-factorisable corrections vanish in the limit $\Gamma_H/m_H \to 0$, that characterises the NWA, such a mixed approach is very well justified when we observe that the following hierarchy is satisfied
\begin{gather*}
 \frac{ \Gamma_W}{m_W}  \,\, > \,\, \, \frac{\Gamma_t}{m_t}\,\,\,   \gg\,\,\,  \frac{\Gamma_H}{m_H} \,,\,\,\,\,\, \\[0.2cm]
  2.6\%  \, >\,  0.8\%  \,  \gg \, 0.003\% \,.
\end{gather*}
In our paper, various decay channels of the Higgs boson are added. Thus, we can analyse fully realistic final states. We study the scale dependence in the decays of the Higgs boson and compare it  to the one in the Higgs boson production. Additionally, various differential distributions of  Higgs boson decay products are investigated in more detail.

The paper is organised as follows. In Section \ref{sec:description}, we briefly describe the \textsc{Helac-Nlo} framework used for the $t\bar{t}H$ process, and the computation of Higgs boson decays, as well as the various cross-checks performed in this calculation. The theoretical setup, consisting of input parameters, event selection, renormalisation and factorisation  scale choices, is described in Section \ref{sec:setup}. In Section \ref{sec:tth-checks} we perform a comparison with the results of Ref. \cite{Denner:2015yca}. Phenomenological results for  $t\bar{t}H$ production at the integrated and differential fiducial level are discussed in detail in Section \ref{sec:tth-int} and Section \ref{sec:tth-diff}. In Section \ref{sec:tth-nwa} we examine the off-shell effects of the top quarks. The contribution from bottom quarks in the initial states and the modifications required in the jet algorithm are discussed in Section \ref{sec:tth-bottom}. The production and decay of the Higgs boson are combined in Section \ref{sec:tth-decay}. We consider Higgs boson decays into $b\bar{b}$, $\tau^+\tau^-$, $\gamma\gamma$ and $e^+e^-e^+e^-$, and present results at the integrated and differential fiducial level. Finally, our results are summarised in Section \ref{sec:sum}.

%
\section{Description of the calculation}
\label{sec:description}
%

\begin{figure}[t!]
  \begin{center}
  \includegraphics[trim= 36 526 36 36, width=\textwidth]{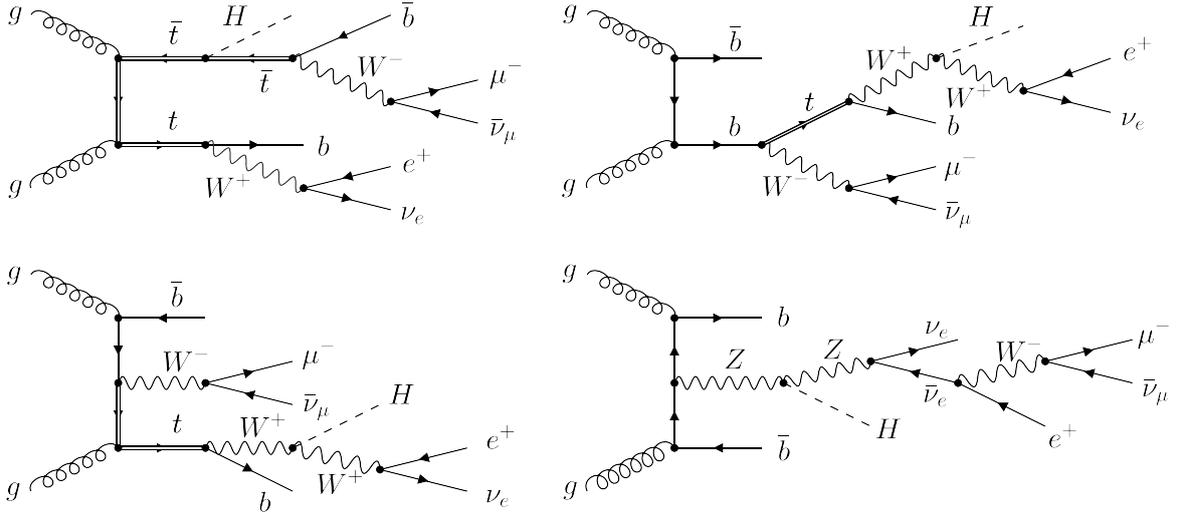}
\end{center}
  \caption{\label{fig:fd-LO} \it  Representative Feynman diagrams of
    the double, single and non-resonant top-quark contributions that
    are part of the matrix element employed in the full off-shell
    computation for the  $pp\to e^+\nu_e\, \mu^-\bar{\nu}_\mu \, b\bar{b}\,H$ process.   
    Feynman diagrams were produced with the help of the
    \textsc{FeynGame} program \cite{Harlander:2020cyh}.}
\end{figure}

In a first step, we study Higgs boson production in association with top quarks while keeping the Higgs boson  stable $(\Gamma_H=0)$. The complex-mass scheme \cite{Denner:1999gp,Denner:2005fg,Bevilacqua:2010qb,Denner:2012yc} is used to include all off-shell effects of the top quarks and massive gauge bosons ($W^{\pm},Z)$ in a gauge invariant way. In particular, we consider the following process
\begin{equation}
pp\to e^+\nu_e\, \mu^-\bar{\nu}_\mu \, b\bar{b}\,H +X
\end{equation}
at NLO QCD, i.e.  at ${\cal O}(\alpha_s^3\alpha^5)$. The LO process at ${\cal O}(\alpha_s^2\alpha^5)$ comprises the  $gg\to e^+\nu_e\, \mu^-\bar{\nu}_\mu \, b\bar{b}\,H$ and $q\bar{q}\to e^+\nu_e\, \mu^-\bar{\nu}_\mu \, b\bar{b}\,H$ sub-processes with $q\in\{u,d,c,s\}$ and $174$ and $76$ Feynman diagrams respectively. A few examples of Feynman diagrams contributing to the LO process are presented in Figure \ref{fig:fd-LO}. They are shown for the dominant $gg$ partonic sub-process. The Cabibbo-Kobayashi-Maskawa (CKM) mixing matrix is kept diagonal. We work in the five-flavour scheme, where the mass and the Yukawa coupling of the bottom quark $(Y_b)$ is set to zero. The effects of a non-vanishing $Y_b$ coupling have been  checked at LO and are smaller than $0.1\%$ for bottom quark induced channels and even less for channels without bottom quarks in the initial state. For now and as a first approach, initial state bottom quarks are neglected due to their suppressed contribution. Nevertheless, this contribution will be discussed in detail at a later stage in Section \ref{sec:tth-bottom}. The matrix elements are computed using the Dyson-Schwinger off-shell iterative algorithm \cite{Draggiotis:1998gr,Draggiotis:2002hm,Papadopoulos:2005ky} instead of the traditional Feynman diagrams approach. This iterative algorithm is implemented in \textsc{Helac-Dipoles} \cite{Czakon:2009ss} based on \textsc{Helac-Phegas} \cite{Kanaki:2000ey,Papadopoulos:2000tt,Cafarella:2007pc}. The phase-space integration is performed and optimised with the help of \textsc{Parni} \cite{vanHameren:2007pt} and \textsc{Kaleu} \cite{vanHameren:2010gg}. The virtual corrections are computed with \textsc{Helac-1Loop} \cite{vanHameren:2009dr}, where the one-loop matrix elements are reduced to scalar integrals at the integrand level using the OPP reduction technique \cite{Ossola:2006us} as implemented in \textsc{CutTools} \cite{Ossola:2007ax}. The scalar integrals are obtained from the library \textsc{OneLOop} \cite{vanHameren:2010cp}. Since the computation of loop diagrams is very time consuming, the number of evaluated loop diagrams is reduced by using only unweighted Born events for the calculation of the virtual corrections as described in Ref. \cite{Bevilacqua:2011xh}. The finite part and $\epsilon$-poles are cross-checked with the module \textsc{MadLoop} \cite{Hirschi:2011pa} in \textsc{MadGraph} \cite{Alwall:2014hca} for a few phase-space points. In addition, the cancellation of the $\epsilon$-poles between the virtual corrections and the $\mathcal{I}\textrm{-operator}$ implemented in \textsc{Helac-Dipoles} is checked. For the real corrections we have to take into account the same sub-processes as above but with an additional gluon in the final state. We have $1162$ Feynman diagrams for the gluon induced $gg\to e^+\nu_e\, \mu^-\bar{\nu}_\mu \, b\bar{b}\,H g$ and $476$ for the quark induced $q\bar{q}\to e^+\nu_e\, \mu^-\bar{\nu}_\mu \, b\bar{b}\,H g$ sub-process. At NLO, we encounter additional sub-processes, namely $gq\, (g\bar{q})\to e^+\nu_e\, \mu^-\bar{\nu}_\mu \, b\bar{b}\,Hq \,(\bar{q})$, which are connected by crossing with the quark induced ones. The calculation of the real corrections is performed via subtraction methods implemented in \textsc{Helac-Dipoles} and is identical from the QCD point of view with the calculations in Ref. \cite{Bevilacqua:2018woc,Bevilacqua:2019cvp,Bevilacqua:2020pzy}. In particular, two independent schemes are implemented, the Catani-Seymour dipole formalism \cite{Catani:1996vz,Catani:2002hc} and the Nagy-Soper subtraction scheme \cite{Bevilacqua:2013iha}. In addition, a phase-space restriction on the subtraction terms is implemented for both schemes as described in Ref. \cite{Bevilacqua:2009zn,Nagy:1998bb,Nagy:2003tz,Czakon:2015cla}. This restriction is used to reduce the number of required subtraction terms and as a cross-check, since the final real emission results should not depend on the restriction. For the full off-shell calculation the Nagy-Soper subtraction scheme is used, while for the NWA case the Catani-Seymour dipole formalism is employed instead. The calculation in the NWA is also performed in the \textsc{Helac-Nlo} framework and details about the implementation are given in Ref. \cite{Bevilacqua:2019quz}. All parts of the off-shell calculation are stored in modified Les Houches Event files (LHEFs) \cite{Alwall:2006yp} as partially unweighted events \cite{Bevilacqua:2016jfk} with additional information based on Ref. \cite{Bern:2013zja}. This allows us to create differential distributions and obtain results for  different PDF sets or other factorisation and renormalisation scale settings without rerunning the entire calculation from scratch.

In the last part, we combine the production and the decay of the Higgs boson  with a $t\bar{t}$ pair. We reuse the LHEFs to model the Higgs boson decays in the NWA.  Consequently, we are able to include the production and decays of the Higgs boson in association with top quarks  using the NWA for the Higgs boson at NLO in QCD, while keeping full off-shell effects for top quarks and for all other unstable particles.  This approach is very well justified  as we argued before. As we consider decays of a scalar particle, no spin information is transferred  from the production to the decay stage which further warrants the use of the LHEFs. The phase-space integration for
Higgs boson decay products is performed using our own implementation of the algorithm outlined in Ref. \cite{Platzer:2013esa}, called \textsc{Rambo} (on diet). In particular,  we generate unweighted events in the rest frame of the Higgs boson to model its  decays. The implementation is cross-checked at LO with the NWA implementation in \textsc{Helac-Dipoles}. All parts of the calculation with Born-like kinematics work exactly in the same way.  In this paper we consider the following decay channels of the Higgs boson
\begin{enumerate}[(i)]
	\item $H \to b\bar{b}$,
	\item $H \to \tau^+\tau^-$,
	\item $H \to \gamma\gamma$,
	\item $H \to Z^*Z^*\to e^+e^-e^+e^-$.
\end{enumerate}
In the last case, the Higgs boson decays into two off-shell $Z$-bosons, which further decay into two $e^+e^-$ pairs \cite{Bredenstein:2006rh}. Since we are interested in the $e^+e^-e^+e^-$ events in the fiducial phase-space regions near the Higgs boson mass peak, this approach is very well justified. For the $4\ell$ events well above the Higgs boson mass peak, i.e. when   $M_{4\ell} \gg m_H$, sizable Higgs boson continuum interference effects might occur that need to be considered as well, see e.g.  Refs.  \cite{Kauer:2012hd,Caola:2013yja}. We do not include leptonic $\tau^\pm$ and $W^\pm$-gauge boson decays, even though the two decay chains seem to be straightforward in the  NWA approach.  The reason is rather technical, as we simply tend to avoid the final states were additional  neutrinos can appear. This makes it easier to directly compare with the results presented in Ref. \cite{Denner:2015yca} where a missing transverse momentum cut has been applied to the two neutrinos from the top quark decays already at the Higgs boson production stage. Of course, had we generated the LHEFs without this cut in the first place, both $H \to \tau^+ \tau^-\to \ell^+  \nu_\ell \, \ell^- \bar{\nu}_\ell\, \nu_\tau \bar{\nu}_\tau$ and $H\to W^+W^-\to \ell^+ \nu_\ell \, \ell^- \bar{\nu}_\ell$ with $\ell^\pm=e^\pm,\mu^\pm$ could have been added as well. To keep all obtained events and to avoid generation of new LHEFs, we decided to focus only on the final states without neutrinos.  It should be noted that generating such a complicated $ 2 \to  7$ final state is very time consuming and requires a dedicated computer cluster. The differential cross section  for the $pp \to t\bar{t}H \to t\bar{t} X$ process, where $t\bar{t}$ symbolises the following $2 \to 6$ process $pp \to e^+\nu_e \, \mu^-\bar{\nu}_\mu \, b\bar{b}$ and $X$ stands for $X=b\bar{b}\,,\tau^+\tau^-\,,\gamma\gamma \,, e^+e^-e^+e^-$, can be written as
\begin{equation}
  \begin{split}
    \label{eq_NWA}
d\sigma&=d\sigma_{t\overline{t}H}\frac{d\Gamma_{H\to X}}{\Gamma_H} \\[0.1cm]
&=d\sigma_{t\overline{t}H}^0\frac{d\Gamma_{H\to X}^0}{\Gamma_H}+d\sigma_{t\overline{t}H}^1\frac{d\Gamma_{H\to X}^0}{\Gamma_H}+d\sigma_{t\overline{t}H}^0\frac{d\Gamma_{H\to X}^1}{\Gamma_H}\,.
\end{split}
\end{equation}
In the second line the Higgs boson production cross section and decay rate are expanded up to NLO in QCD, while $\Gamma_H$ is kept fixed. Specifically, $d\sigma^0_{t\bar{t}H}$, $d\Gamma^0_{H\to X}$ represent LO contributions as calculated with the NLO input parameters and $d\sigma^1_{t\bar{t}H}$, $d\Gamma^1_{H\to X}$ describe the NLO  corrections. The last term is only present in the case of the Higgs boson decay into a $b\bar{b}$ pair \footnote{ We note that terms like $d\sigma^1_{t\bar{t}H} \,
d\Gamma^1_{H\to X}$ are not included in this defintion since such
terms would be of the order of $\mathcal{O}(\alpha_s^4)$ and are part
of NNLO corrections. With this definition, the total NLO cross section
is not equal to the NLO production cross section times the NLO
branching ratio. On the other hand, this relation can be obtained by
simply rescaling the terms in Eq. \eqref{eq_NWA} by appropriate
factors. However, such a definition might lead to underestimated
theoretical uncertainties and even affect the scale dependence of the
production process as outlined for example in Ref. \cite{Gauld:2019yng} (see appendix B) for the $pp \to VH+X \to \ell^+ \ell^- b\bar{b}$ process, where
$V=W^\pm,Z$.}. For $H\to b\bar{b}$ also the $\mathcal{I}$-operator and the virtual corrections are calculated using LHEFs. We use analytical results for the virtual corrections and the $\mathcal{I}$-operator where the latter one is explicitly cross-checked with the $\mathcal{I}$-operator implemented in \textsc{Helac-Dipoles}. The Catani-Seymour dipole formalism with phase-space restriction on the dipoles, as implemented in \textsc{Helac-Dipoles}, is used for the calculation of the real corrections in this decay chain. For the $H\to \gamma\gamma$ decay, NLO QCD corrections are neglected as these would require two-loop integrals. Furthermore, it is well know that the NLO corrections relative to the lowest order are small in general as shown in Ref. \cite{Spira:1995rr}. As we previously mentioned, the five flavour scheme is used. Thus,  the bottom quark mass is still neglected throughout the calculation.  This allows us to resum large initial state collinear logarithms into PDFs. In order to keep a non-zero coupling between the Higgs boson and bottom quarks, we keep $Y_b$ non-zero in the decay stage and perform the renormalisation in the $\overline{\rm MS}$ scheme. Moreover, we distinguish between the renormalisation scale in the production $\mu_R$ and decay $\mu_{R,\,dec}$. The decay width into a bottom quark pair at NLO is checked with the result in Ref. \cite{Spira:2016ztx} by setting $\mu_{R,\,dec}=m_H$. In $H\to \tau^+ \tau^-$ decays, we do not neglect the mass of the $\tau$ lepton in the matrix element or the phase space. For the loop-induced decay into two photons we use the matrix element at LO as given in Ref. \cite{Shifman:1979eb,Marciano:2011gm}. This matrix element comprises two types of contributions: from $W$ gauge bosons and from massive fermions. In the latter case  we consider massive fermion loops from top quarks, bottom quarks and $\tau$ leptons. However, the contribution from the latter two is negligible.

%
\section{LHC setup}
\label{sec:setup}
%

We consider the  $pp\to e^+\nu_e\, \mu^-\bar{\nu}_\mu \, b\bar{b}\,H +X$ process at the LHC with $\sqrt{s}=13$ TeV. We use the same input parameters as in Ref. \cite{Denner:2015yca}, as briefly summarised in the following. The number of active flavours is set to $N_f=5$ and the contribution from bottom quarks in the initial state is neglected at first. We will return to this contribution later in this article.  The CT10NLO \cite{Lai:2010vv} PDF set is used at LO and NLO QCD for the comparison, while newer PDF sets are employed for the rest of the paper. Specifically, as our default PDF set we use  NNPDF3.1 NLO (LO) \cite{Ball:2017nwa} with $\alpha_s(m_Z)=0.118$. To quantify the differences between various PDF sets, that  originate among others from the choice of the data used and the theoretical assumptions made for the global fit, we also provide numerical results at NLO QCD for  CT18 \cite{Hou:2019efy} and MSHT20 \cite{Bailey:2020ooq}.  The running of the strong coupling constant $\alpha_s$, with two-loop (one-loop) accuracy at NLO (LO) is provided by the LHAPDF interface \cite{Buckley:2014ana}. The electromagnetic coupling $\alpha$, is derived from the Fermi constant in the $G_\mu-$scheme
\begin{equation}
	\alpha =\frac{\sqrt{2}}{\pi} \,G_\mu \, m_W^2\,\left(1-\frac{m_W^2}{m_Z^2}\right)\,,
	~~~~~~~~~~~~~~~~~~~~~
	G_{ \mu}=1.16637 \cdot 10^{-5} \textrm{ GeV}^{-2}\,.
\end{equation}
The values for the masses and widths are given by
\begin{equation}
	\begin{array}{lll}
		m_{t}=173 \textrm{ GeV} \,, &\quad \quad \quad
		&m_{H}=126 \textrm{ GeV} \,, 
		\vspace{0.2cm}\\
		m^{\textrm{OS}}_{W}= 80.385  \textrm{ GeV} \,, &
		&\Gamma^{\textrm{OS}}_{W} =  2.0850  \textrm{ GeV}\,, 
		\vspace{0.2cm}\\
		m^{\textrm{OS}}_{Z}=91.1876 \textrm{ GeV} \,, &
		&\Gamma^{\textrm{OS}}_{Z} = 2.4952 \textrm{ GeV}\,,
	\end{array}
\end{equation}
where $m_V^{\rm OS}$ and $\Gamma_V^{\rm OS}$ with $V = W, Z$ stand for  the measured on-shell values (OS) for the masses and widths of the $W$ and $Z$ boson. They are converted into pole values according to  Ref. \cite{Bardin:1988xt}
\begin{equation}
	m_V=\frac{M^{\textrm{OS}}_V}{\sqrt{1+\left(\Gamma^{\textrm{OS}}_V/m^{\textrm{OS}}_V\right)^2}} \,, \quad \quad \quad
	\Gamma_V=\frac{\Gamma^{\textrm{OS}}_V}{\sqrt{1+\left(\Gamma^{\textrm{OS}}_V/m^{\textrm{OS}}_V\right)^2}}\,.
\end{equation}
The latter are used as input parameters for this calculation. The masses of all other quarks and leptons are set to zero. A Higgs boson mass of $m_H=126$ GeV is used to facilitate the comparison with results from Ref. \cite{Denner:2015yca}. Even though the current value is closer to $m_H = 125$ GeV, our conclusions do not depend on the specific value used \footnote{
 We have  checked that for the integrated LO cross section changing the
Higgs boson mass to $m_H=125$ GeV  amounts to $2\%$.  Indeed, for the
NNPDF3.1 PDF set  using $\mu_0=H_T/2$ one would obtain $\sigma_{LO} =
2.2618(4)$ fb instead of $\sigma_{LO} =2.2130(2)$ fb.}. Furthermore, the top quark width is derived from the equations in Ref. \cite{Denner:2012yc} using NLO corrections as calculated in Ref.  \cite{Jezabek:1988iv}.  The top quark width is treated as a fixed parameter throughout this work. Its value corresponds to a fixed scale $\mu = m_t$, that characterises the top quark decays. The $\alpha_s(m_t)$ parameter is independent of $\alpha_s(\mu_0)$ that enters the matrix element calculations as well as PDFs, since the latter describes the dynamics of the whole process. The value of $\alpha_s(m_t)$ is taken from the CT10NLO PDF set. In this way we obtain the following values for the top quark width in the full off-shell case
\begin{equation}
	\Gamma_{t}^{\textrm{LO}} =  1.472886 \textrm{ GeV}\,, \quad \quad \quad
	\Gamma_{t}^{\textrm{NLO}} =  1.346449 \textrm{ GeV}\,,
\end{equation}
whereas for the NWA case they are given by
\begin{equation} \label{eq:width_nwa}
  \Gamma_{t,\,\textrm{NWA}}^{\textrm{LO}} =  1.495948 \textrm{ GeV}\,,
  \quad \quad \quad
	\Gamma_{t,\,\textrm{NWA}}^{\textrm{NLO}} =  1.367547 \textrm{ GeV}\,.
\end{equation}
Final state partons with pseudo-rapidity $|\eta|<5$ are clustered into jets using the IR-safe {\it anti}-$k_T$ jet algorithm \cite{Cacciari:2008gp} with the jet-resolution parameter $R=0.4$. We require two charged leptons, two $b$-jets, missing transverse momentum and one stable Higgs boson. All final states have to pass the following experimental cuts
\begin{equation}
	\begin{array}{lllll}
          p_{T,\,b}\,>\,25 \textrm{ GeV\,,}&\quad&\left|y_b\right|\,<\,2.5\,,
          &\quad&p_{T,\,miss}\,>\,20\textrm{ GeV}\,,
		\vspace{0.2cm}\\
          p_{T,\,\ell}\,>\,20 \textrm{ GeV\,,}
          &&\left|y_\ell\right|\,<\,2.5\,,&&
	\end{array}
\end{equation}
where $\ell=e,\,\mu$. The factorisation, $\mu_F$, and renormalisation, $\mu_R$, scale are set to a common value $\mu_F=\mu_R=\mu_0$. For $\mu_0$  we use the two scale settings defined in Ref. \cite{Denner:2015yca} 
\begin{equation}
	\mu_{fix}= m_t+\frac{m_H}{2}=236\textrm{ GeV}
\end{equation}
and
\begin{equation}
  \mu_{dyn}=\left(m_{T,\,t}\,m_{T,\,\bar{t}}\,m_{T,\,H}\right)^{\frac{1}{3}}
  \quad \quad \quad \quad
  \textrm{with}
  \quad \quad \quad \quad 
  m_T=\sqrt{m^2+p_T^2}.
\end{equation}
In addition, we use a third scale  setting $\mu_0=H_T/2$ with 
\begin{equation}
	H_T=p_{T,\,b_1}+p_{T,\,b_2}+p_{T,\,e^+}+p_{T,\,\mu^-}+p_{T,\,miss}+p_{T,\,H} \,,
\end{equation}
where $b_1$ and $b_2$ are the two bottom-flavoured jets with highest transverse momentum \footnote{
 Had we used  $m_{T, \, H}$ instead of $p_ {T, \, H}$ in the definition of $\mu_0= H_T/2$, our
theoretical predictions would be very similar. At the NLO level, both scales give almost identical results. The LO cross section is slightly smaller, which affects the ${\cal K}$-factor for the process. Moreover, the LO and NLO theoretical uncertainties from the scale variation are alike for both scale settings.}. This specification becomes relevant only when the contributions induced by the bottom-quark parton density are included as well. In that case, we can have up to three $b$-jets in the final state. On the other hand, $p_{T,\,miss}$ is the total missing transverse momentum from escaping neutrinos defined by
\begin{equation}
    p_{T,\,miss} = |\vec{p}_{T,\,\nu_e} + \vec{p}_{T,\,\bar{\nu}_\mu}|\,.
\end{equation}
The additional scale setting is chosen such that information about the underlying resonant nature of the process is not used at all. It seems a more natural choice for the process at hand where full off-shell effects are included, thus,  also Feynman diagrams with single- and non-resonant top-quark  contributions are present at the matrix element level. 
  As differential cross sections extend up to large energy scales, single- and non-resonant contributions might become significant, sometimes even dominant, in the specific phase-space regions, see e.g. Ref. \cite{Bevilacqua:2019quz}. Thus, a $H_T$ based scale setting is better motivated, especially for dimenionful observables, than any scale choice which assumes the double-resonant nature of the process for each phase-space point.  Furthermore, $\mu_0=H_T/2$ does not require  reconstruction of the $t\bar{t}$ pair which is not free from ambiguities. 
Nevertheless, for the integrated fiducial cross section, which is dominated by the double resonant top-quark contributions, $\mu_0=H_T/2$ leads to similar results as $\mu_0=\mu_{dyn}$ and $\mu_0=\mu_{fix}$. Theoretical uncertainties arising from missing higher orders are estimated using a $7$-point scale variation, in which the factorisation and renormalisation scales are varied independently in the range
\begin{equation}
	\frac{1}{2} \, \mu_0  \le \mu_R\,,\mu_F \le  2 \,  \mu_0\,, \quad
	\quad
	\quad \quad 
	\quad \quad \quad \quad \quad \quad \frac{1}{2}  \le
	\frac{\mu_R}{\mu_F} \le  2 \,.
\end{equation}
In practice, this amounts to considering the following pairs
\begin{equation}
	\label{scan}
	\left(\frac{\mu_R}{\mu_0}\,,\frac{\mu_F}{\mu_0}\right) = \Big\{
	\left(2,1\right),\left(0.5,1  
	\right),\left(1,2\right), (1,1), (1,0.5), (2,2),(0.5,0.5)
	\Big\} \,.
\end{equation}
As it is usually done, we  search for the minimum and maximum of the resulting cross sections.   For the PDF error, we use the  prescription of the NNPDF3.1 group to provide the $68\%$ confidence level (C.L.) PDF uncertainties. Specifically, the NNPDF3.1 group uses a Monte Carlo sampling method in conjunction with neural networks, and the PDF uncertainties are obtained with the help of the replica method. On the other hand, a Hessian representation is used for CT18 and MSHT20, where the PDF uncertainties of CT18 have to be re-scaled to obtain the $68\%$ C.L. since they are originally provided at $90\%$ C.L.

In the following, we summarise the additional parameters required for the decay of the Higgs boson. We use the masses given in Ref. \cite{deFlorian:2016spz} for the $\tau$ lepton and the bottom quark in the on-shell and $\overline{\textrm{MS}}$-scheme 
\begin{equation}
	m_{\tau}=1.77682\textrm{ GeV}, \qquad\quad
	m_b^{OS}=4.92\textrm{ GeV}, \qquad\quad
	\overline{m}_b(\overline{m}_b)=4.18\textrm{ GeV}.
\end{equation}
The Higgs width is calculated using the parameters described above and the program \textsc{Hdecay} \cite{Djouadi:1997yw,Djouadi:2018xqq},  which leads to
\begin{equation}
	\Gamma_H=4.226\cdot 10^{-3}\textrm{ GeV}.
\end{equation}
For the Yukawa coupling of the bottom quark, we need the $\overline{\textrm{MS}}$ mass for different scale choices. To this end, the running of the bottom quark mass with two-loop (one-loop) accuracy at NLO (LO) is used, see e.g. Ref. \cite{Harlander:2003ai}. Using the strong coupling constant from the NNPDF3.1 PDF set at NLO (LO), we obtain the following values for $\overline{m}_b$ for all scales required in this calculation
\begin{equation}
	\notag\overline{m}_b(m_H/2)=3.160804\textrm{ GeV}\,,\quad \overline{m}_b(m_H)=2.999774\textrm{ GeV}\,,\quad
	\overline{m}_b(2m_H)=2.860548\textrm{ GeV}\,,
\end{equation}
at LO and
\begin{equation}
	\notag\overline{m}_b(m_H/2)=2.977119\textrm{ GeV}\,,\quad \overline{m}_b(m_H)=2.805836\textrm{ GeV}\,,\quad
	\overline{m}_b(2m_H)=2.660844\textrm{ GeV}\,,
\end{equation}
at NLO, where we have used $\overline{m}_b(\overline{m}_b)$ as initial value. We use the same kinematical cuts as already discussed for all Higgs boson decay products, including $\tau$ leptons. In addition, we require the following cuts for the two photons
\begin{equation}
	\begin{array}{lllll}
          R_{\gamma\gamma}\,>\,0.3\,,
          &\quad&R_{\gamma \ell}\,>\,0.3\,,&\quad&R_{ \gamma b}\,>\,0.3\,,
		\vspace{0.2cm}\\
          p_{T,\,\gamma}\,>\,25 \textrm{ GeV\,,}
          &&\left|y_{\gamma}\right|\,<\,2.5\,,&&
	\end{array}
\end{equation}
and apply Frixione's photon-isolation condition as defined in Ref. \cite{Frixione:1998jh} before the jet algorithm. In particular,  for each parton $i$ we evaluate $R_{\gamma i}$ between this parton and the photon. We reject the event unless the following condition is fulfilled
\begin{equation}
	\sum_i E_{T,\, i} \,\Theta (R-R_{\gamma i})\leq \epsilon_{\gamma} \,E_{T,\, \gamma}\left(\frac{1-\textrm{cos}(R)}{1-\textrm{cos}(R_{\gamma, \,j})}\right)^n \quad\quad\quad\quad \forall R\leq R_{\gamma , \,j}\,,
\end{equation}
where  we set  $\epsilon_{\gamma}=1$, $n=1$ and $R_{\gamma, \,j}=0.3$. Furthermore, $E_{T,\,i}$  is the transverse energy of the parton $i$  and $E_{T,\gamma}$ is the transverse energy of the photon.  Jets reconstructed inside the cone size $R_{\gamma,\,j}$ are not subject to additional selection criteria. We note that the condition has to be  satisfied for both photons from the $H\to \gamma \gamma$ decay.

For Higgs boson decays we still use the following scale setting $ \mu_0 = H_T / 2$ for $\mu_R $ and $\mu_F $. However, now the transverse momentum of the Higgs boson  needs to be reconstructed from its decay products. In the case of $H\to \gamma \gamma, \tau^+\tau^-, e^+e^-e^+e^-$, the Higgs boson can be reconstructed exactly since we have no corrections in the decays. Only for the $H\to b\bar{b}$ decay do we need a Higgs reconstruction prescription. In particular, we minimise the following $Q_{i,j}$ value
\begin{equation}
\label{Higgs}
	Q_{i,j}= \left| M_{b_ib_j}-m_H \right|\,,
\end{equation}
where $i,j$ run over all $b$-jets  that are present, to find the two that originate from the Higgs boson decay. Morover, in Eq. \eqref{Higgs} $M_{b_ib_j}$ stands for the reconstructed Higgs mass and $m_H$ for the Higgs mass.  The remaining two $b$-jets are assigned to the top quark decays. We choose to set and vary the renormalisation scales independently for the production and decay parts. We evaluate the cross section for a total of $21$ different scale settings as described in Ref. \cite{Gauld:2019yng}. The latter are obtained from all possible combinations of
\begin{equation}
 \mu_F= \xi  \cdot \, \frac{H_T}{2}\,,    \quad \quad \quad \quad  \mu_{R} =  \xi \cdot \, \frac{H_T}{2}\,,    \quad \quad \quad \quad \mu_{R,\, dec.} =  \xi \cdot  \, m_H\,,
\end{equation}
where $\xi\in [0.5,1,2]$ with the additional constraint 
\begin{equation}
    \frac{1}{2} \le \frac{\mu_{R}}{\mu_{F}} \le 2 \,,
    \end{equation}
following the conventional $7$-point scale variation for the Higgs boson production process.

%
\section{Numerical checks}
\label{sec:tth-checks}
%
\begin{table}[t!]
\begin{center}
	\begin{tabular}{|cccc|}
		\hline
		$\mu_0$&& \textsc{Helac-NLO} & DF  \\ \hline
		&&&\\[-0.4cm]
		$\mu_{dyn}$&$\sigma_{\textrm{LO}}$ $[$fb$]$&$2.2659(2)^{+30.8\%}_{-22.0\%}$&$2.2656(1)^{+30.8\%}_{-22.0\%}$\\[0.2cm]
		&$\sigma_{\textrm{NLO}}$ $[$fb$]$&$2.654(2)^{+0.9\%}_{-4.7\%}$&$2.656(3)^{+0.9\%}_{-4.6\%}$\\[0.1cm]
		&$\mathcal{K}$&$1.171(1)$&$1.172(1)$\\ \hline
		&&&\\[-0.4cm]
		$\mu_{fix}$&$\sigma_{\textrm{LO}}$ $[$fb$]$&$2.2402(2)^{+31.0\%}_{-22.0\%}$&$2.2401(1)^{+31.0\%}_{-22.0\%}$\\[0.2cm]
		&$\sigma_{\textrm{NLO}}$ $[$fb$]$&$2.633(2)^{+0.6\%}_{-5.0\%}$&$2.633(3)^{+0.6\%}_{-5.0\%}$\\[0.1cm]
		&$\mathcal{K}$&$1.175(1)$&$1.176(1)$\\ \hline
	\end{tabular}
\end{center}
\caption{\label{tab:checks} \it
Integrated cross section at LO and NLO QCD for the $pp\to e^+\nu_e\mu^-\bar{\nu}_{\mu}b\bar{b}\,H$ process at the LHC with $\sqrt{s}=13\textrm{ TeV}$. Results are evaluated using $\mu_0=\mu_{dyn}$ and $\mu_0=\mu_{fix}$ for the CT10NLO PDF set. Also given is a comparison between \textsc{Helac-Nlo} and the results from Ref. \cite{Denner:2015yca} (indicated by \textrm{DF}).}
\end{table}

To verify our results for the $pp\to e^+\nu_e\mu^-\bar{\nu}_{\mu}b\bar{b}\,H$ process, we perform a comparison of our predictions with the results from Ref. \cite{Denner:2015yca} at the integrated and differential fiducial level. Since this calculation was performed with a completely independent computational framework, agreement of the results provides a strong indication of the correctness of the two calculations. In Table \ref{tab:checks}, the integrated fiducial cross section and the $\mathcal{K}\textrm{-factor}$, defined as $\mathcal{K}=\sigma_{\textrm{NLO}}/\sigma_{\textrm{LO}}$, is shown for both calculations at LO and NLO QCD. Theoretical predictions are provided  for $\mu_0=\mu_{dyn}$ and $\mu_0=\mu_{fix}$ for the CT10NLO PDF set. Also reported are theoretical uncertainties as obtained by $7$-point scale variation together with  Monte Carlo errors. We find good agreement between the two calculations at LO and NLO QCD for both scale settings. In particular, the $\mathcal{K}\textrm{-factor}$ and the scale uncertainties are also in agreement within the statistical uncertainties. The scale uncertainty comparison is an indirect comparison of different scale choices implied by the scale variation.
\begin{figure}[htbp]
	\begin{center}
		\includegraphics[width=0.45\textwidth]{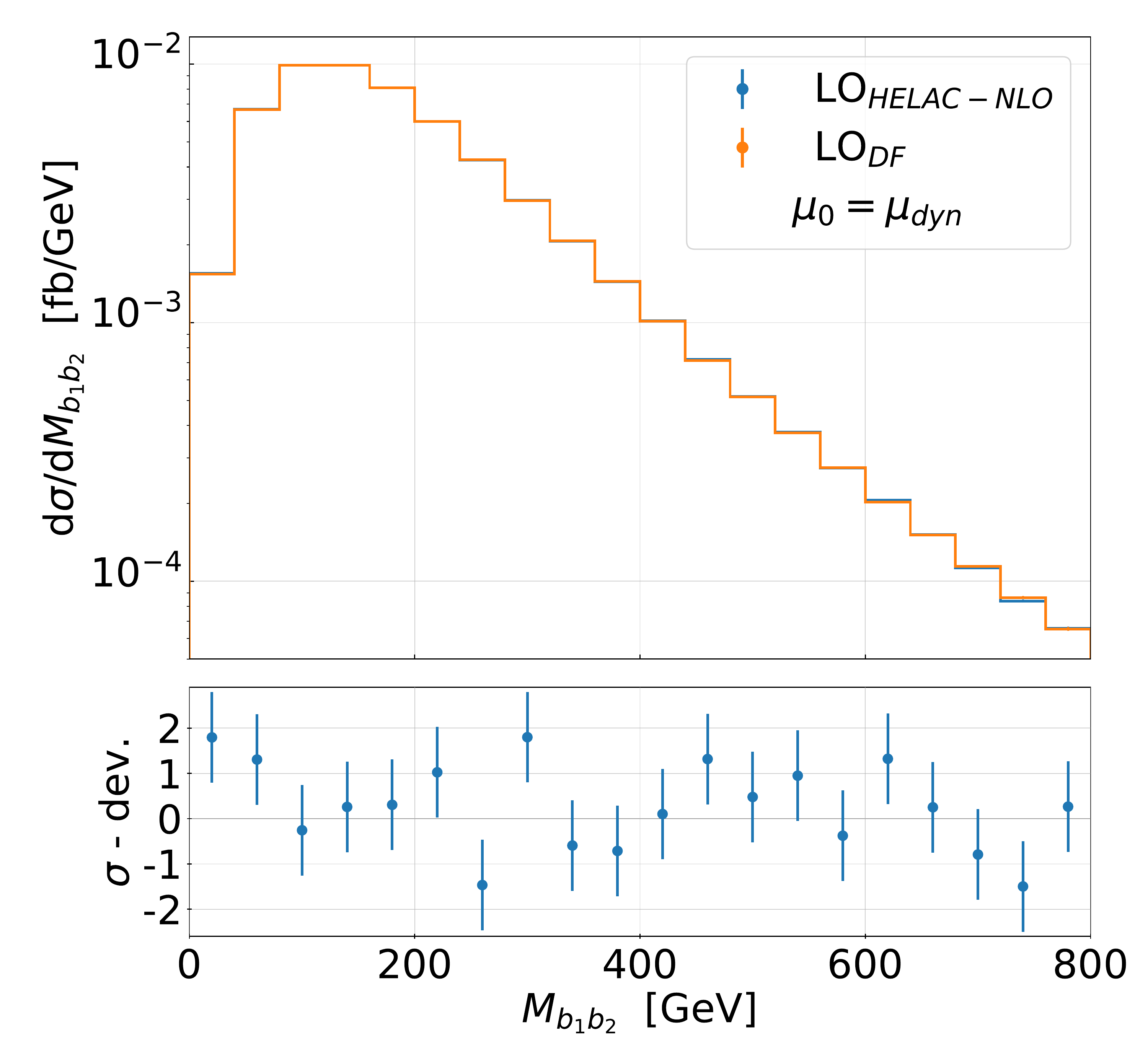}
		\includegraphics[width=0.45\textwidth]{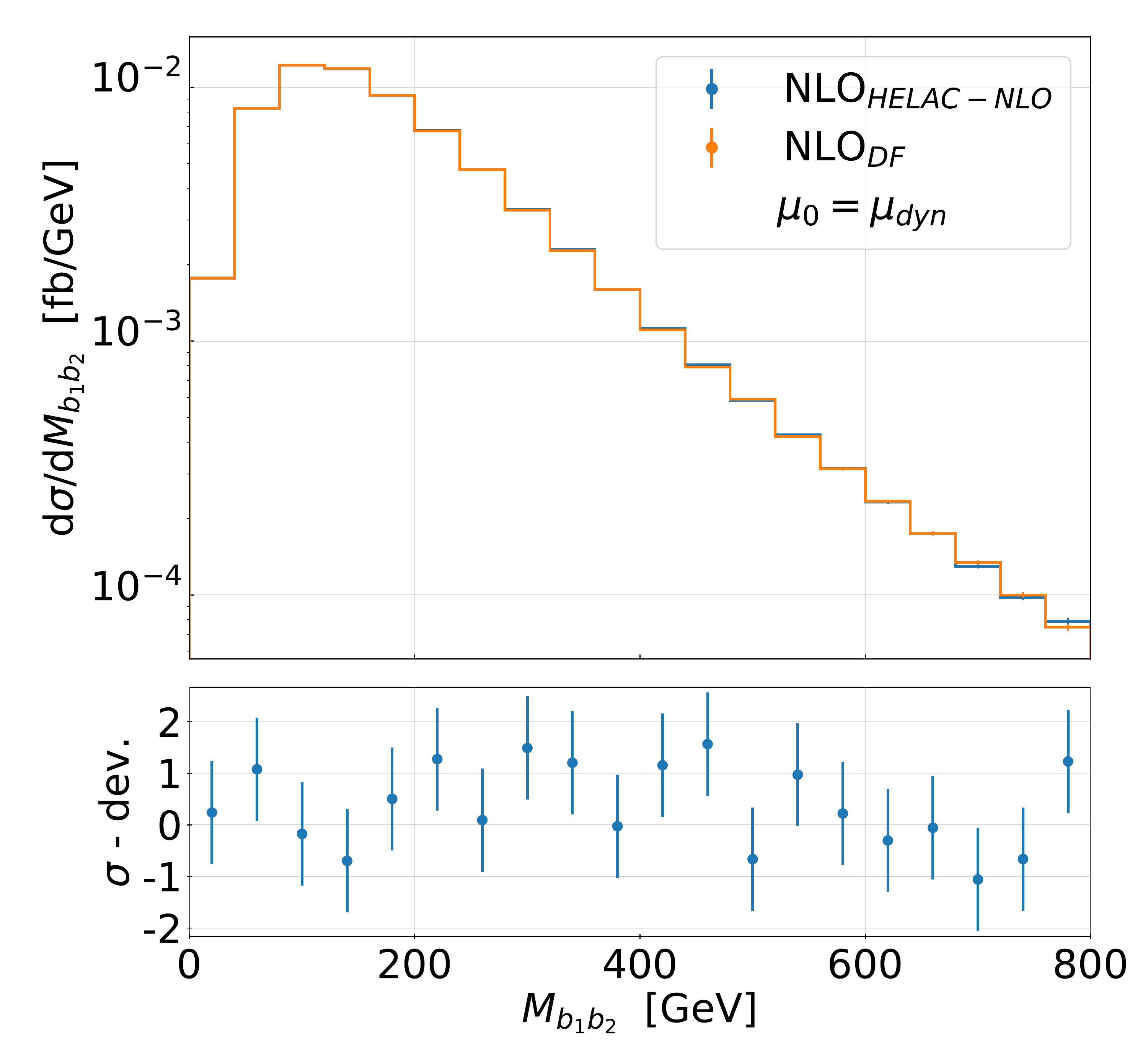}
		\includegraphics[width=0.45\textwidth]{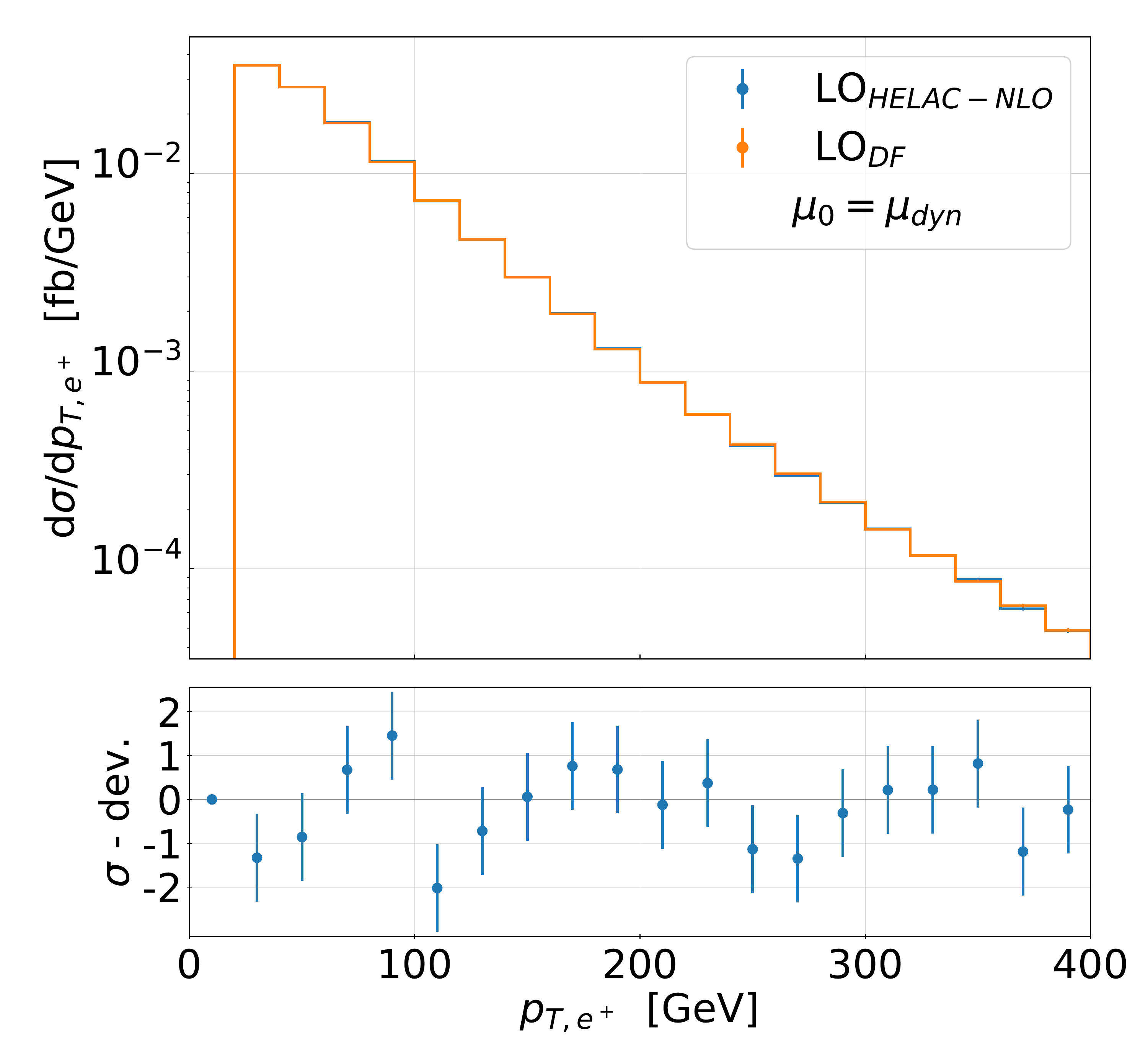}
		\includegraphics[width=0.45\textwidth]{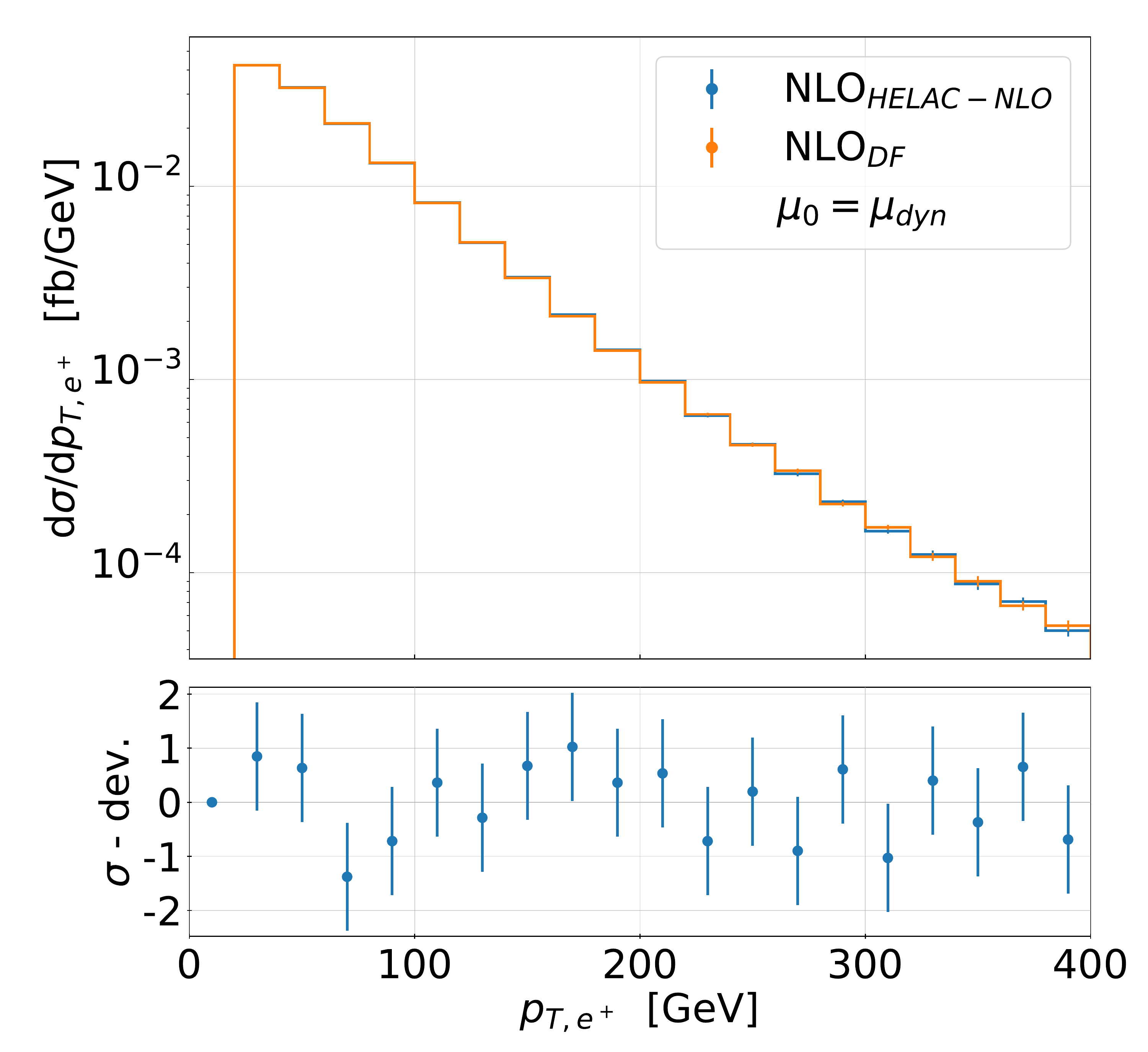}
		\includegraphics[width=0.45\textwidth]{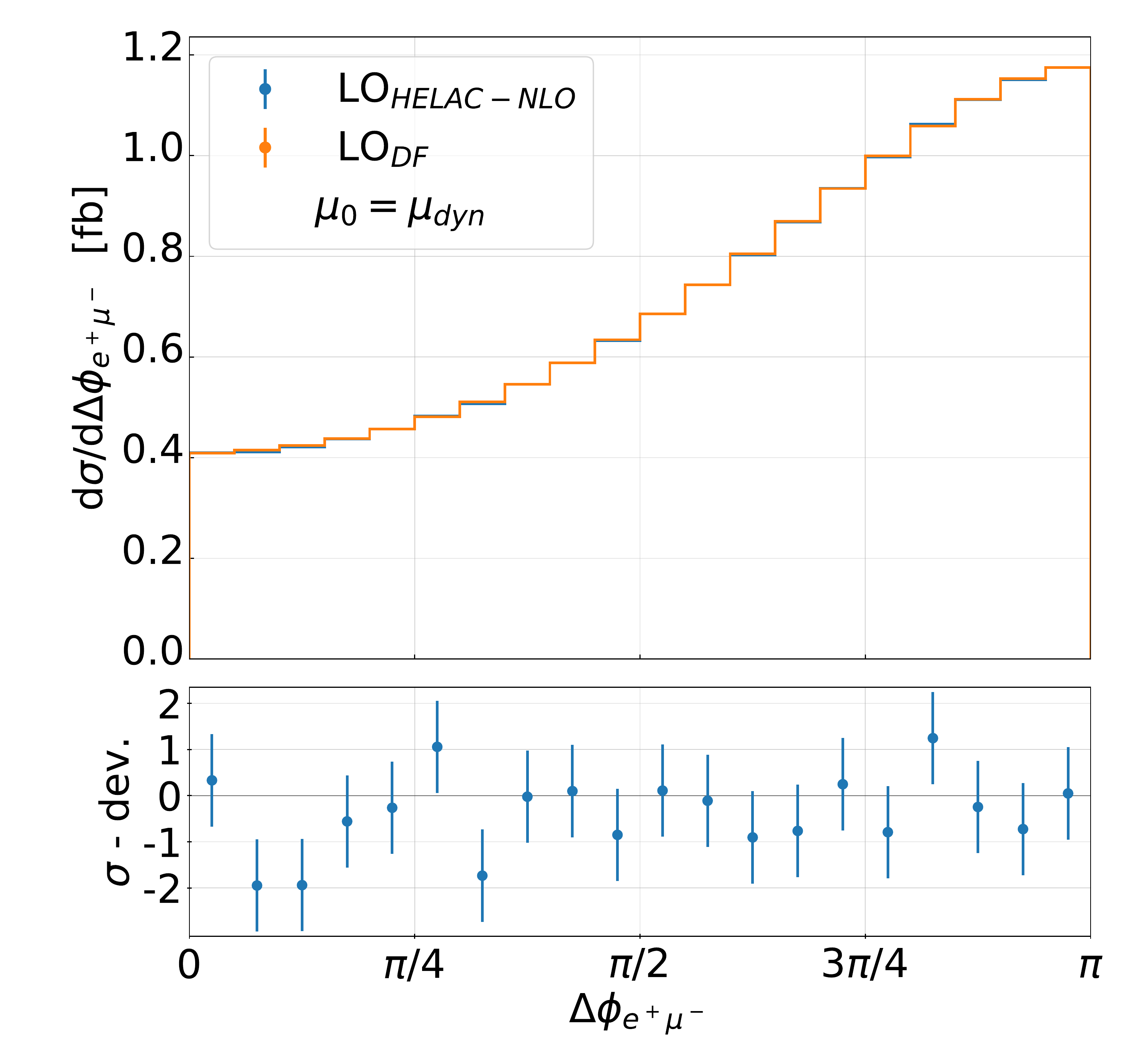}
		\includegraphics[width=0.45\textwidth]{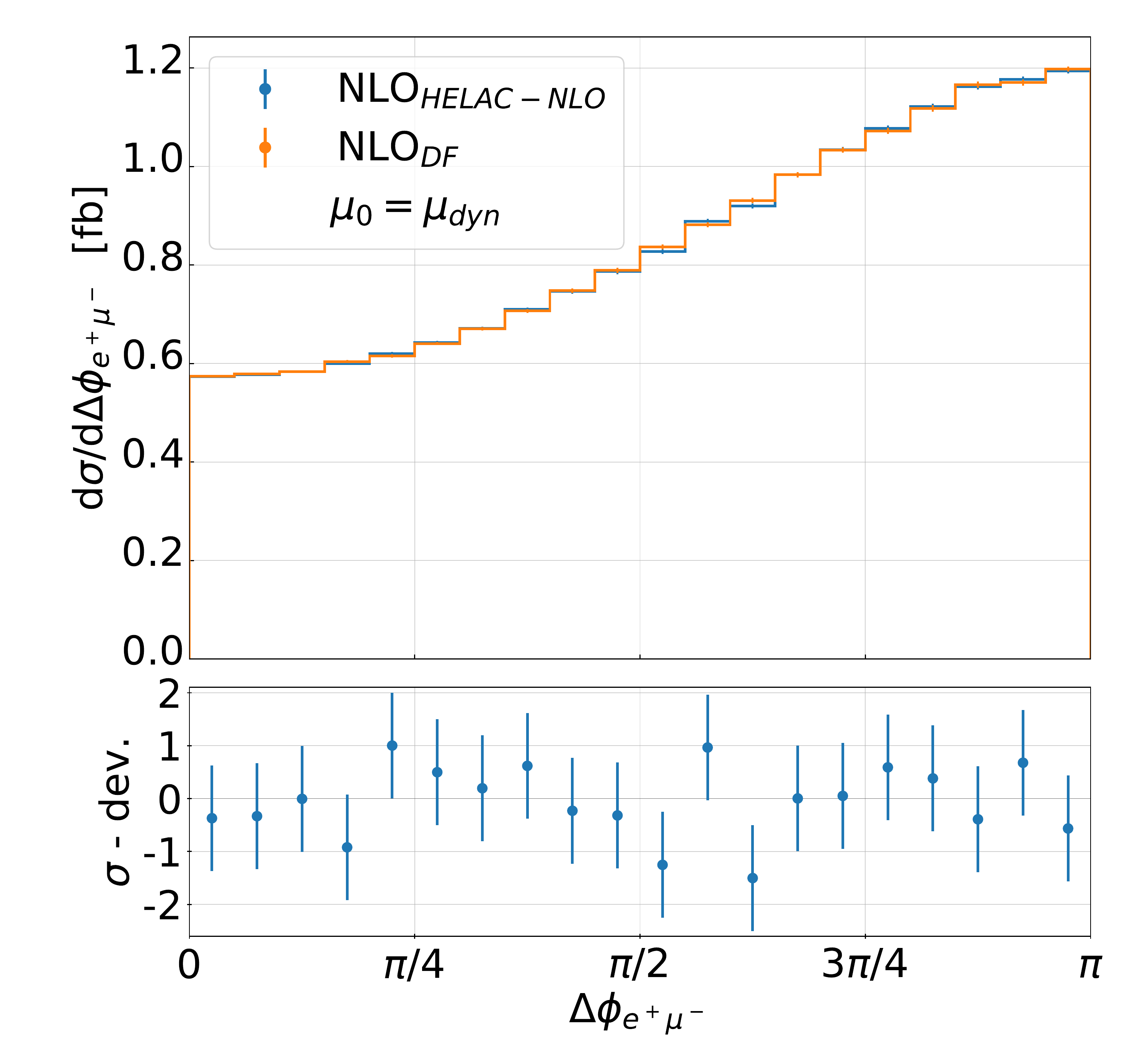}
	\end{center}
\caption{\label{fig:checks} \it
Differential distributions for the observables $M_{b_1b_2}$, $p_{T,e^+}$ and $\Delta\phi_{e^+\mu^-}$ at LO and NLO QCD for the $pp\to e^+\nu_e\mu^-\bar{\nu}_{\mu}b\bar{b}\,H$ process at the LHC with $\sqrt{s}=13\textrm{ TeV}$. Results are evaluated using  $\mu_0=\mu_{dyn}$ and the CT10NLO PDF set. Also given is a comparison between \textsc{Helac-Nlo} and the results from Ref. \cite{Denner:2015yca} (indicated by \textrm{DF}). Lower panels display the difference between both calculations in standard deviations (denoted as $\sigma\textrm{-dev.}$).}
\end{figure}

In Figure \ref{fig:checks}, we perform a similar comparison at the differential fiducial level for the invariant mass of the $b\bar{b}$ system, $M_{b_1b_2}$, the transverse momentum of the positron, $p_{T,e^+}$ and the difference in
azimuthal angle between the two charged leptons, $\Delta \phi_{e^+\mu^-}= |\phi_{e^+}-\phi_{\mu^-}|$ at LO (left) and NLO (right). We employ $\mu_0=\mu_{dyn}$ and the CT10NLO PDF set. For each observable, bin-by-bin, we display in the lower panels the difference between the two calculations in standard deviations (denoted as $\sigma\textrm{-dev.}$). The numerical values of Ref. \cite{Denner:2015yca} were obtained using the online tool \textsc{WebPlotdigitizer} \cite{Rohatgi2020}. This tool can be used for the semi-automatic extraction of numerical values from images of data visualisations which makes an extraction of a large number of data points fairly easy. Potential uncertainties due to the extraction are neglected and the same Monte Carlo errors as in our calculation are assumed. Again we find overall good agreement between the two calculations at LO and NLO. 

%
\section{Integrated fiducial cross sections}
\label{sec:tth-int}
%

In this section, we first analyze the two main sources of theoretical uncertainties resulting from the scale dependence and the PDF uncertainties at the integrated fiducial level. We extend the discussion of the scale dependence of Ref. \cite{Denner:2015yca} by an additional dynamical scale setting $\mu_0=H_T/2$ and investigate the differences between a fixed and dynamical scale choice. Second, we present results for the integrated cross section for more recent PDF sets and discuss the internal PDF uncertainties and the differences between the results for various PDF sets.
\begin{table}[t!]
\begin{center}
	\begin{tabular}{|ccccc|}
		\hline
		$\mu_0$&PDF&$\sigma_{\textrm{LO}}$&$\sigma_{\textrm{NLO}}$&$\mathcal{K}$\\
		&&$[$fb$]$&$[$fb$]$&\\ \hline
		&&&&\\[-0.4cm]
		$H_T/2$&NNPDF3.1&$2.2130(2)^{+30.1\%}_{-21.6\%}$&$2.728(2)^{+1.1\%}_{-4.7\%}$&$1.23$\\ [0.1cm] 
		\hline
		&&&&\\[-0.4cm]
		$\mu_{fix}$&NNPDF3.1&$2.3005(2)^{+30.8\%}_{-21.9\%}$&$2.731(2)^{+0.6\%}_{-5.4\%}$&$1.19$\\ [0.1cm] 
		\hline
		&&&&\\[-0.4cm]
		$\mu_{dyn}$&NNPDF3.1&$2.3320(2)^{+30.7\%}_{-21.9\%}$&$2.754(2)^{+0.9\%}_{-5.1\%}$&$1.18$\\ [0.1cm]  
		\hline
	\end{tabular}
\end{center}
\caption{\label{tab:scale} \it
Integrated fiducial cross section at LO and NLO QCD for the $pp\to e^+\nu_e\mu^-\bar{\nu}_{\mu}b\bar{b}\,H$ process at the LHC with $\sqrt{s}=13\textrm{ TeV}$. Results are presented  for $\mu_0=H_T/2$, $\mu_0=\mu_{fix}$ and $\mu_0=\mu_{dyn}$. The NNPDF3.1 PDF set is employed.}
\end{table}

In Table \ref{tab:scale}, we show the integrated cross section at LO and NLO QCD together with the corresponding $\mathcal{K}\textrm{-factor}$ for all three scale choices for the NNPDF3.1 PDF set. The deviations between the three scales are up to $5\%$ at LO and less than $1\%$ at NLO. The deviations are well within the scale uncertainties at the specific order in perturbation theory. Estimating the theoretical uncertainties by using the maximum of the lower and upper bounds from the scale variation, we find that they are very similar in size for all three scale settings. The scale dependence reduces drastically from about $30\%$ at LO to $5\%$ at NLO, which corresponds to a reduction by a factor of $6$ \footnote{ In the case of truly asymmetric uncertainties the errors are often symmetrised. After symmetrisation the scale uncertainty at LO does not change substantially, i.e. it is reduced down to $26\%$. However, at the NLO in QCD the reduction is considerable as far as $3\%$. Therefore, in this case by going from LO to NLO we have reduced the theoretical error by a factor of $9$.}. Had we varied the factorisation and the renormalisation scale simultaneously instead, the maximum of the lower and upper bound of the scale variation would not change. The $\mathcal{K}\textrm{-factor}$ ranges from $1.18$ to $1.23$. The largest value is observed for $\mu_0=H_T/2$ and is caused by the smaller value of the integrated cross section at LO. In summary, all three scales lead to very similar results at the integrated level and, in particular, we do not observe any differences between a fixed and a dynamical scale setting, which is not surprising as the latter should mainly play an important role at the differential level.
\begin{figure}[t!]
	\begin{center}
		\includegraphics[width=0.48\textwidth]{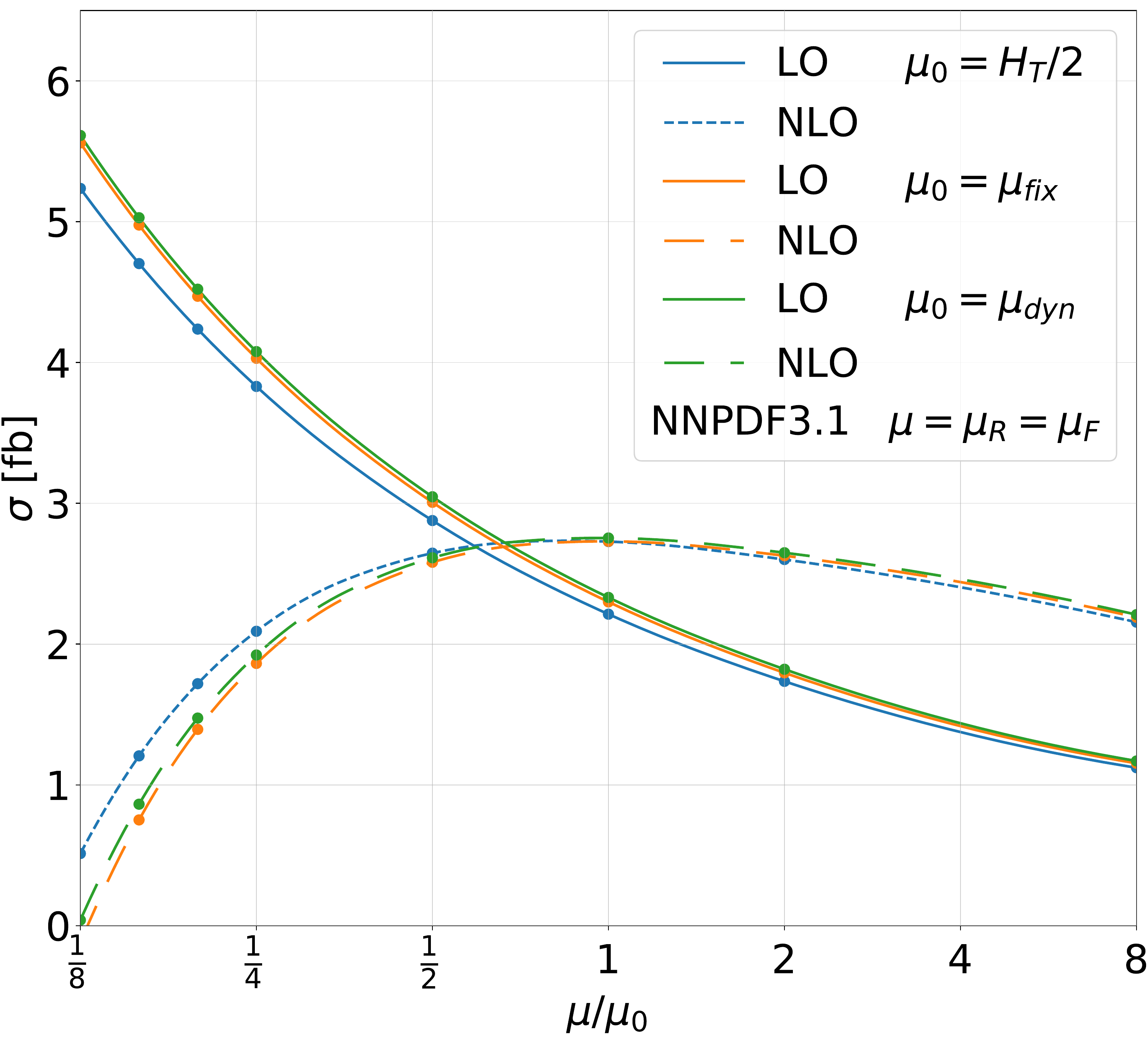}
		\includegraphics[width=0.48\textwidth]{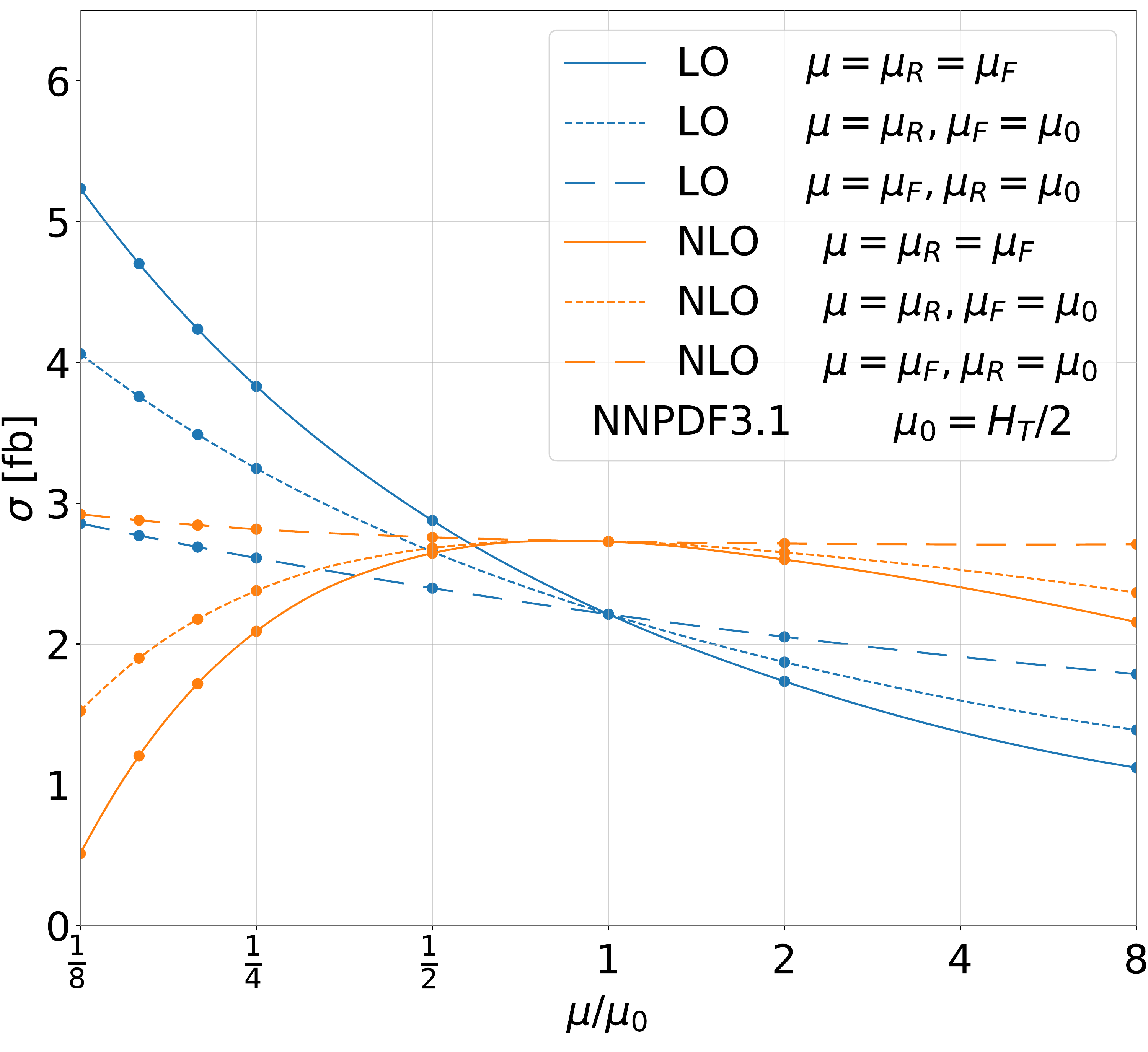}
		\includegraphics[width=0.48\textwidth]{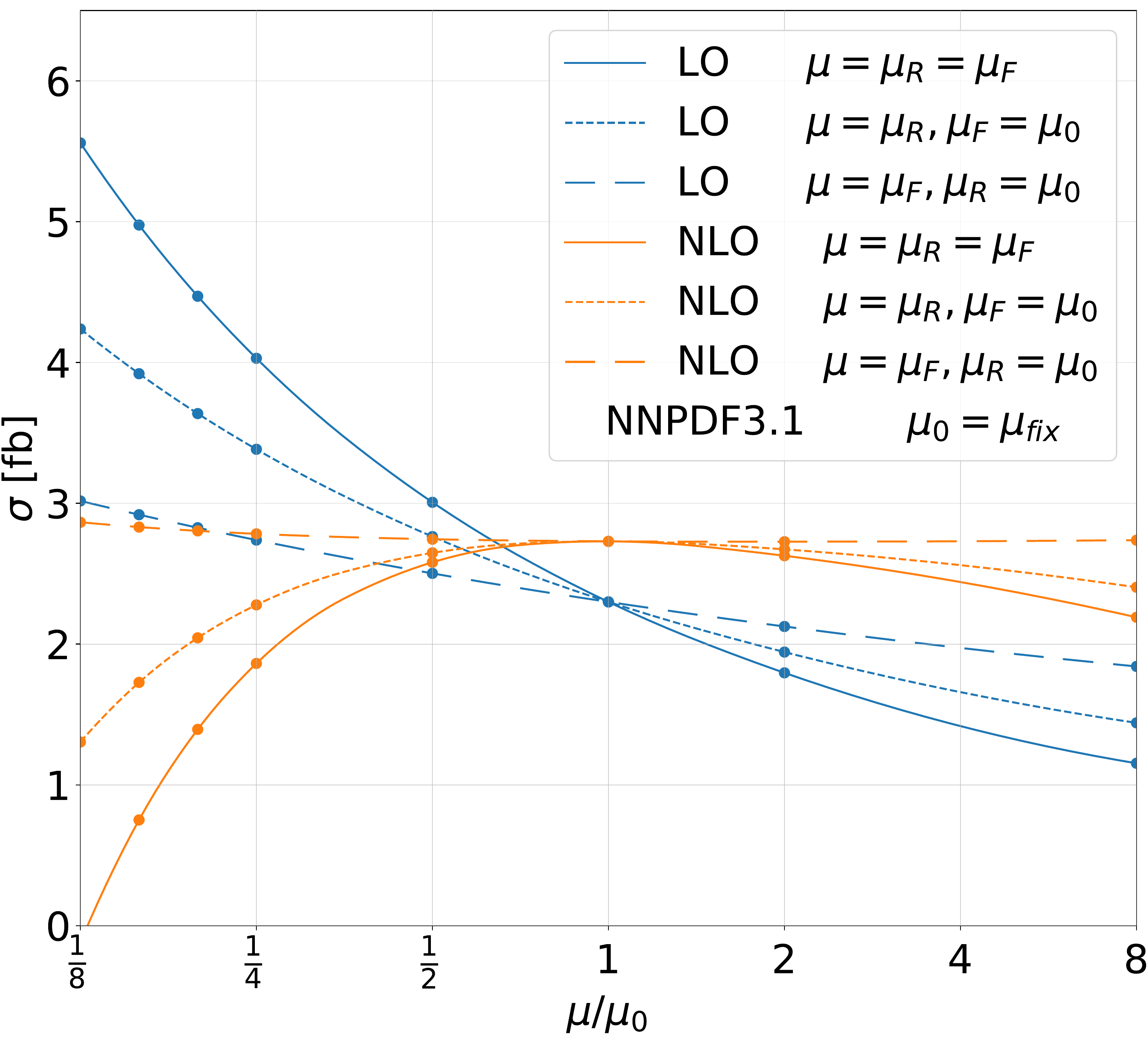}
		\includegraphics[width=0.48\textwidth]{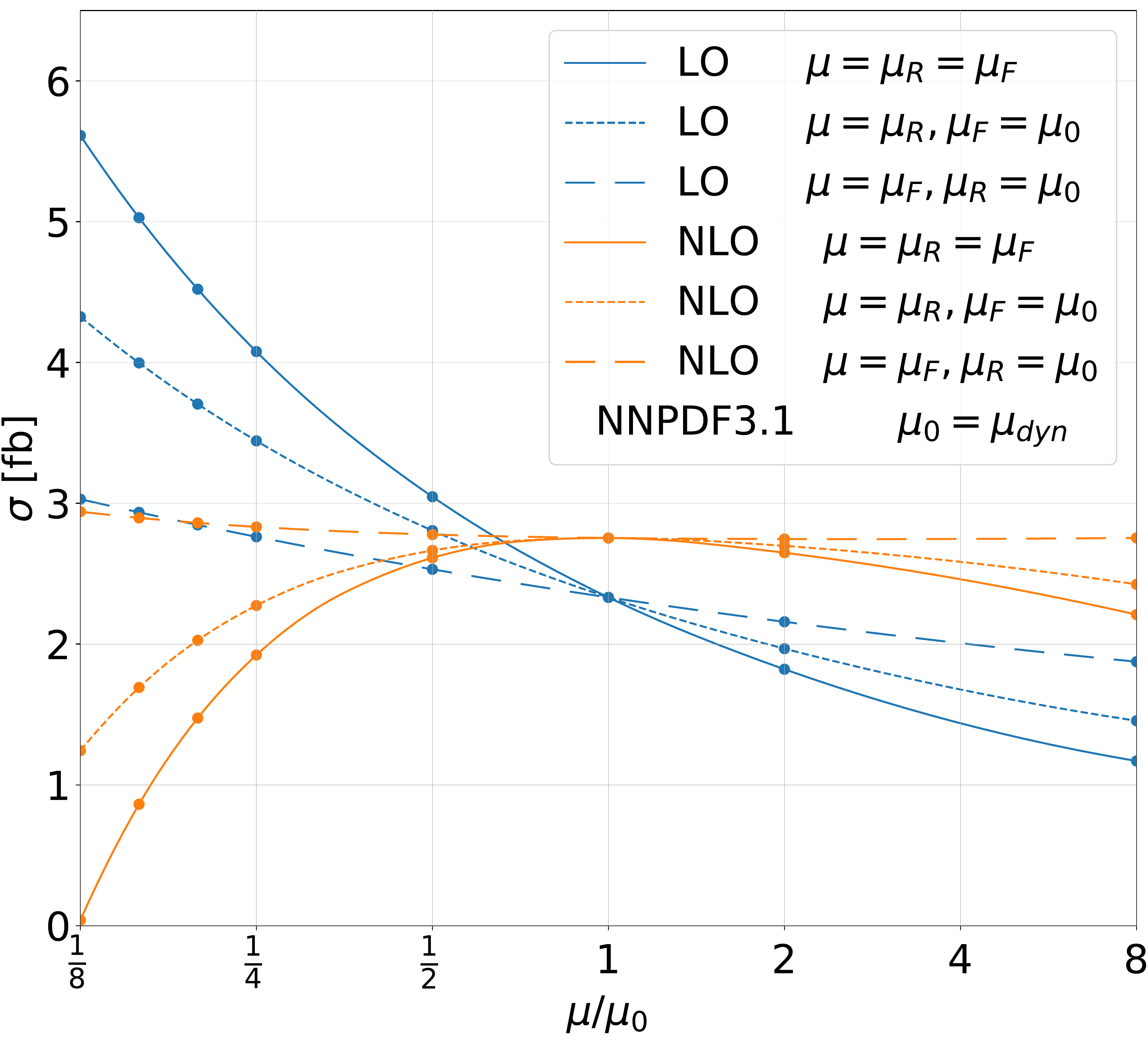}
	\end{center}
	\caption{\label{fig:int-scale} \it
		Scale dependence of the integrated fiducial cross section at LO and NLO QCD for the $pp\to e^+\nu_e\mu^-\bar{\nu}_{\mu}b\bar{b}\,H$ process at the LHC with $\sqrt{s}=13\textrm{ TeV}$. Results are given  for $\mu_0=H_T/2$, $\mu_0=\mu_{fix}$ and $\mu_0=\mu_{dyn}$ as well as for the NNPDF3.1 PDF set.}
\end{figure}

In Figure \ref{fig:int-scale}, the integrated fiducial cross section at LO and NLO in QCD is shown as a function of the factorisation/renormalisation scale for all three scale choices for the NNPDF3.1 PDF set. In the first plot, the integrated cross section is shown for the simultaneous variation of the factorisation and renormalisation scale. The remaining three plots display the simultaneous variation and the variation of one scale, factorisation or renormalisation, while the other scale is kept fixed at the corresponding central value, $\mu_0$. When both scales are varied at the same time, we observe that for $\mu_0=\mu_{fix}$ and $\mu_0=\mu_{dyn}$, the scale dependence is almost identical over the entire range both at LO and NLO. For the fixed scale, we observe that the integrated cross section becomes negative for small values at NLO. However, the choice of the  PDF set plays a role since this behaviour is not observed for CT10NLO. For $\mu_0=H_T/2$, our default scale choice, we find that at LO this scale setting leads to smaller values of the integrated cross section, where the deviations to the other two scales decrease for larger values. At NLO, all three scales are very similar in the range from $0.5$ to $2$ where  the theoretical uncertainties arising from the scale dependence are actually evaluated. We can further notice that in this range, the dependence on the factorisation scale is constant. Thus, the scale variation is driven solely by the changes in $\mu_R$. Consequently, the 3-point and 7-point scale variation would lead to essentially the same results.
\begin{figure}[t!]
	\begin{center}
		\includegraphics[width=0.98\textwidth]{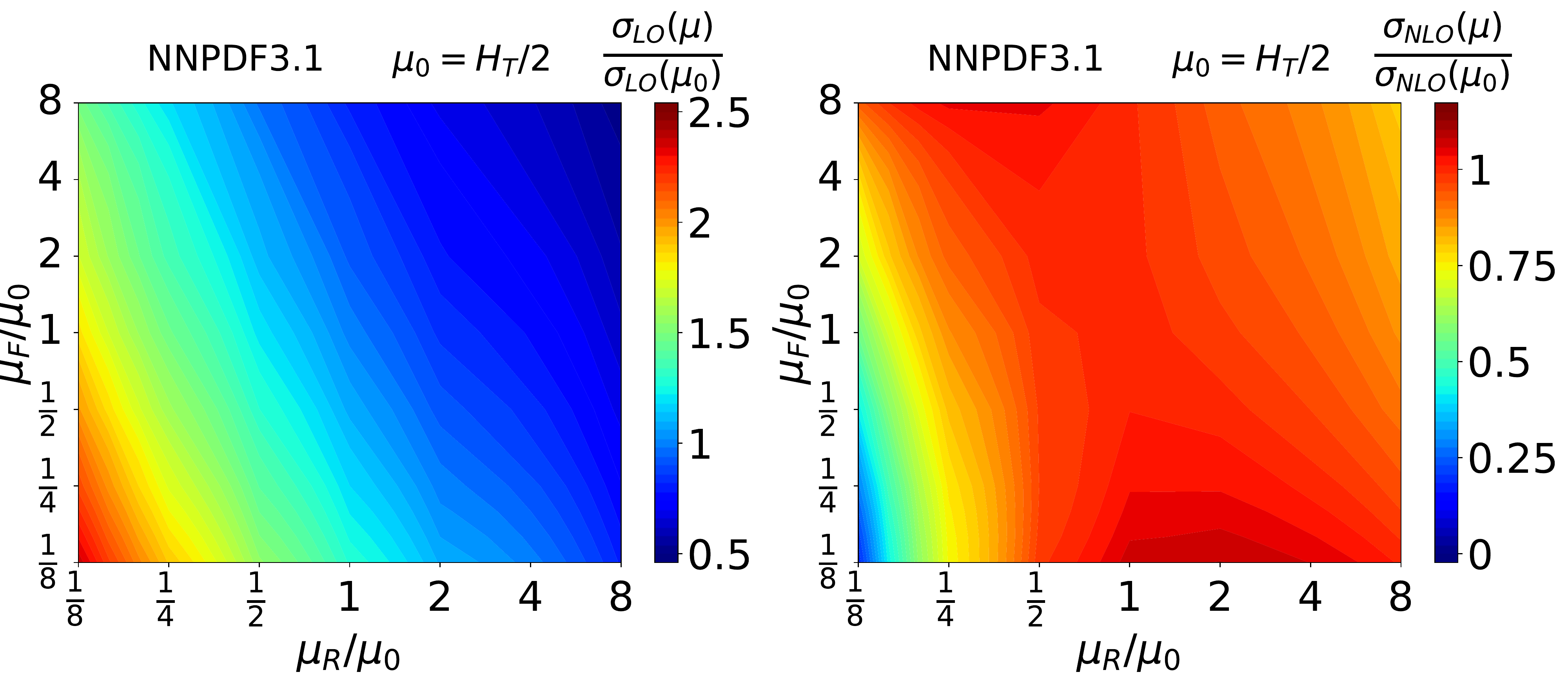}
		\includegraphics[width=0.98\textwidth]{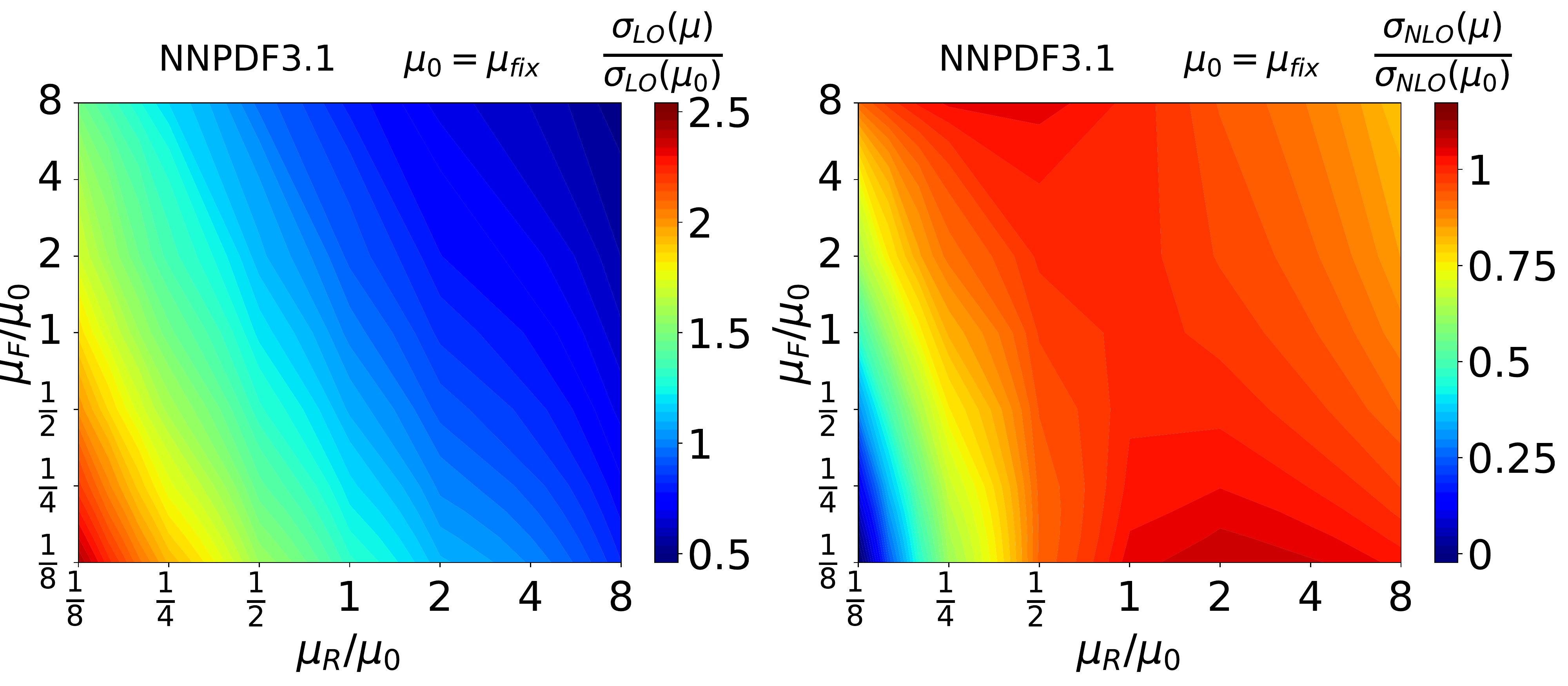}
	\end{center}
	\caption{\label{fig:int-scale2} \it
		Contour plots for the scale dependence in the $\mu_R - \mu_F$ plane at LO (left) and NLO (right) for the $pp\to e^+\nu_e\mu^-\bar{\nu}_{\mu}b\bar{b}\,H$ process at the LHC with $\sqrt{s}=13\textrm{ TeV}$. Results are given for $\mu_0=H_T/2$ and $\mu_0=\mu_{fix}$ as well as for the NNPDF3.1 PDF set.}
\end{figure}

In Figure \ref{fig:int-scale2}, the scale dependence is shown for the normalised integrated cross section at LO and NLO in the $\mu_R-\mu_F$ plane as a contour plot for $\mu_0=H_T/2$ and $\mu_0=\mu_{fix}$. As the scale dependence of $\mu_0=\mu_{fix}$ and $\mu_0=\mu_{dyn}$ is almost identical, the latter one is not shown again. This graphical visualisation allows a more detailed look at the scale dependence and, in particular, at the interplay between the changes in $\mu_R$ and $\mu_F$. At LO and NLO, we again observe that changes in the renormalisation scale dominate the scale dependence. The latter is reduced significantly when moving from LO to NLO. In addition, at LO, we find only minor differences in the scale dependence between the two scales, while at NLO, larger effects can be found, especially for small values of the scales. 
\begin{table}[t!]
\begin{center}
	\begin{tabular}{|cccc|}
		\hline
		PDF set&$\sigma_{\rm NLO}$ $[$fb$]$&$\delta_{scale}$&$\delta_{\rm PDF}$\\ \hline
		&&&\\[-0.4cm]
		NNPDF3.1&2.728(2)&${}^{+0.030\, (1.1\%)}_{-0.127\, (4.7\%)}$&${}^{+0.030\, (1.1\%)}_{-0.030\, (1.1\%)}$\\[0.2cm] 
		CT18&2.660(2)&${}^{+0.029\, (1.1\%)}_{-0.121\, (4.6\%)}$&${}^{+0.065\, (2.4\%)}_{-0.062\, (2.3\%)}$\\[0.2cm]
		MSHT20&2.689(2)&${}^{+0.030\, (1.1\%)}_{-0.123\, (4.6\%)}$&${}^{+0.050\, (1.9\%)}_{-0.042\, (1.6\%)}$\\[0.1cm] \hline
	\end{tabular}
\end{center}
\caption{\label{tab:pdf} \it
	Integrated fiducial cross section at NLO in QCD for the $pp\to e^+\nu_e\mu^-\bar{\nu}_{\mu}b\bar{b}\,H$ process at the LHC with $\sqrt{s}=13\textrm{ TeV}$. Results are given  for $\mu_0=H_T/2$ as well as for NNPDF3.1, CT18 and MSHT20 PDF sets.}
\end{table}

In addition to the scale uncertainties, it is important to study the second main source of theoretical uncertainties, that comes from the parameterization of PDF sets. We should note that there has been a lot of progress in determining the PDFs in the last few years. For example, various new NNLO processes have  recently been included in PDF fits. In addition, all groups included large sets of LHC data, which resulted in significant changes to the global PDF fits. With increased statistical precision of many measurements, challenges have arisen in fitting some of these new data sets. Moreover, differences in the fitting methodology, e.g. due to the treatment of correlated systematics,  had a significant impact on the PDFs. {Due to these advances, differences might emerge in both the central values and the uncertainties of the PDFs extracted by different groups. Consequently, it is vital to provide results for various PDF sets  and study their internal uncertainties. In this way,  it is possible to show the size of these differences, if any, and determine whether they are significant.  In Table \ref{tab:pdf} we provide the integrated fiducial cross section at NLO in QCD for  NNPDF3.1, CT18 and MSHT20. The results are given for $\mu_0=H_T/2$. Also shown are the corresponding scale and PDF uncertainties. We observe that the NNPDF3.1 PDF set yields the smallest PDF uncertainties of about $1.1\%$ compared to $2.4\%$ for CT18 and $1.9\%$ for MSHT20. The PDF uncertainties are about a factor of $2-4$ smaller than the corresponding scale uncertainties. Thus, scale uncertainty is still the dominant source of theoretical error. Furthermore, all three PDF sets lead to very similar predictions at the integrated level. Indeed, when comparing $\sigma_{\rm NLO}$ for NNPDF3.1, CT18 and MSHT20, we find the largest differences between NNPDF3.1 and CT18 of about $2.5\%$. We conclude that the PDF's internal uncertainties are similar in size to the differences between the three PDF sets.
\begin{figure}[t!]
	\begin{center}
		\includegraphics[width=0.90\textwidth]{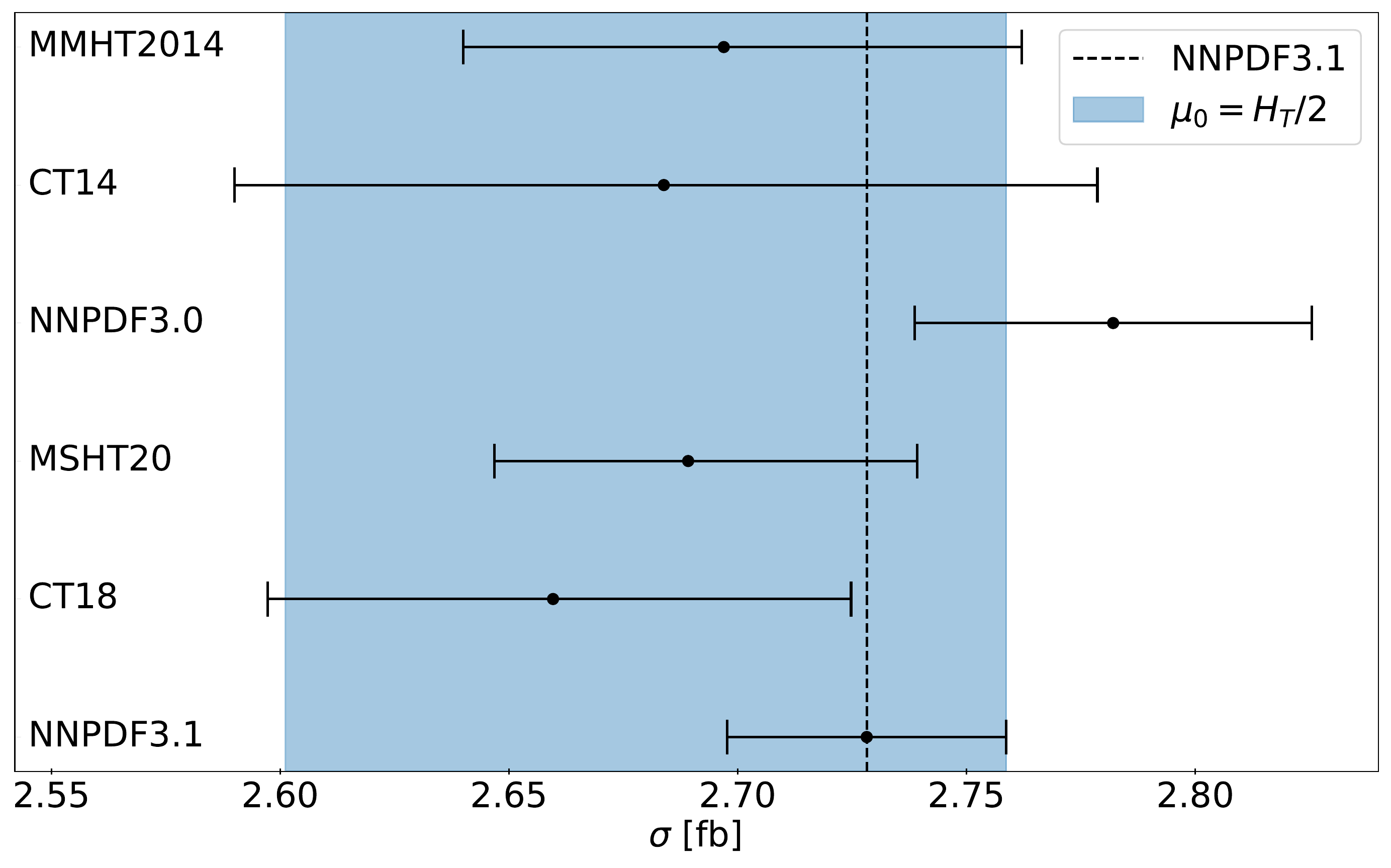}
	\end{center}
	\caption{\label{fig:int-pdf} \it
		Integrated fiducial cross section at NLO in QCD for the $pp\to e^+\nu_e\mu^-\bar{\nu}_{\mu}b\bar{b}\,H$ process at the LHC with $\sqrt{s}=13\textrm{ TeV}$. Results are presented  for $\mu_0=H_T/2$ as well as for various PDF sets. Also given are internal PDF uncertainties. Scale uncertainties are reported for the NNPDF3.1 PDF set and are shown as a blue band.}
\end{figure}

In Figure \ref{fig:int-pdf}, the integrated cross section is shown at NLO QCD for $\mu_0=H_T/2$ for the three previously used PDF sets. In addition, we provide theoretical results for their former versions, NNPDF3.0 \cite{Ball:2014uwa}, CT14 \cite{Dulat:2015mca} and MMHT14 \cite{Harland-Lang:2014zoa}. For each PDF set internal uncertainties are shown in black. For NNPDF3.1 the scale uncertainties are also shown in blue as reference. We find for all three groups that the former and current PDF sets agree well  within their corresponding PDF uncertainties. The latter are reduced by about $1\%$ for CT18 and by about $0.5\%$ for MSHT20 and NNPDF3.1. We find overall good agreement between the different PDF sets, with the largest deviations between CT18 and NNPDF3.0 of about $1.6\sigma$ or $4.4\%$, which is about the same size as the scale uncertainties and thus not negligible. The largest PDF uncertainties are found for CT14 at about $3.5\%$, which is still smaller but of comparable size to the scale uncertainties. 

%
\section{Differential fiducial cross sections and
PDF uncertainties}
\label{sec:tth-diff}
%

We start this section with a comparison between the different scale settings at the differential level. We then continue the discussion of NLO QCD corrections and perturbative stability. At the integrated fiducial level, we have concluded that the scale choice has only a minor impact on the cross section. At the differential level, however, a fixed scale choice cannot handle the dynamics of the process in certain phase-space regions, leading to perturbative instabilities. It is well known that this problem can be solved with a judicious choice of a dynamical scale. This aspect is discussed below for various observables. We end this section with a comparison of the two main theoretical uncertainties.
\begin{figure}[t!]
	\begin{center}
		\includegraphics[width=0.49\textwidth]{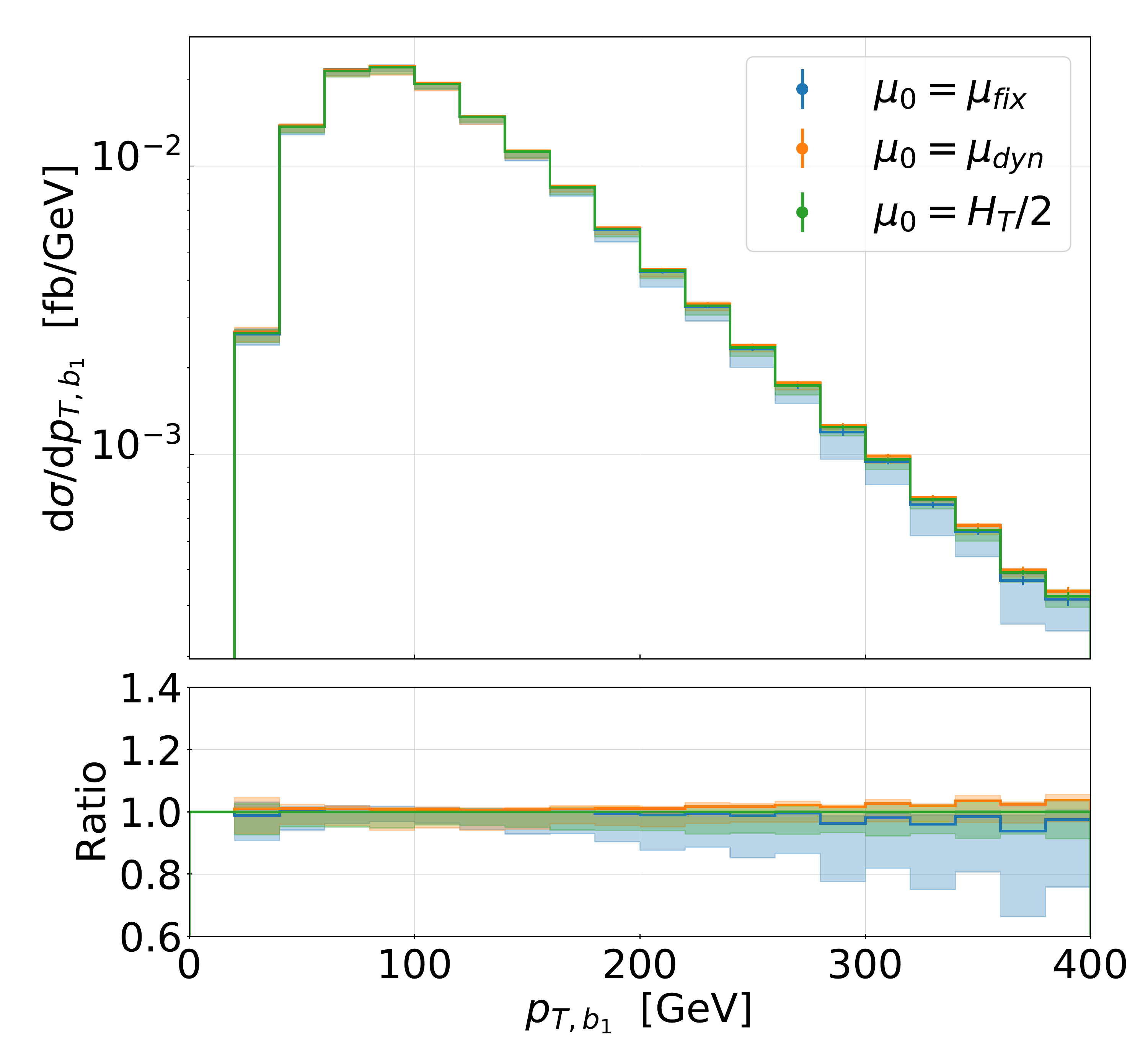}
		\includegraphics[width=0.49\textwidth]{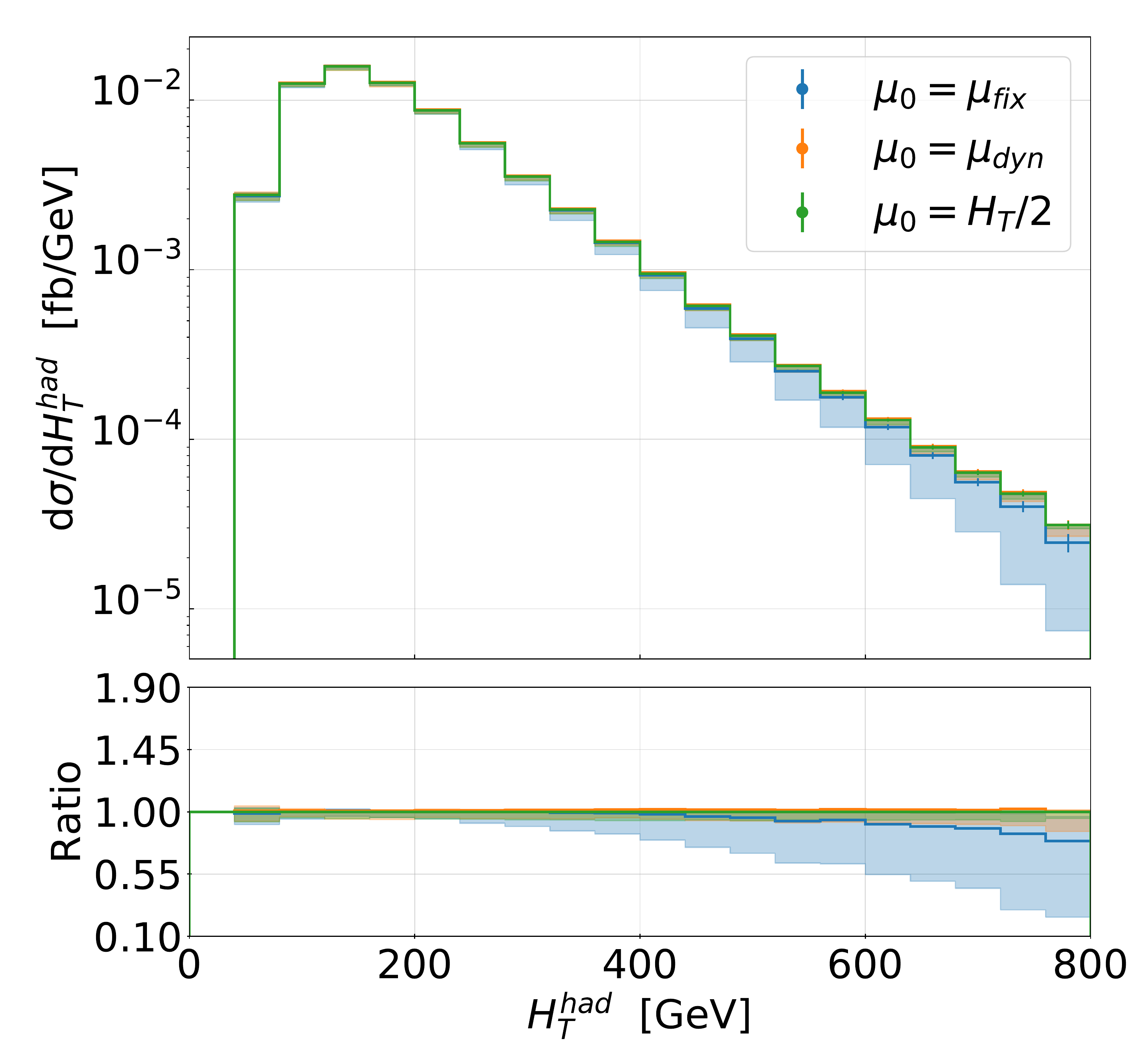}
		\includegraphics[width=0.49\textwidth]{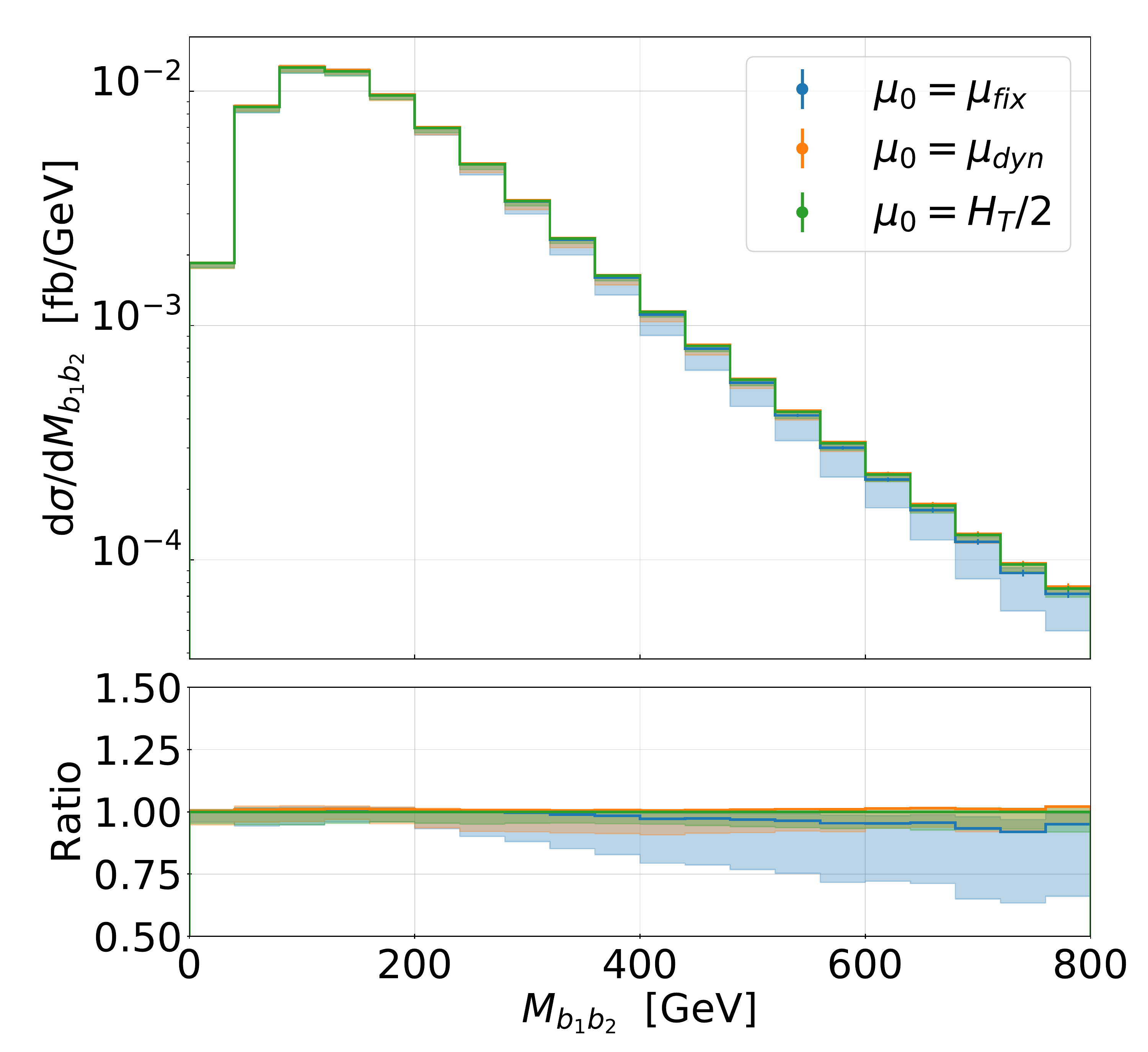}
		\includegraphics[width=0.49\textwidth]{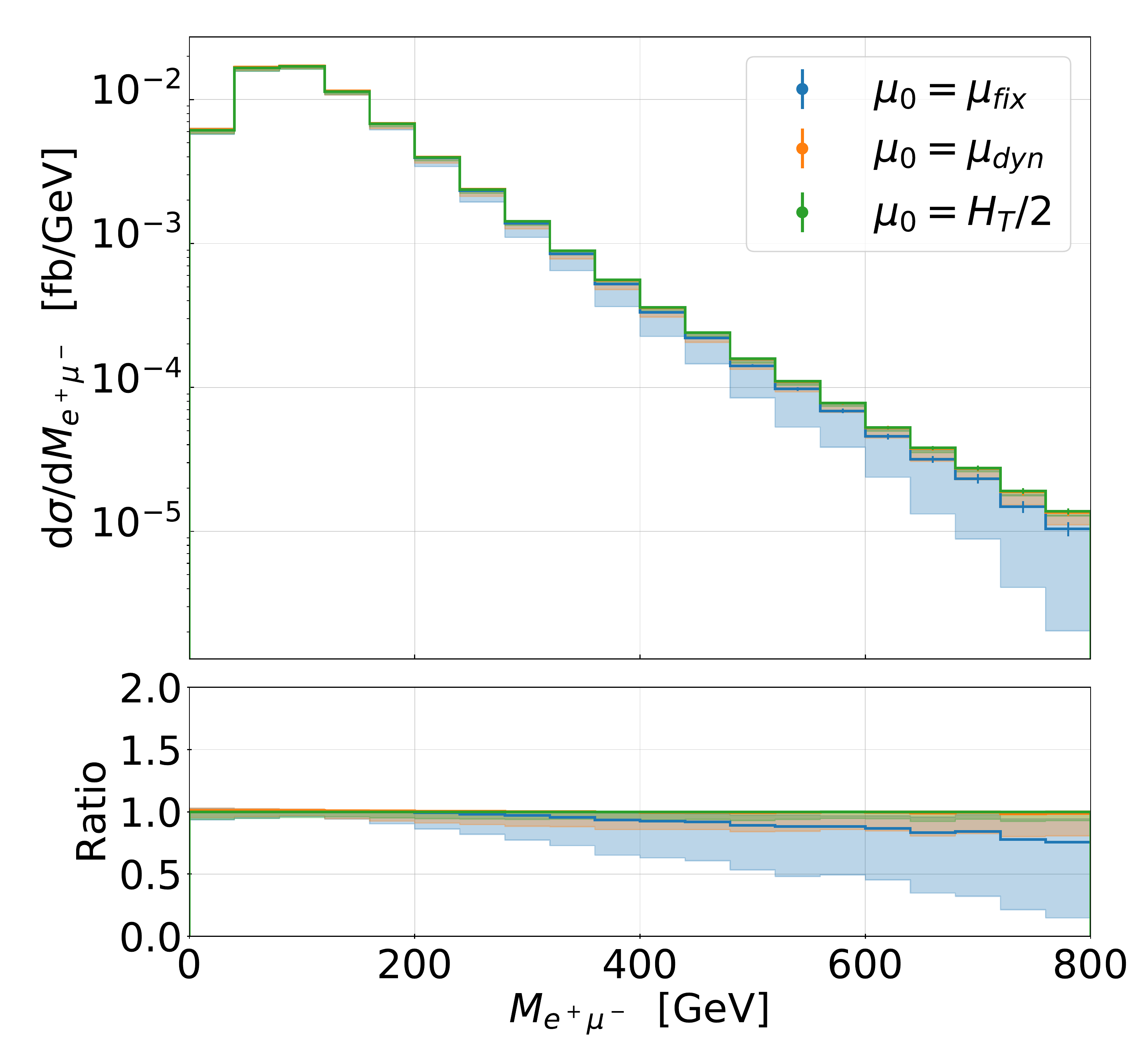}
	\end{center}
	\caption{\label{fig:diff-scales1} \it
		Differential distributions at NLO QCD for the observables $p_{T,b_1}$, $H_T^{had}$, $M_{b_1b_2}$ and $M_{e^+\mu^-}$ for the $pp\to e^+\nu_e\mu^-\bar{\nu}_{\mu}b\bar{b}\,H$ process at the LHC with $\sqrt{s}=13\textrm{ TeV}$. Results are obtained  for $\mu_0=\mu_{fix}$, $\mu_0=\mu_{dyn}$ and $\mu_0=H_T/2$. The NNPDF3.1 PDF set is employed. The upper panels show absolute predictions together with corresponding uncertainty bands resulting from scale variations. Also given are Monte Carlo integration errors. The lower panels display the ratio to $\mu_0=H_T/2$.}
\end{figure}
\begin{figure}[t!]
	\begin{center}
		\includegraphics[width=0.49\textwidth]{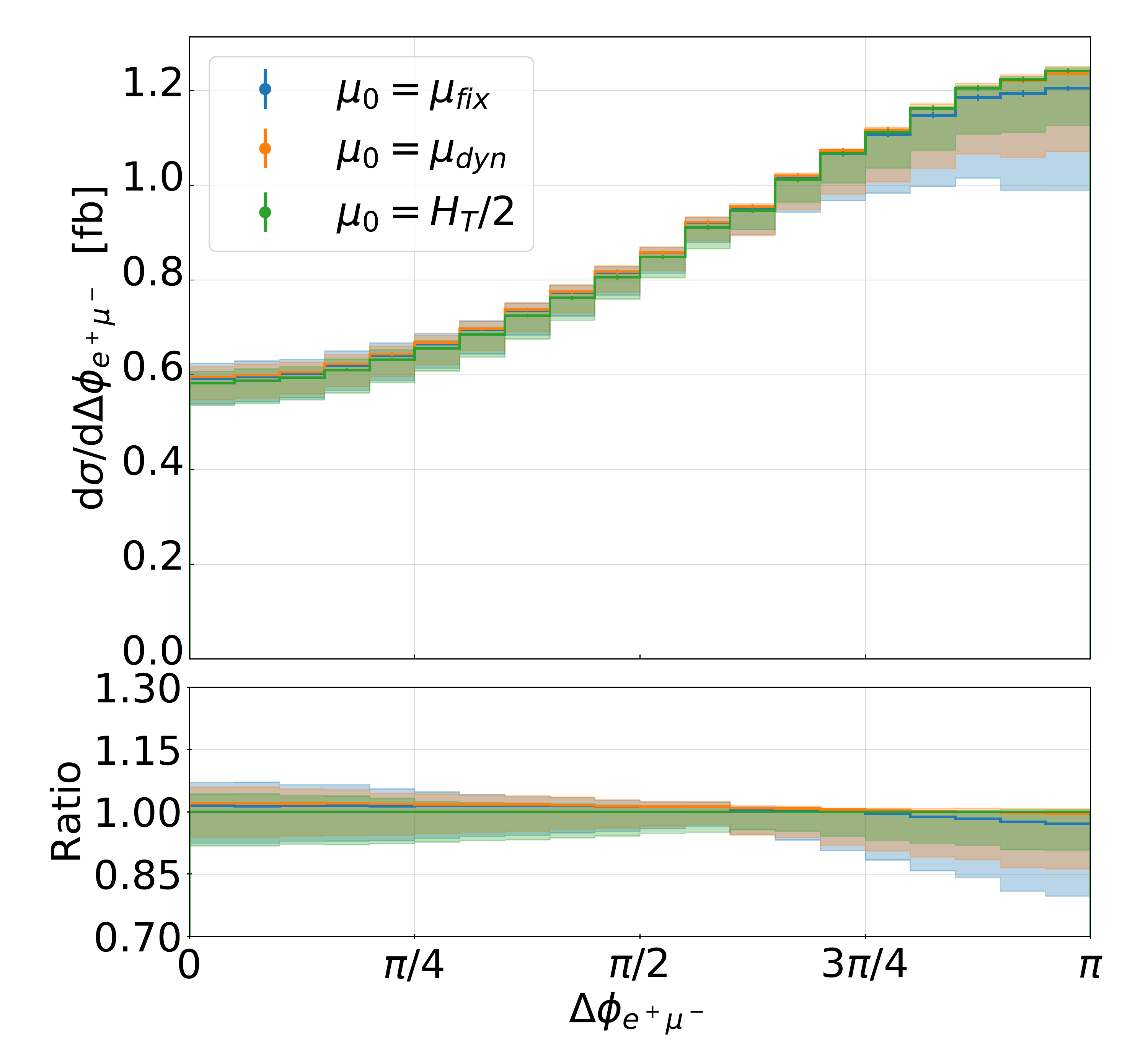}
		\includegraphics[width=0.49\textwidth]{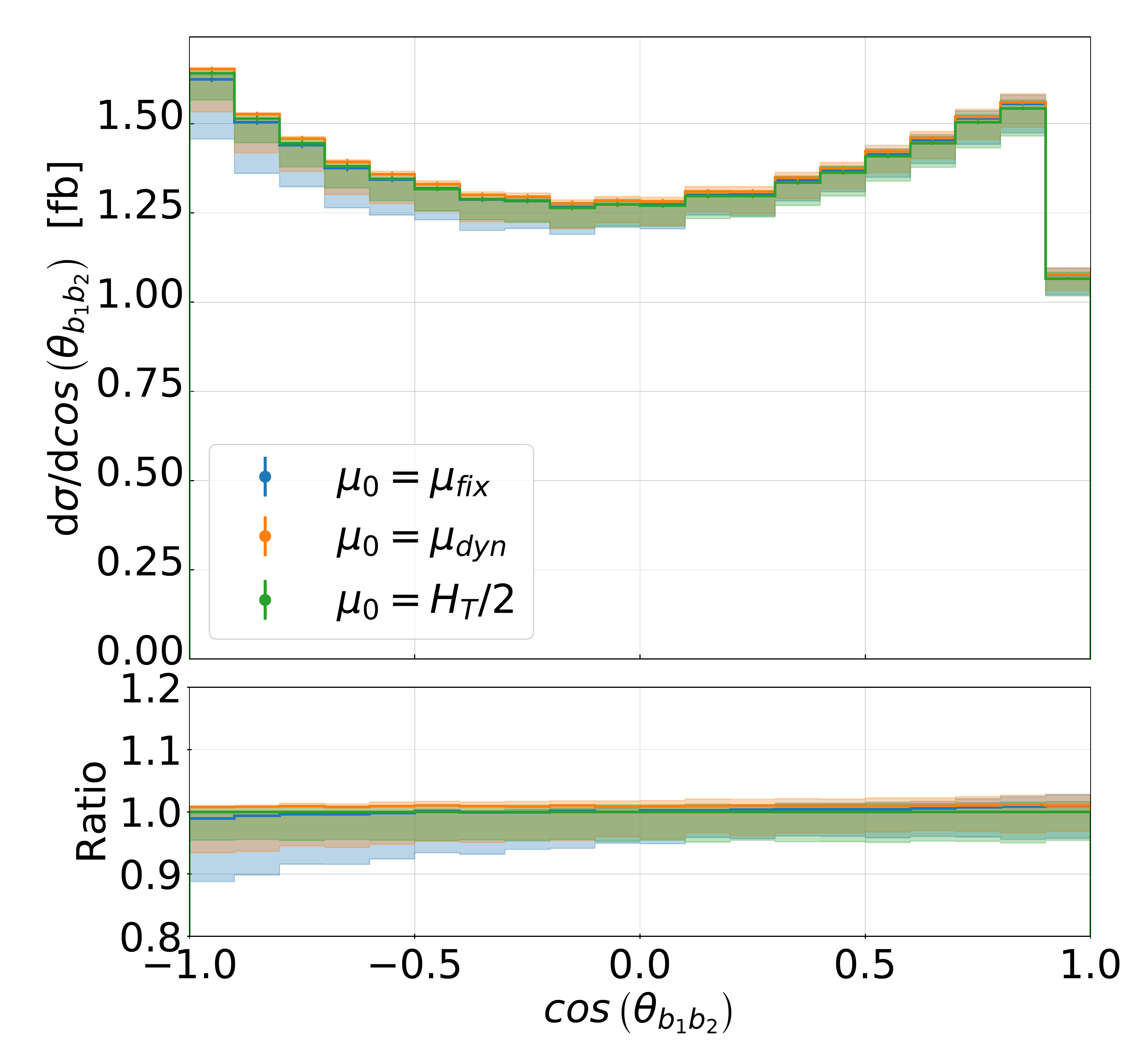}
		\includegraphics[width=0.49\textwidth]{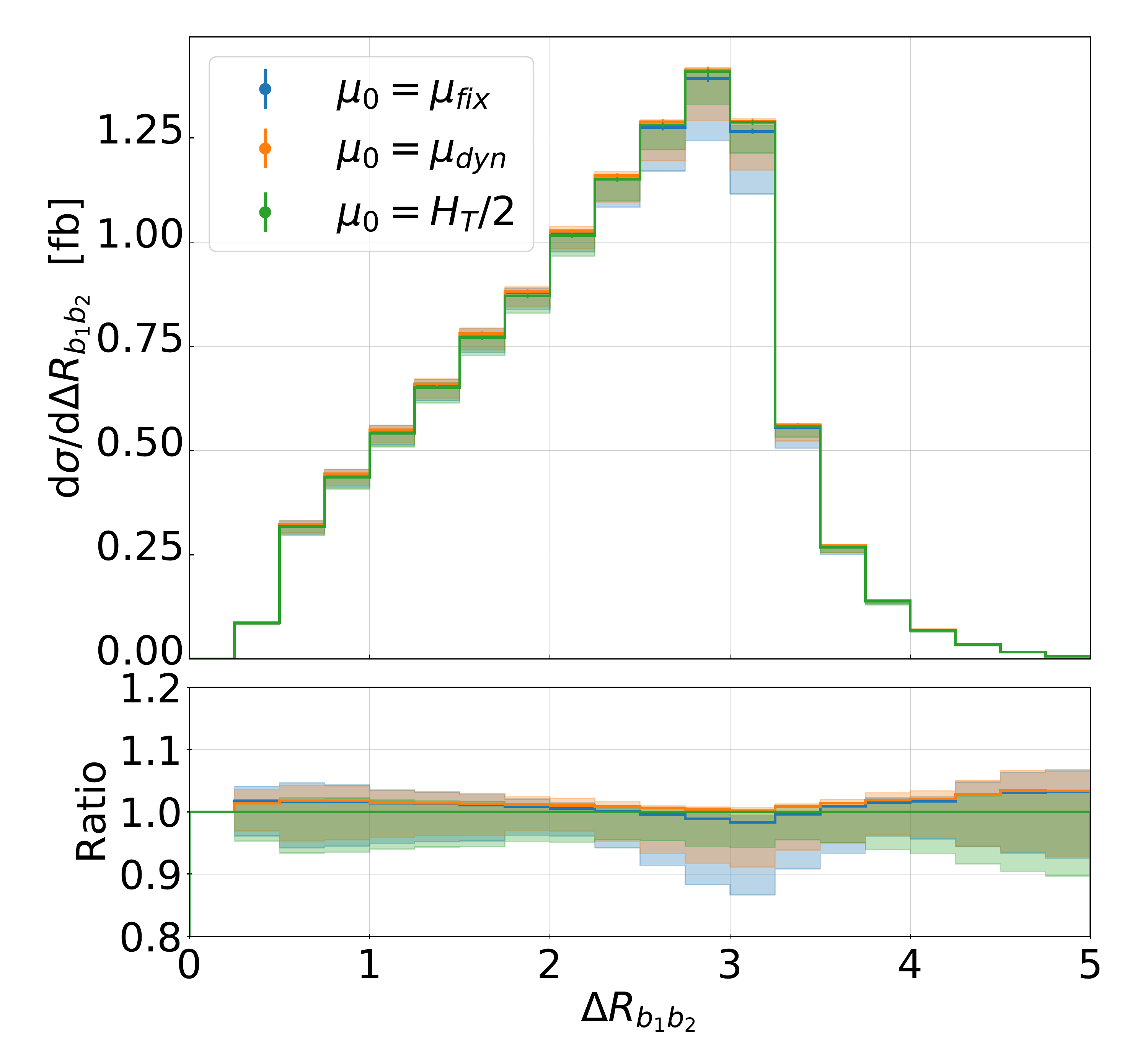}
		\includegraphics[width=0.49\textwidth]{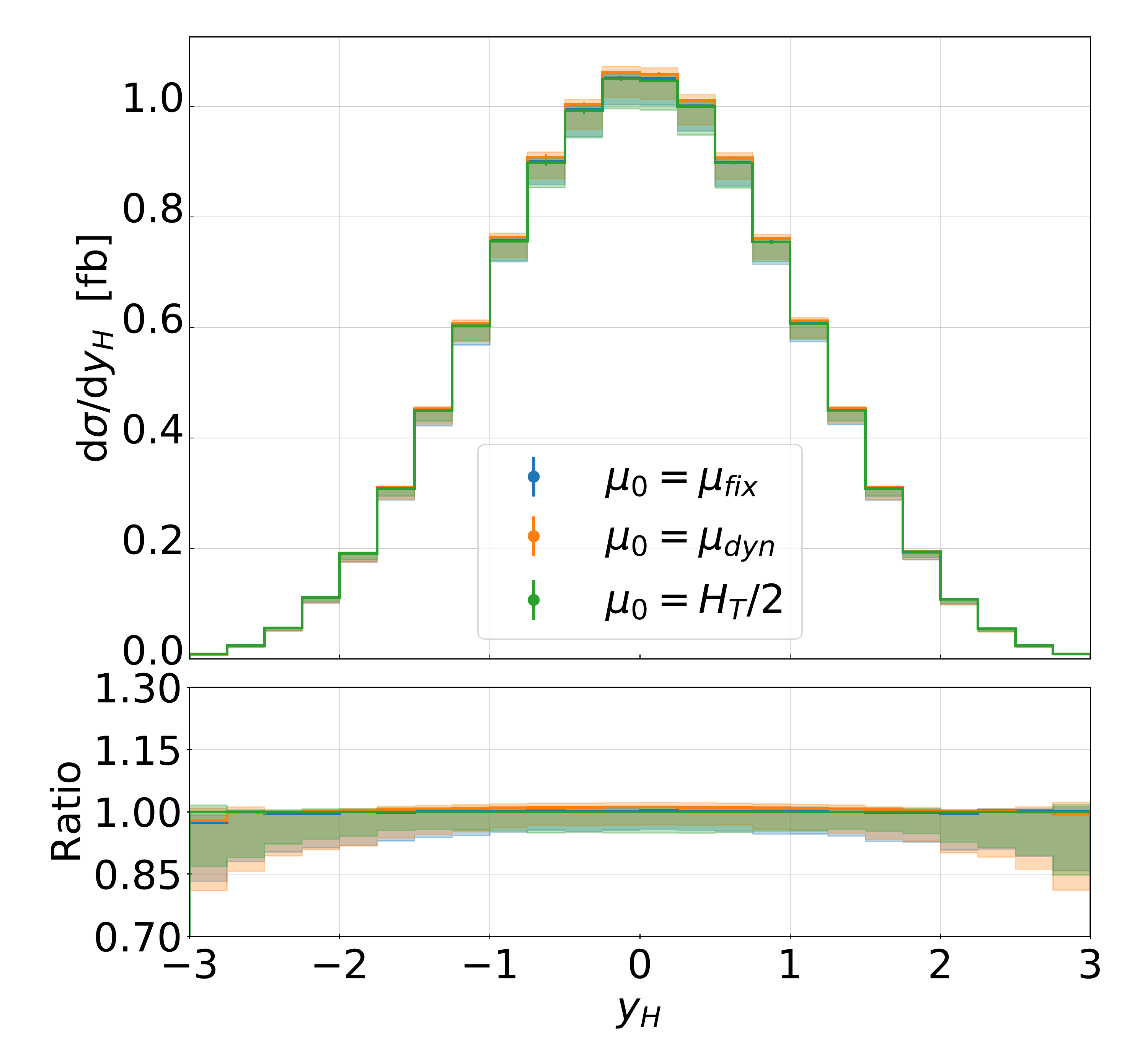}
	\end{center}
	\caption{\label{fig:diff-scales2} \it
		Differential distributions at NLO QCD for the observables $\Delta \phi_{e^+\mu^-}$, $\cos\left(\theta_{b_1b_2}\right)$, $\Delta R_{b_1b_2}$ and $y_H$ for the $pp\to e^+\nu_e\mu^-\bar{\nu}_{\mu}b\bar{b}\,H$ process at the LHC with $\sqrt{s}=13\textrm{ TeV}$. Results are presented for $\mu_0=\mu_{fix}$, $\mu_0=\mu_{dyn}$ and $\mu_0=H_T/2$. The NNPDF3.1 PDF set is employed. The upper panels show absolute predictions together with corresponding uncertainty bands resulting from scale variations. Also given are Monte Carlo integration errors. The lower panels display the ratio to $\mu_0=H_T/2$.}
\end{figure}

We start in Figure \ref{fig:diff-scales1} with the differential distributions for the transverse momentum of the hardest $b$-jet $(p_{T,\,b_1})$, the scalar sum of the transverse momenta of the two $b$-jets ($H_T^{had}$, where  $H_T^{had}= p_{T,\,b_1}+p_{T,\,b_2}$), the invariant mass of the first and second hardest $b$-jet $(M_{b_1b_2})$ and the invariant mass of the two charged leptons $(M_{e^+\mu^-})$. NLO QCD results are shown for all three scale choices, $\mu_0=\mu_{fix}$, $\mu_0=\mu_{dyn}$ and $\mu_0=H_T/2$. Moreover,  the NNPDF3.1 PDF set is employed. Scale uncertainties are displayed as uncertainty bands and the lower panels show the ratio to the results obtained for the  $\mu_0=H_T/2$ scale setting. For the transverse momentum of the hardest $b$-jet ($p_{T,\,b_1}$), we find that the results for the three scale choices are comparable and differ in the tails only up to $5\%$. While the scale uncertainties for the two dynamical scales are rather constant and stay below $10\%$, the fixed scale setting leads to significantly larger effects of $25\%$ or more. For the observable $H_T^{had}$, $M_{b_1b_2}$ and $M_{e^+\mu^-}$ we find that the two dynamical scales lead to similar results. For $\mu_0=\mu_{fix}$ larger deviations  are present in the tails when comparing  to $\mu_0=\mu_{dyn}$ and $\mu_0=H_T/2$. Specifically, the differences are up to $20\%$ for $H_T^{had}$, $5\%$ for $M_{b_1b_2}$ and $25\%$ for $M_{e^+\mu^-}$. Furthermore, for $\mu_0=\mu_{fix}$ the scale uncertainties increase up to even $70\%-80\%$. For $\mu_0=\mu_{dyn}$ and $\mu_0=H_T/2$ they are below $20\%$ and $10\%$ respectively. Thus, we conclude that the two dynamical scales are alike for $\mu_0$. The fixed scale setting, however, is quite different already for the central value of the scale. Furthermore, it results  in significantly larger scale uncertainties in the tails of these dimensionful distributions. 

In Figure \ref{fig:diff-scales2} we show the azimuthal angle in the transverse plane between the two charged leptons $(\Delta\phi_{e^+\mu^-})$, the cosine of the angle between the two $b$-jets $(\cos\left(\theta_{b_1b_2}\right))$, the distance in the azimuthal angle rapidity plane between the two $b$-jets $(\Delta R_{b_1b_2})$ and the rapidity of the Higgs boson $(y_H)$. For $\Delta\phi_{e^+\mu^-}$ we find that the differences between the three scales are below $3\%$ over the entire range. The scale uncertainties are very similar for $\mu_0=\mu_{fix}$, $\mu_0=\mu_{dyn}$ and  $\mu_0=H_T/2$, and only for large values of $\Delta\phi_{e^+\mu^-} \approx \pi$, do we observe a reduction of the scale uncertainties for dynamic scale settings. In these regions, they are reduced from $18\%$ for $\mu_0=\mu_{fix}$ to $13\%$ for $\mu_0=\mu_{dyn}$ and to $9\%$ for $\mu_0=H_T/2$. For $\cos\left(\theta_{b_1b_2}\right)$, we find that the differences among the various scale choices are small, below $2\%$. Again, only for large values of $\theta_{b_1b_2}$, i.e. small values of $\cos\left(\theta_{b_1b_2}\right)$, can the scale uncertainties be reduced by a special scale choice. In particular, they are reduced from $10\%$ for $\mu_0=\mu_{fix}$ to $7\%$ for $\mu_0=\mu_{dyn}$ and down to $5\%$ for $\mu_0=H_T/2$. This kinematical region corresponds to the case where the two $b$-jets (the two charged leptons) can be found almost back-to-back, which is typical for a produced top-quark pair and, to large extent, for its decay products. For $\Delta R_{b_1b_2}$ and  $y_H$, we again observe small differences among the three scale settings. We notice also in this case that for $\Delta R_{b_1b_2}$, the scale dependence around $\Delta R_{b_1  b_2} \approx \pi$ is improved for $\mu_0=\mu_{dyn}$ and $\mu_0=H_T/2$.  For $y_H$, the scale uncertainties are below $5\%$ in the central rapidity region while they increase up to $15\%-20\%$ in the forward-backward regions. This observation is independent of the choice made for $\mu_R$ and $\mu_F$. In conclusion, for angular distributions, which we have examined, the results are not affected by the scale choice. This is  in contrast to the case of dimensionful observables, which we have  discussed before. On the other hand, we find improvements in the scale dependence for dimensionless observables when we use a dynamical scale setting. In particular, for the $\mu_R=\mu_F=\mu_0=H_T/2$ scale choice, the smallest theoretical error is obtained.  
\begin{figure}[t!]
	\begin{center}
		\includegraphics[width=0.49\textwidth]{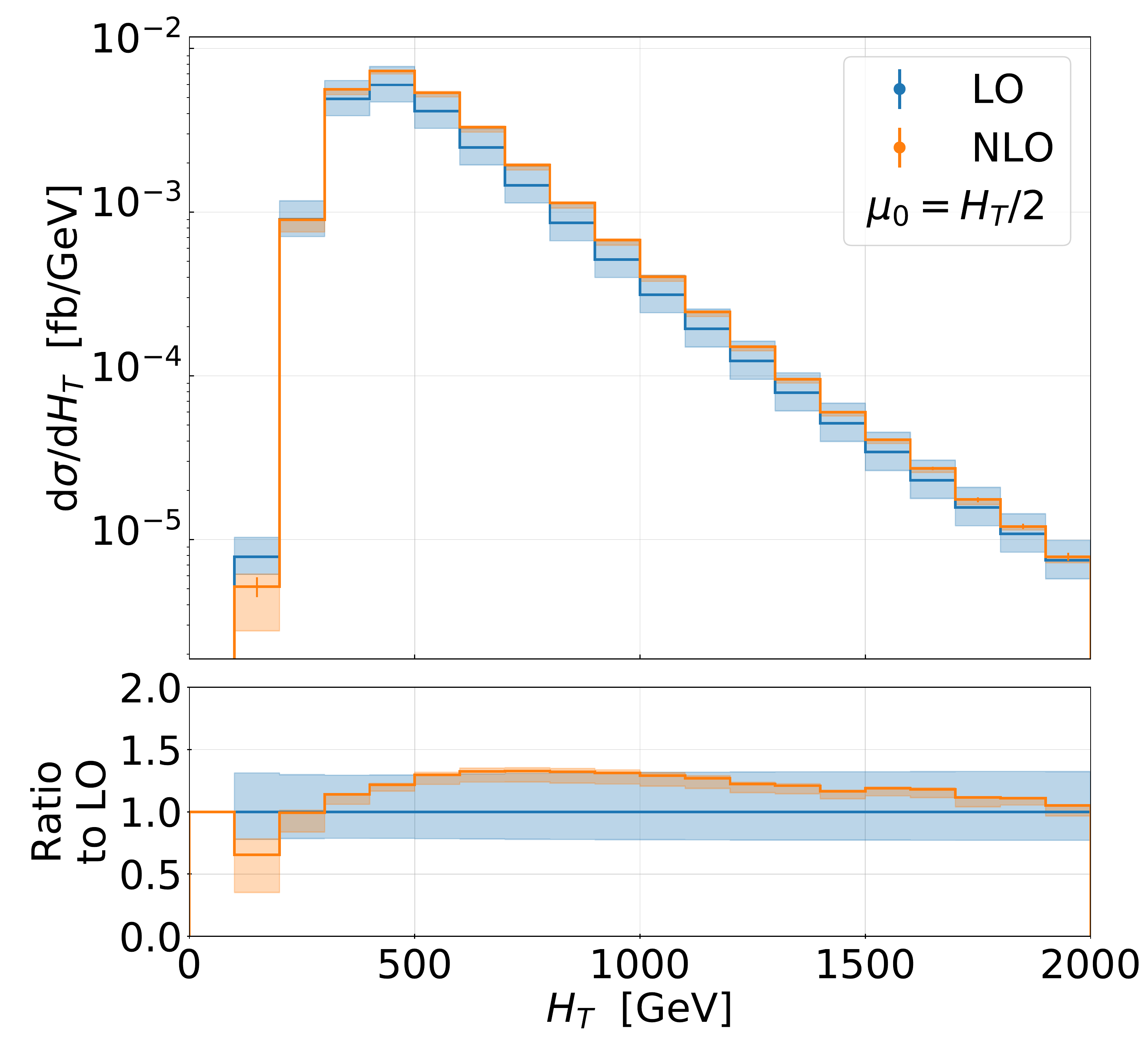}
		\includegraphics[width=0.49\textwidth]{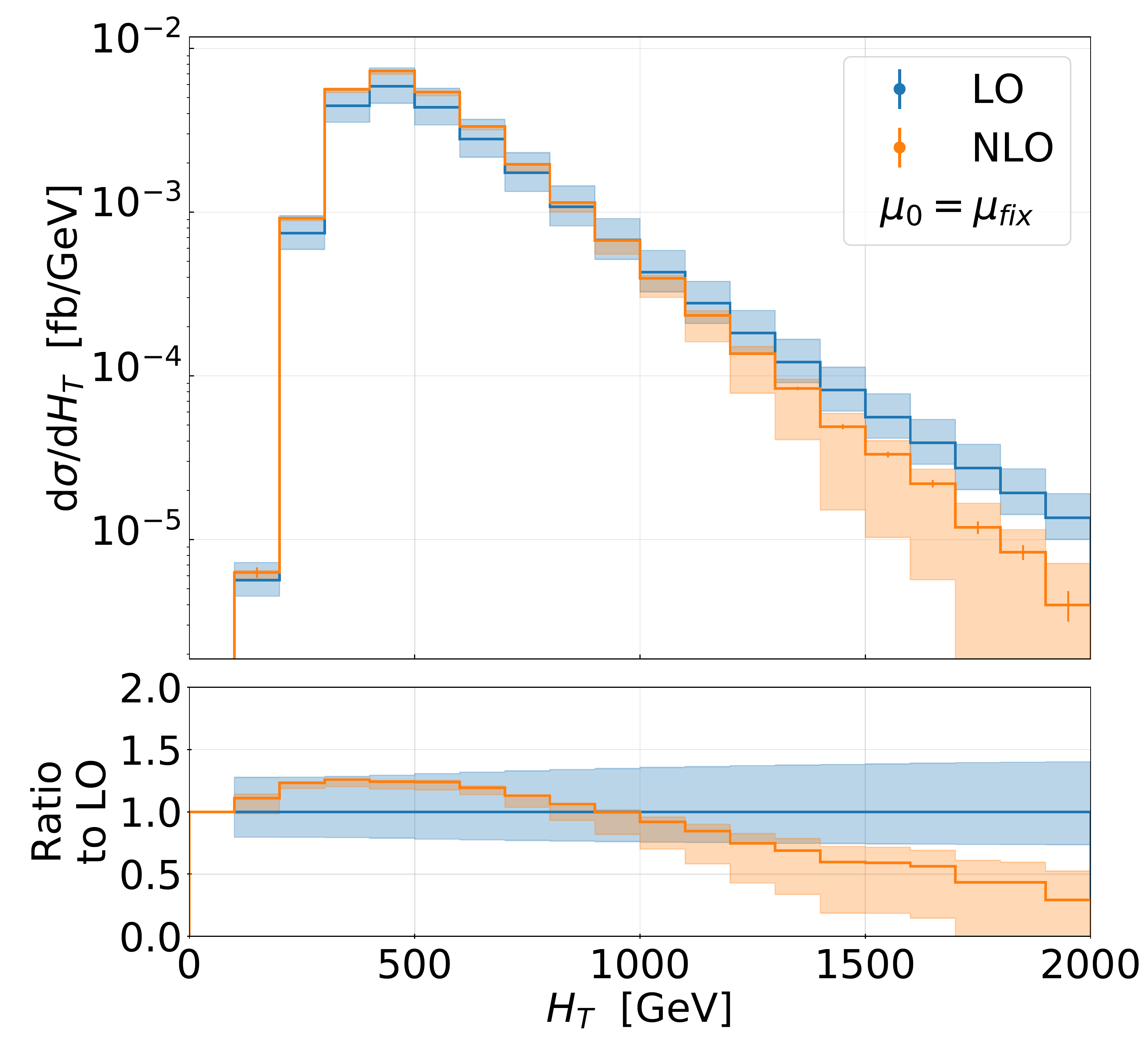}
		\includegraphics[width=0.49\textwidth]{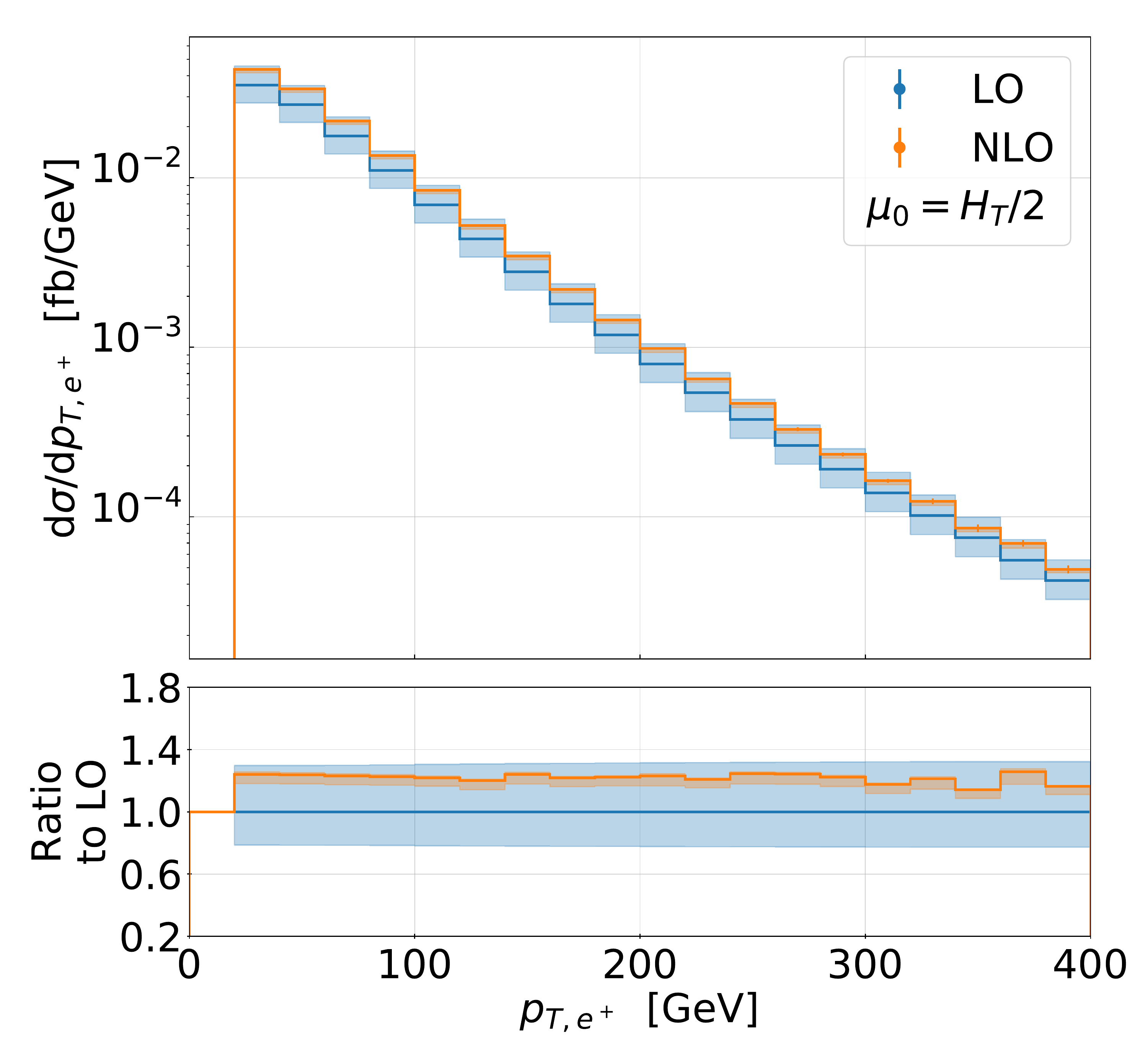}
		\includegraphics[width=0.49\textwidth]{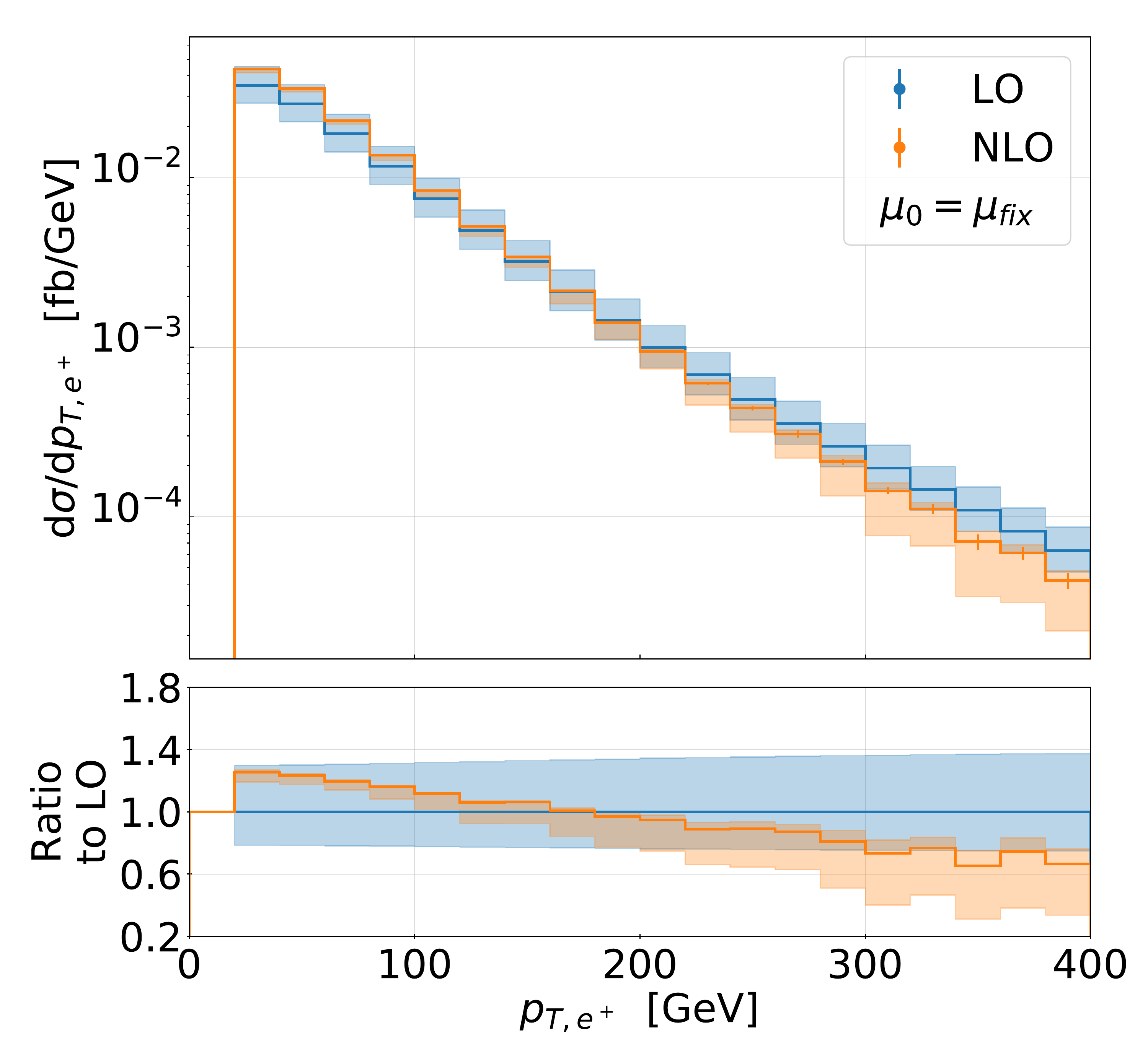}
	\end{center}
	\caption{\label{fig:kfac1} \it
		Differential distributions at LO and NLO QCD with the corresponding uncertainty bands for the observables $H_T$ and $p_{T,\,e^+}$ for the $pp\to e^+\nu_e\mu^-\bar{\nu}_{\mu}b\bar{b}\,H$ process at the LHC with $\sqrt{s}=13\textrm{ TeV}$. Results are presented  for $\mu_0=H_T/2$ and $\mu_0=\mu_{fix}$. The NNPDF3.1 PDF set is employed. Also given are Monte Carlo integration errors. The lower panels display the differential $\mathcal{K}\textrm{-factor}$.}
\end{figure}

We continue the discussion at the differential level with  the differential  $\mathcal{K}\textrm{-factor}$ defined as  $\mathcal{K}=d\sigma_{\rm NLO}/d\sigma_{\rm LO}$ to asses the size of the NLO QCD corrections and the perturbative stability of the two scales $\mu_0=H_T/2$ and $\mu_0=\mu_{fix}$. For the discussion of the differences between a dynamical and a fixed scale, only $\mu_0=H_T/2$ is used in the following, since the conclusions are rather similar for $\mu_0=H_T/2$ and $\mu_0=\mu_{dyn}$.  We start with Figure \ref{fig:kfac1} where the following two observables $H_T$ and $p_{T,\,e^+}$ are shown. We present the LO and NLO predictions in the upper panels of the plot for $\mu_0=H_T/2$ (left) and $\mu_0=\mu_{fix}$ (right). The lower panels display the differential  $\mathcal{K}$-factor and the corresponding uncertainty bands due to the scale variation. For $H_T$, we notice that, already at LO, the scale uncertainties are reduced from $40\%$ for $\mu_0=\mu_{fix}$ to $32\%$ for $\mu_0=H_T/2$. At NLO, we find that, for the fixed scale choice, scale uncertainties can be as high as $100\%$ and even negative values are included in the uncertainty bands. On the other hand, for $\mu_0=H_T/2$, they are only of the order of $8\%$. In addition, we find that the NLO and LO predictions do not fit within the uncertainty bands in the tails for $\mu_0=\mu_{fix}$ while for $\mu_0=H_T/2$, the NLO predictions are completely included in the uncertainty bands of the LO prediction. The higher order QCD corrections are not constant in both cases and reach up to $35\%$ even for the dynamical scale setting. Thus, the NLO QCD corrections are necessary for a precise prediction for $H_T$. For the second observable, $p_{T,\,e^+}$, the dynamical scale leads to a better perturbative stability in the tails where the NLO and LO predictions for $\mu_0=\mu_{fix}$ are clearly separated from each other. The scale uncertainties are reduced from about $30\%$ at LO to $5\%$ at NLO for $\mu_0=H_T/2$. Instead, for the fixed scale choice they can exceed $50\%$ at NLO, which is even more than the corresponding LO scale uncertainty for that scale setting. Finally, the NLO QCD corrections for $p_{T,\,e^+}$ with $\mu_0=H_T/2$ are rather constant and of the order of $20\%$ over the whole plotted range. The behavior of most of the dimensionful observables, that we have examined, is similar to that of $H_T$ and  $p_{T,\,e^+}$. 
\begin{figure}[t!]
	\begin{center}
		\includegraphics[width=0.49\textwidth]{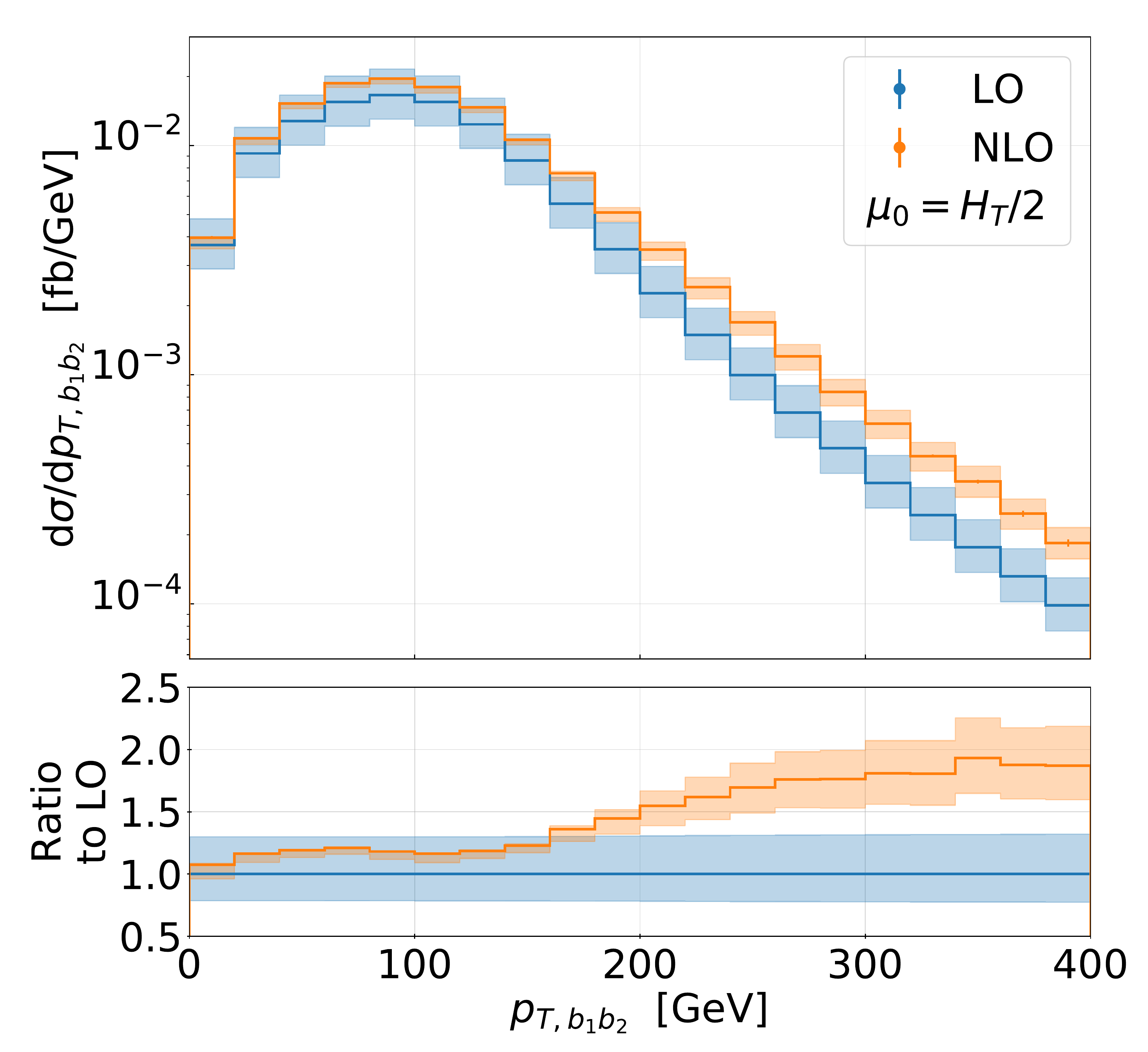}
		\includegraphics[width=0.49\textwidth]{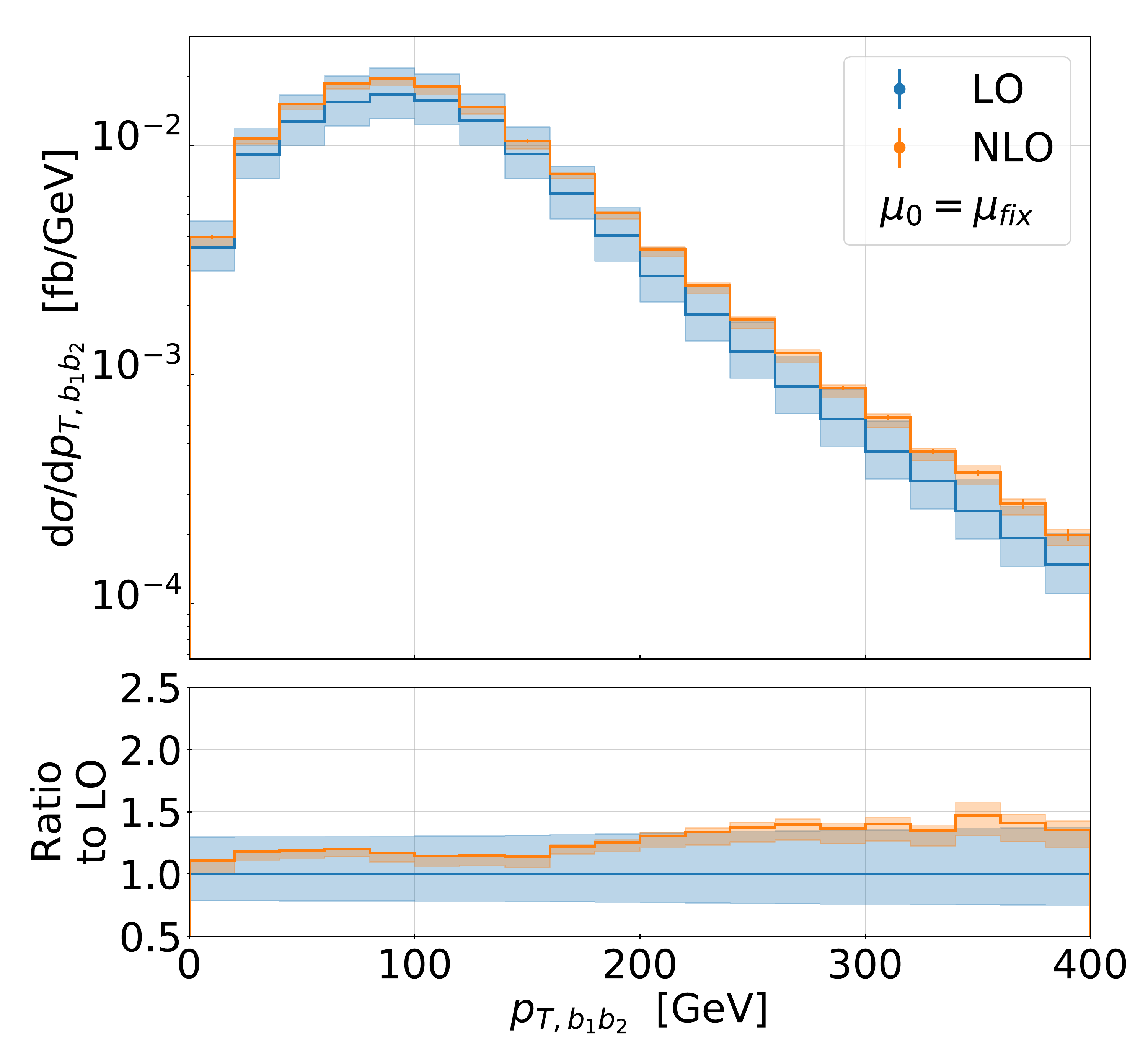}
		\includegraphics[width=0.49\textwidth]{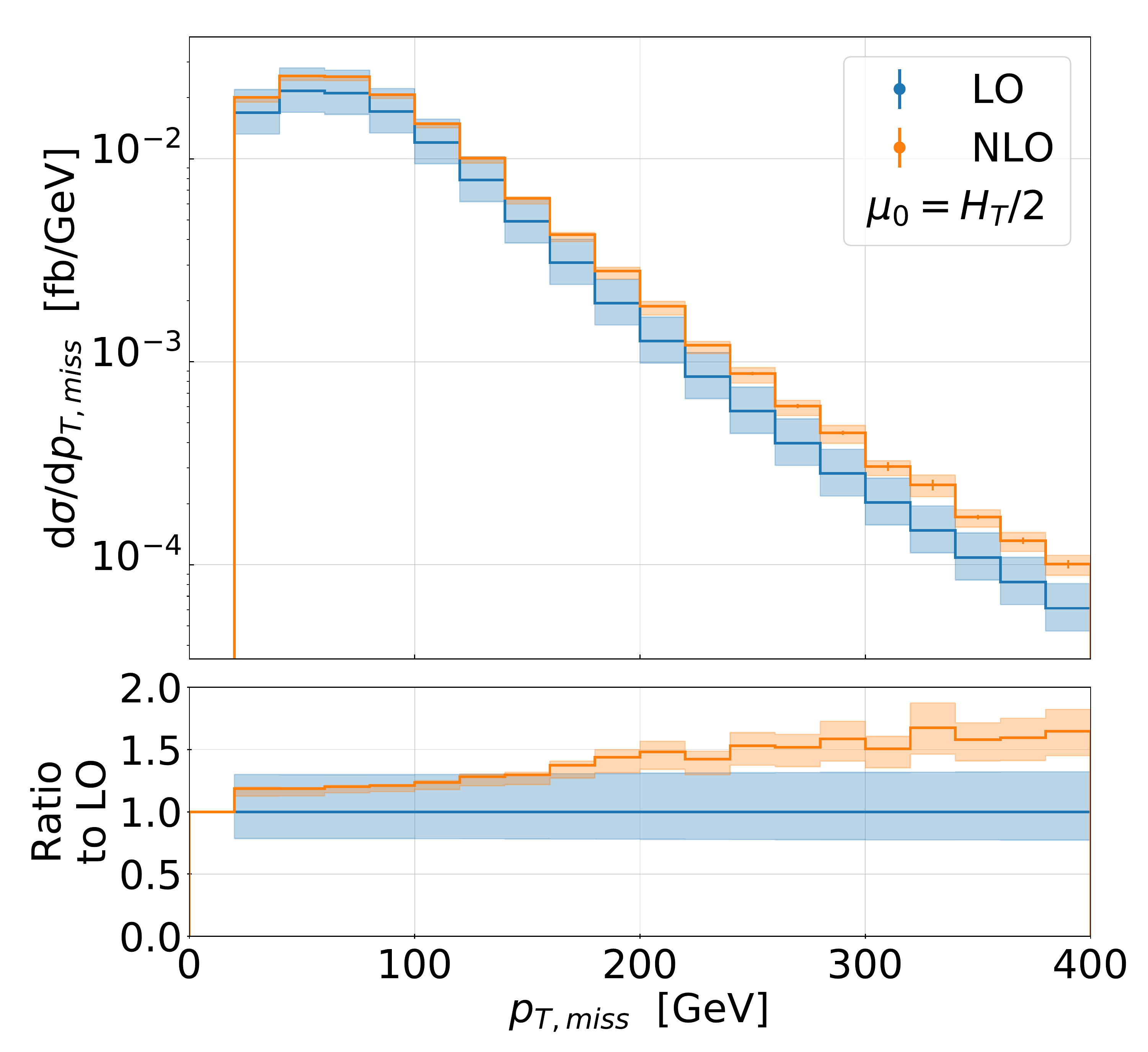}
		\includegraphics[width=0.49\textwidth]{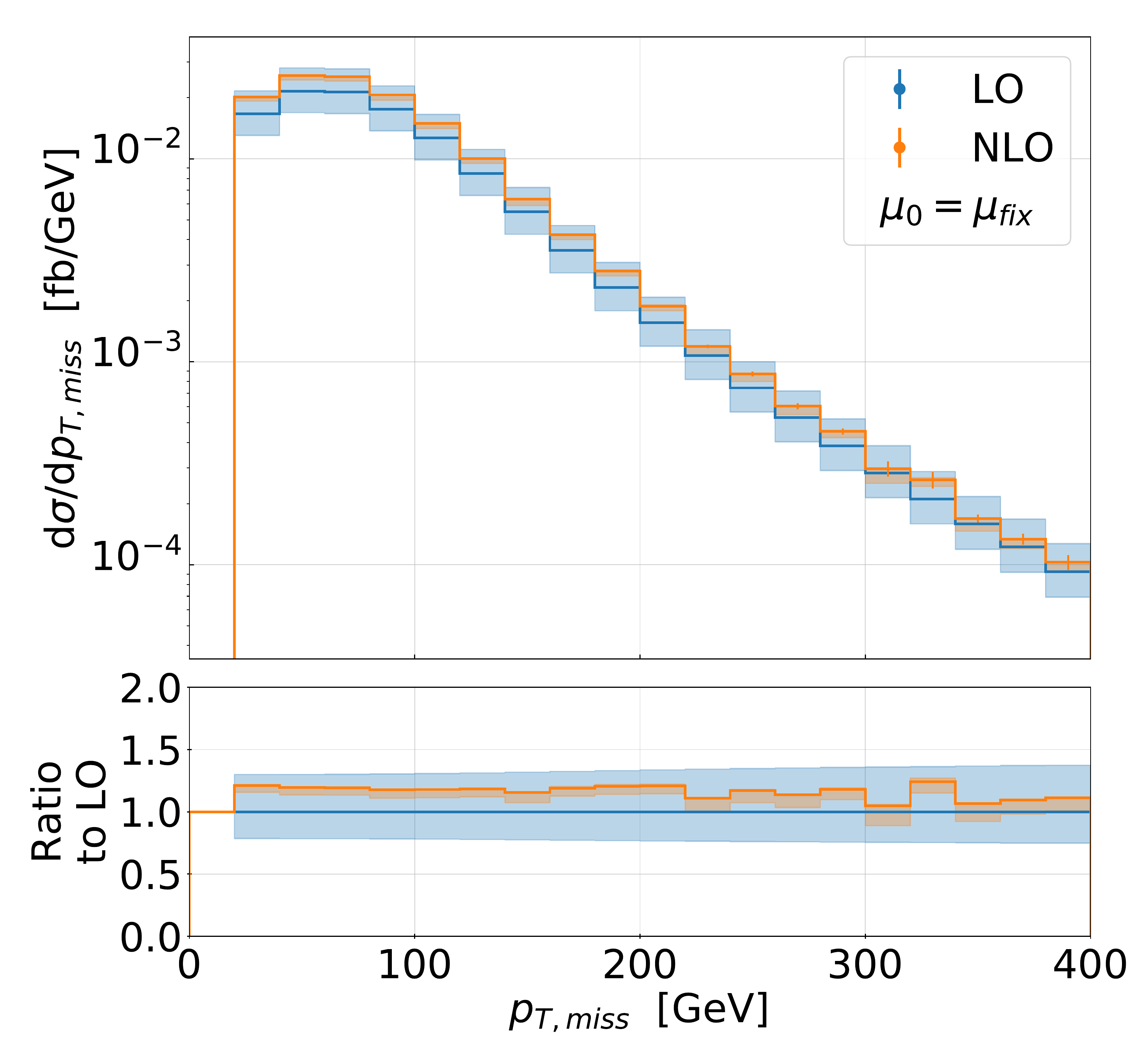}
	\end{center}
	\caption{\label{fig:kfac2} \it
		Differential distributions at LO and NLO QCD with the corresponding uncertainty bands for the observables $p_{T,\,b_1b_2}$ and $p_{T,\,miss}$ for the $pp\to e^+\nu_e\mu^-\bar{\nu}_{\mu}b\bar{b}\,H$ process at the LHC with $\sqrt{s}=13\textrm{ TeV}$. Results are presented for $\mu_0=H_T/2$ and $\mu_0=\mu_{fix}$. The NNPDF3.1 PDF set is employed. Also given are Monte Carlo integration errors. The lower panels display the differential $\mathcal{K}\textrm{-factor}$.}
\end{figure}

In the following, we present two observables that are rather special. Specifically, in Figure \ref{fig:kfac2}, the observables $p_{T,\, b_1b_2}$ and $p_{T, \,miss}$ are shown.  We can notice that up to around $150$ GeV for both observables, NLO QCD corrections are of the order of $20\%-30\%$. Above  this value, however, they increase substantially and can reach $90\%$ for $p_{T,\,b_1b_2}$ and $60\%$ for $p_{T,\,miss}$. Because of these huge QCD corrections, the LO and NLO predictions do not overlap anymore for the transverse momentum values above $150$ GeV. Such effects are well known and discussed (also for $p_{T, \,\ell\ell}$, that is not shown here), e.g. in Refs. \cite{Denner:2012yc,Denner:2015yca,Bevilacqua:2019cvp}, where NLO QCD calculations for top quark pair production with and without heavy bosons are presented. In Ref. \cite{Czakon:2020qbd}, similar findings are described for the $t\bar{t}$ production process in the di-lepton top quark decay channel at next-to-next-to-leading order (NNLO) in QCD. In this case the full NWA with corrections to top-quark production and subsequent top-quark decay is employed. Furthermore a comparison to CMS and ATLAS data is carried out. We note that such huge effects are not observed for the fix scale setting. However, since at NLO in QCD the differences between the two scale choices, $\mu_0=\mu_{fix}$ and $\mu_0=H_T/2$, are about $2\%$ for $p_{T, \,miss}$ and up to $10\%$ in the tails for $p_{T,\,b_1b_2}$ we can attribute these discrepancies to the corresponding LO differential cross sections. Indeed, the LO cross section, which in general is much more sensitive to the variation of the scale, changes more rapidly than the NLO curve, which drives the NLO/LO ratio in the phase-space region above $150$ GeV. Observables that are constructed from the decay products of both top quarks exhibit a strong suppression of the $t\bar{t}$ cross section at LO above $150$ GeV. This can be understood by analyzing $t\bar{t}$ production in the NWA. In that case the top quark decay products are boosted via their parents that have opposite transverse momenta. Thus, although the transverse momentum of the individual decay products  can be quite large, the $p_T$ of the reconstructed $bb$, $e^+ \mu^-$ and  $\nu_e \bar{\nu}_\mu$ system only acquires very small values. Consequently, the transverse momenta of these systems are suppressed for $p_T \ge 150$ GeV. At NLO, due to additional radiation, this kinematical constraint is partially lifted resulting in an  enhancement of the cross section and a large ${\cal K}$-factor, see e.g. Refs. \cite{Denner:2012yc,Czakon:2020qbd}. For $t\bar{t}H$ production, acquiring a non-vanishing transverse momentum by the Higgs boson relaxes the kinematical constraint already at LO, which leads to much smaller but still substantial NLO QCD corrections for $p_{T,\, b_1b_2}$, $p_{T,\, e^+ \mu^-}$ and $p_{T,\,miss}=|\vec{p}_{T,\,\nu_e} + \vec{p}_{T, \,\bar{\nu}_\mu} |$, see e.g. Ref. \cite{Denner:2015yca}. On the contrary, the scale uncertainties for $p_{T,\,b_1b_2}$ ($p_{T,\,miss}$) only decrease from $17\%$ ($12\%$) for $\mu_0=H_T/2$ to $11\%$ ($10\%$) for $\mu_0=\mu_{fix}$. 
\begin{figure}[t!]
	\begin{center}
		\includegraphics[width=0.49\textwidth]{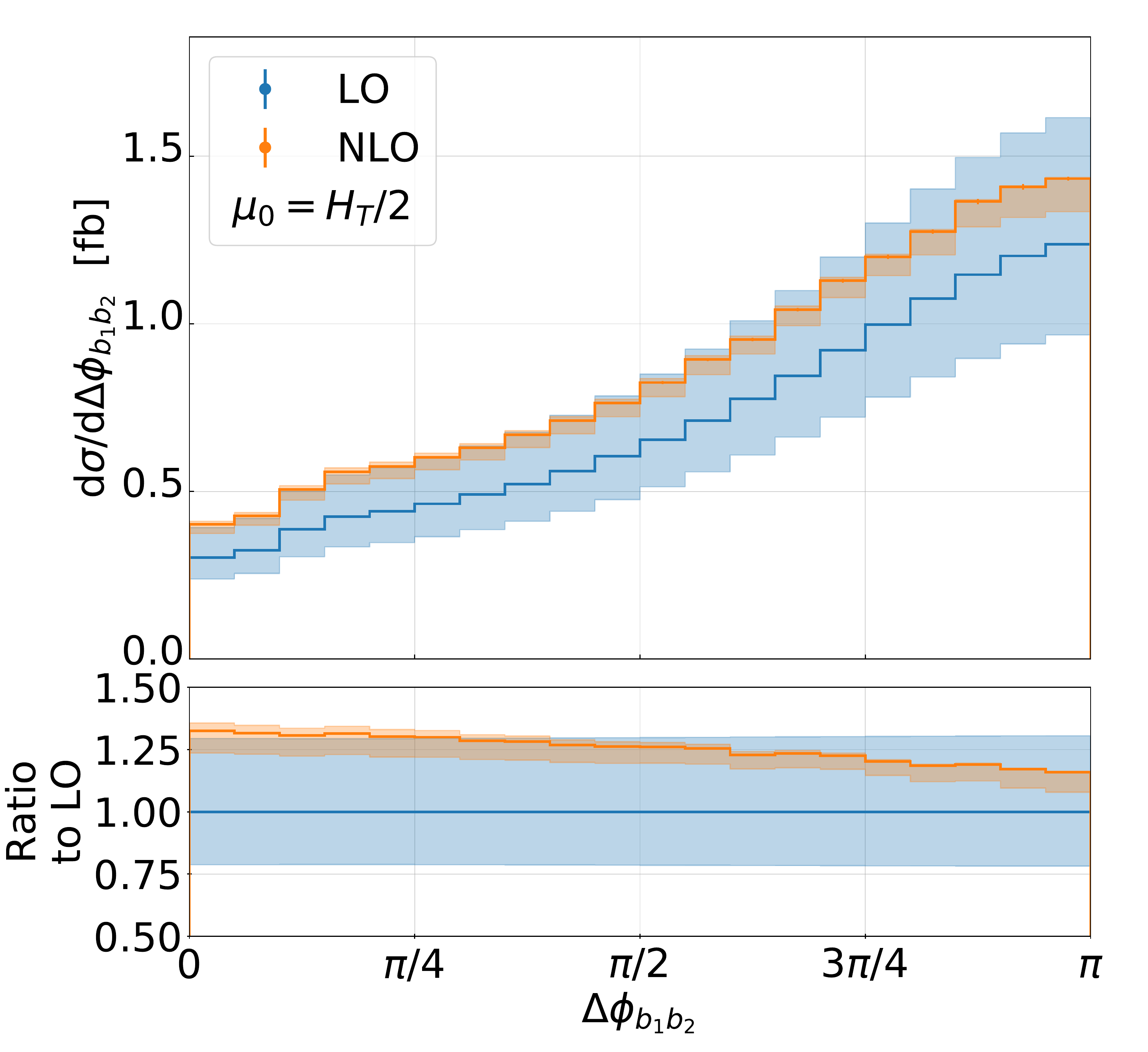}
		\includegraphics[width=0.49\textwidth]{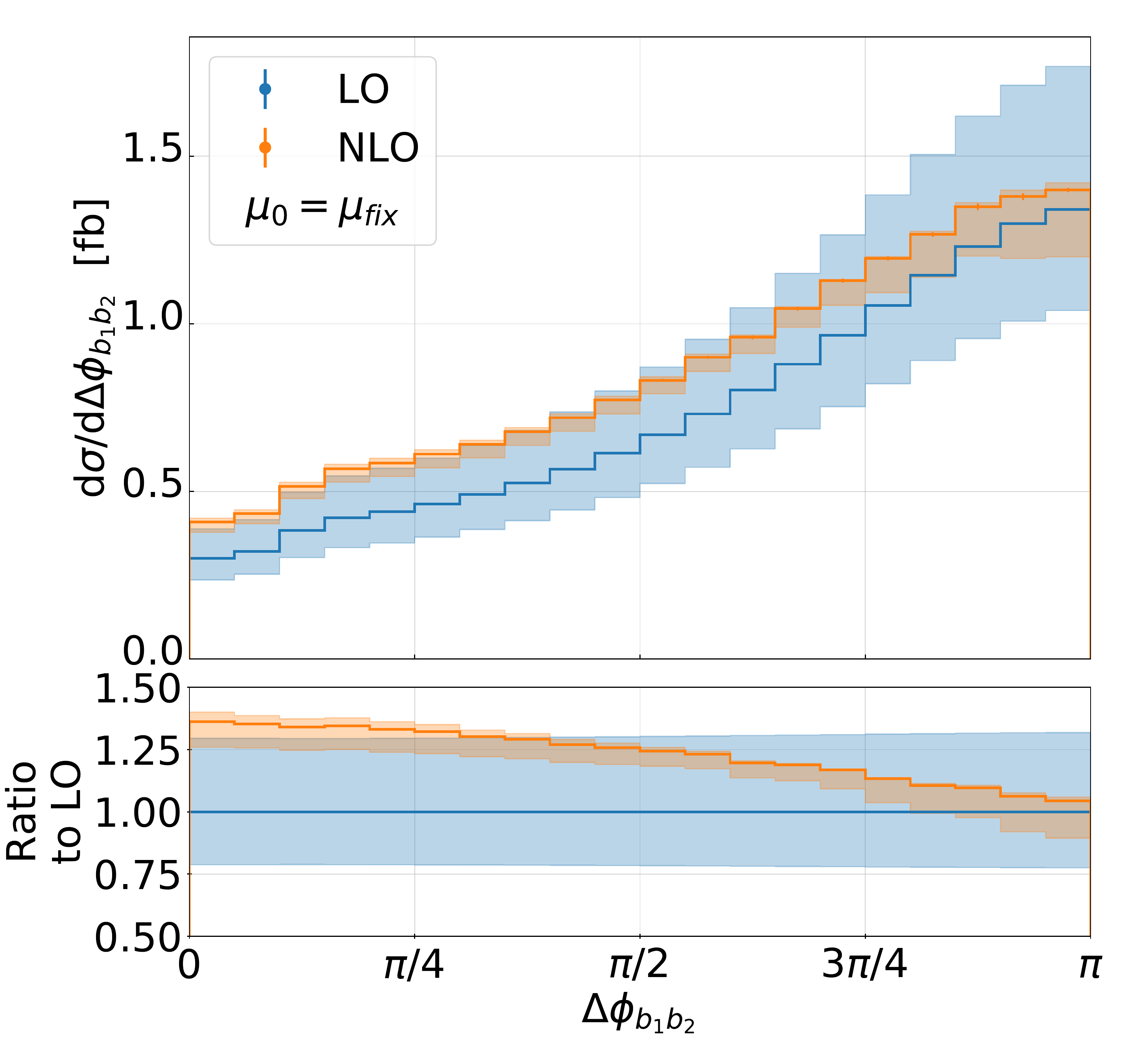}
		\includegraphics[width=0.49\textwidth]{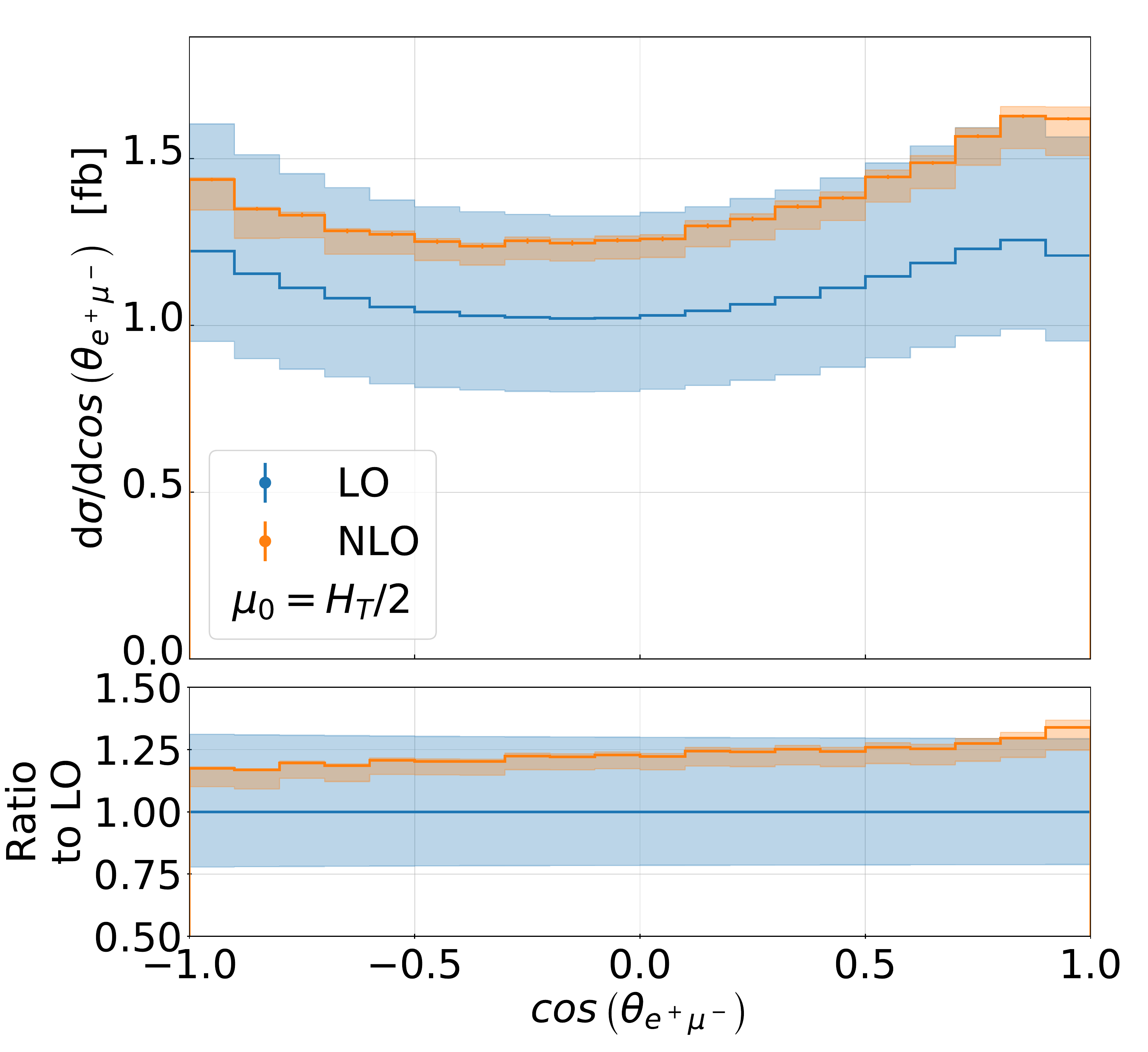}
		\includegraphics[width=0.49\textwidth]{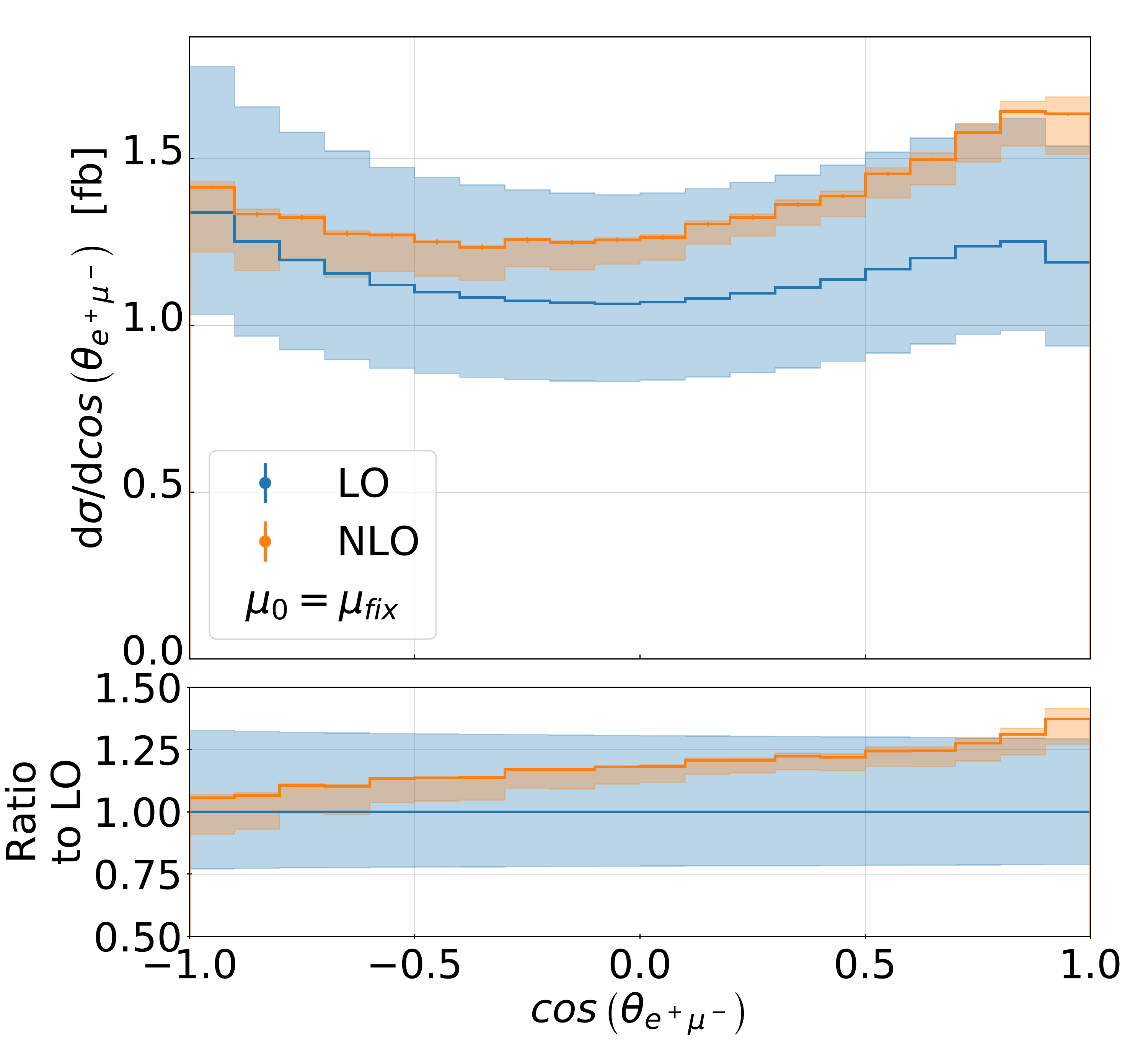}
	\end{center}
	\caption{\label{fig:kfac3} \it
		Differential distributions at LO and NLO QCD with the corresponding uncertainty bands for the observables $\Delta\phi_{b_1b_2}$ and $\cos\left(\theta_{e^+\mu^-}\right)$ for the $pp\to e^+\nu_e\mu^-\bar{\nu}_{\mu}b\bar{b}\,H$ process at the LHC with $\sqrt{s}=13\textrm{ TeV}$. Results are presented for $\mu_0=H_T/2$ and $\mu_0=\mu_{fix}$. The NNPDF3.1 PDF set is employed. Also given are Monte Carlo integration errors. The lower panels display the differential $\mathcal{K}\textrm{-factor}$.}
\end{figure}
\begin{figure}[t!]
	\begin{center}
		\includegraphics[width=0.49\textwidth]{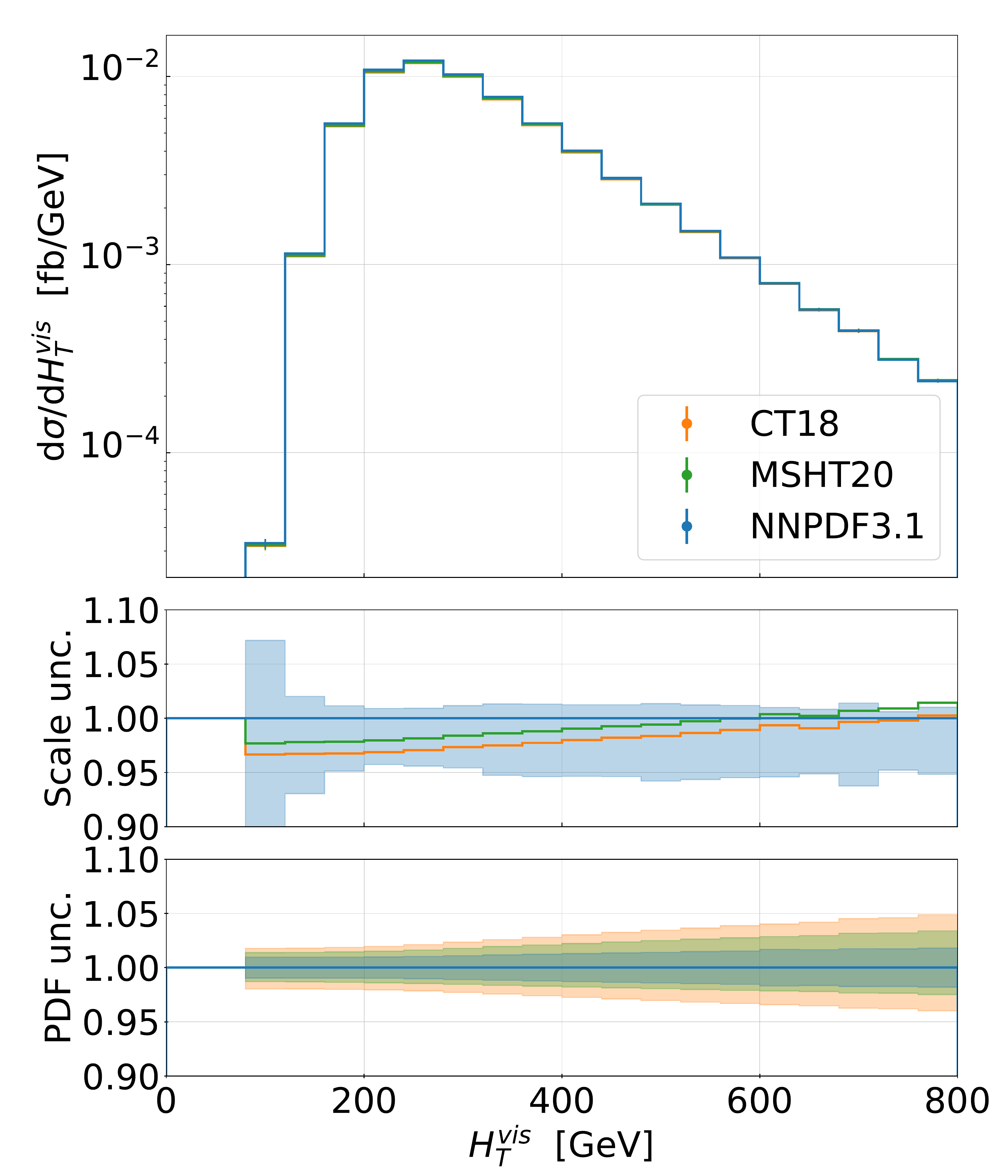}
		\includegraphics[width=0.49\textwidth]{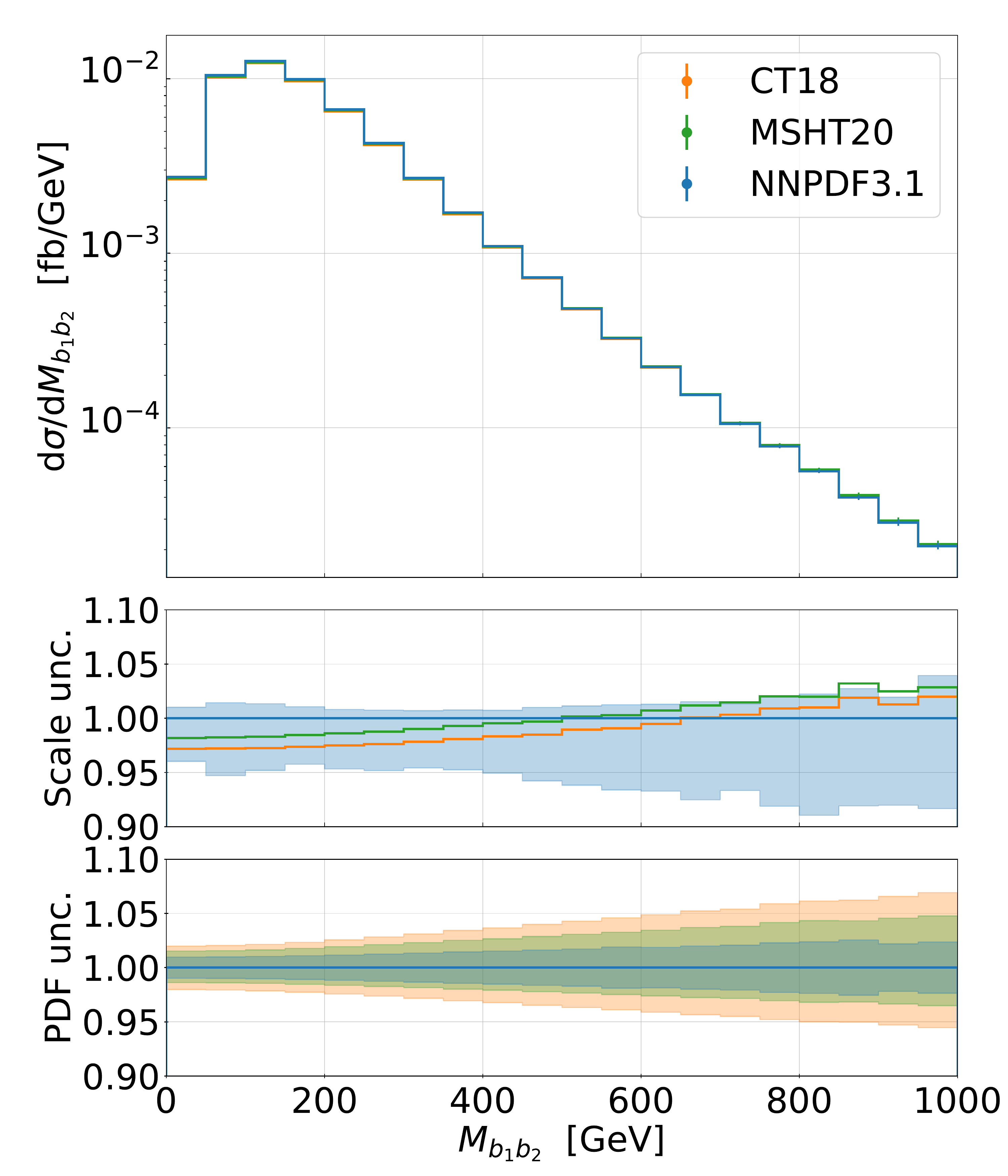}
		\includegraphics[width=0.49\textwidth]{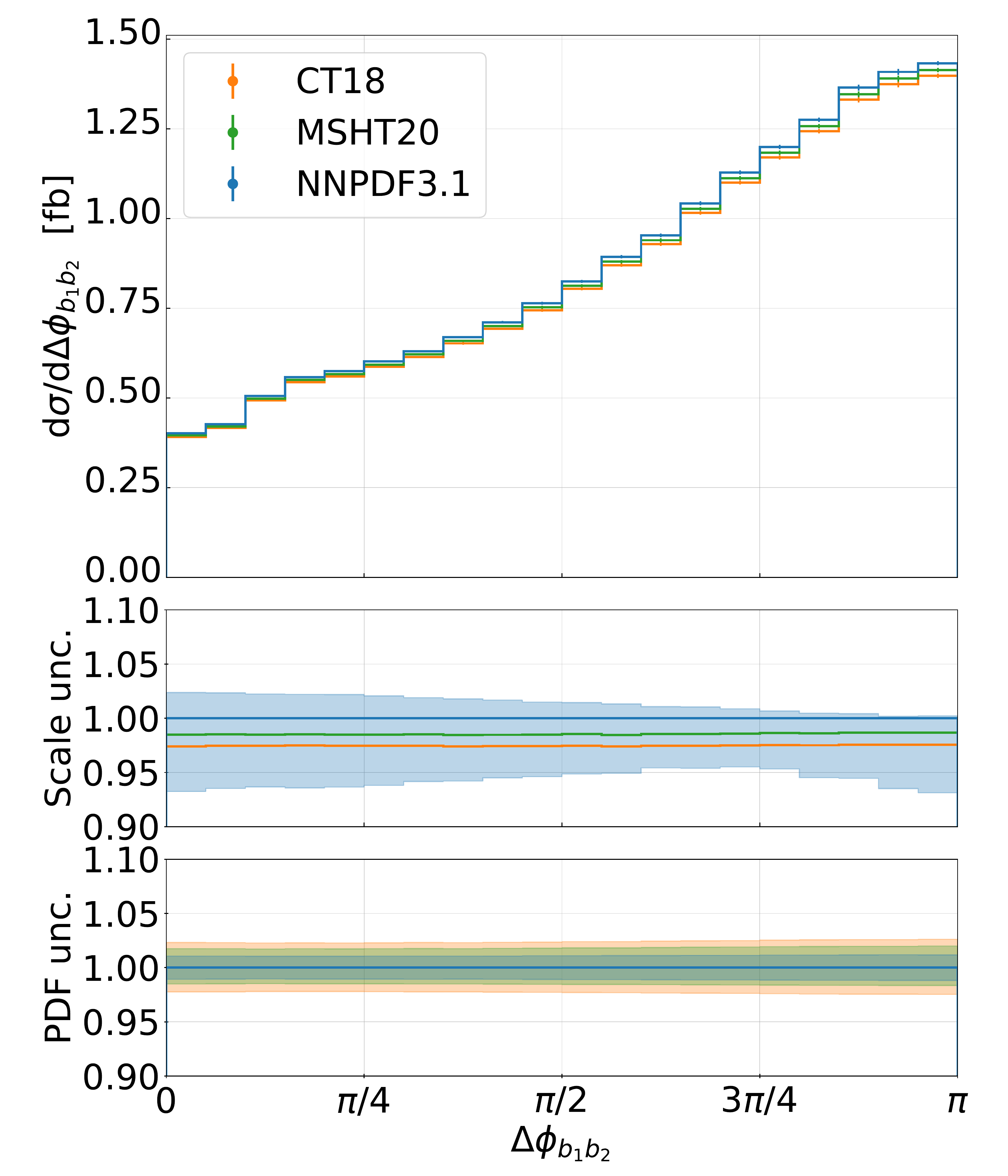}
		\includegraphics[width=0.49\textwidth]{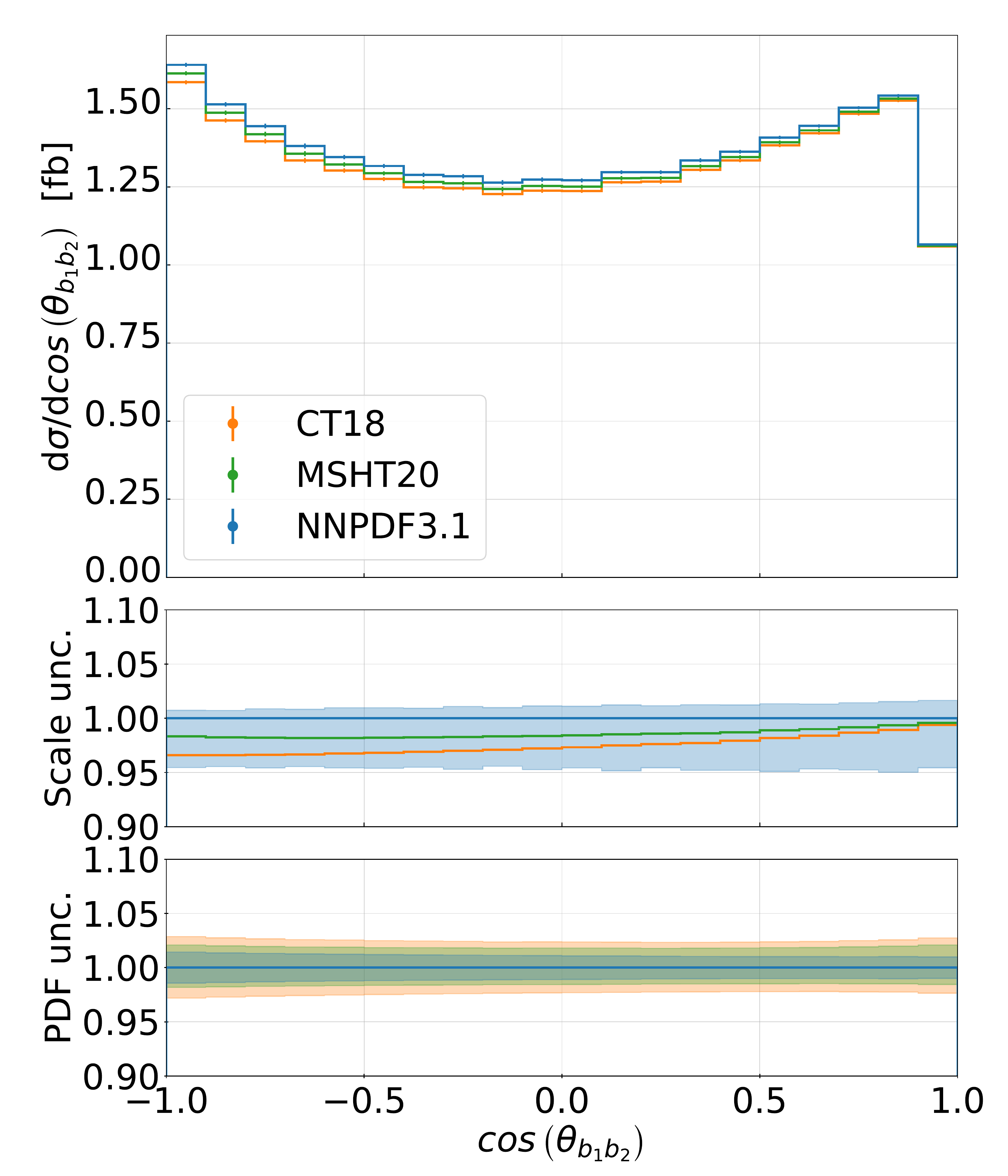}
	\end{center}
	\caption{\label{fig:diff-pdfs} \it
		Differential distributions at NLO QCD for the observables $H_{T}^{vis}$, $M_{b_1b_2}$, $\Delta\phi_{b_1b_2}$ and $\cos\left(\theta_{b_1b_2}\right)$ for the $pp\to e^+\nu_e\mu^-\bar{\nu}_{\mu}b\bar{b}\,H$ process at the LHC with $\sqrt{s}=13\textrm{ TeV}$.  The upper panels show the NLO prediction for three different PDF sets, NNPDF31, MSHT20 and CT18 and with $\mu_0=H_T /2$. Also given are Monte Carlo integration errors. The middle panels display the ratio to the NNPDF3.1 set and its scale dependence. The lower panels give the PDF uncertainties obtained for each PDF set separately.}
\end{figure}

At last, we show in Figure \ref{fig:kfac3} two examples for dimensionless observables. In particular, we display the
azimuthal angle between the two $b$-jets, $\Delta\phi_{b_1b_2}$, and the cosine of the angle between the positron and the muon, $\cos\left(\theta_{e^+\mu^-}\right)$. At LO, for both scale choices, the scale uncertainties are of the order of $30\%$. We find that at the NLO level in QCD for large angles, $\mu_0=H_T/2$ leads to smaller scale uncertainties $(7\%)$ than $\mu_0=\mu_{fix}$ $(14\%)$.  For both scales, the LO and NLO predictions fit within the uncertainty bands and have similarly large QCD corrections for small angles of about $30\%$. For larger angles, the NLO QCD corrections are reduced to about $15\%$ for $\mu_0=H_T/2$ and $5\%$ for $\mu_0=\mu_{fix}$. Therefore, the dynamical scale leads to smaller shape distortions,  i.e. flatter differential ${\cal K}$-factors. We conclude that, for angular distributions, both scale settings can be safely used. However, the dynamical scale choice leads to a reduction of scale uncertainties for the back-to-back phase-space region.

At the integrated level, the scale dependence clearly dominates the total theoretical uncertainties. Unquestionably, this is true at the differential level. Nevertheless, it is important to investigate the size of PDF uncertainties, since it cannot be excluded that they can be enhanced in certain phase-space regions and play a larger role than at the integrated level. To this end, in Figure \ref{fig:diff-pdfs}, differential distributions for the observables $H_{T}^{vis}$, $M_{b_1b_2}$, $\Delta \phi_{b_1b_2}$ and $\cos\left(\theta_{b_1b_2}\right)$ are shown for the dynamical scale setting  $\mu_0=H_T/2$ and for the  following three PDF sets NNPDF3.1, CT18 and MSHT20. Each figure comprises three parts; the upper panels show the NLO prediction for three different PDF sets for $\mu_0=H_T/2$, the middle panels display the ratio to the NNPDF3.1 set and its scale dependence, whereas the lower panels give the internal PDF uncertainties obtained for each PDF set separately. For the observable $H_{T}^{vis}$, that  is defined as $H_T^{vis}=p_{T,\, e^+} + p_{T,\, \mu^-} + p_{T,\, b_1} + p_{T,\,b_2}$, we find that the PDF uncertainties are growing quite fast towards $p_T$ tails. They are at the $5\%$ level for CT18, $3\%$ for MSHT20 and of the order of $2\%$ for NNPDF3.1. For comparison, the scale uncertainties in this region are about $5\%-6\%$. Thus, they are comparable in size to the PDF uncertainties of the CT18 PDF set. The differences between the three PDF sets are similar to the individual PDF uncertainties. The largest variation, around $3 \% $, can be found when comparing CT18 and NNPDF3.1. For $M_{b_1b_2}$ the situation is rather similar. The PDF uncertainties increase towards the tails where they become comparable to the scale uncertainties. The latter are of the order of $8\%$.  Also in this case, the largest differences are found between the CT18 and NNPDF3.1 PDF sets. For $\Delta\phi_{b_1b_2}$ the differences between the PDF sets and the PDF uncertainties are constant over the entire range. For $\cos\left(\theta_{b_1b_2}\right)$ the same is true for the PDF uncertainties. However, the differences between the three PDF sets increase at larger angles. Nevertheless, theoretical predictions for all PDF sets agree well within the corresponding PDF uncertainties, that are of the order of  $3\%$ for CT18, $2\%$ for MSHT20 and $1\%$ for NNPDF3.1. Furthermore, they are smaller than the scale uncertainties, which are at the level of $5\%$ for $\cos\left(\theta_{b_1b_2}\right)$ and $7\%$ for $\phi_{b_1b_2}$. 

To conclude this part of the paper, the PDF uncertainties are more important at the differential than at the integrated level. They play an essential role in the tails of dimensionful observables, where they become similar in size to the corresponding scale uncertainties. For angular distributions we find that the PDF and scale uncertainties are similar in size to those at the integrated fiducial level.

%
\section{Off-shell vs. on-shell top quark modelling}
\label{sec:tth-nwa}
%

In this section we examine the size of full off-shell effects in the modelling of the top quarks at the integrated and differential fiducial level. So far, we have concentrated on the $pp\to e^+\nu_e\, \mu^-\bar{\nu}_\mu \, b\bar{b}\,H +X$ process at NLO in QCD. Specifically, our calculation is based on the full matrix elements for the final state $e^+\nu_e\, \mu^-\bar{\nu}_\mu \, b\bar{b}\,H$, where  all Feynman diagrams of  ${\cal O}(\alpha_s^3\alpha^5)$  are incorporated and no approximations are used.  Thus, all factorizable and non-factorizable NLO QCD corrections to a $2 \to 7$ process are incorporated  together with all the spin correlations. We note here, that non-factorizable NLO QCD corrections imply a cross-talk between production and decays of top quarks. Although we discuss Higgs production in association with a top quark pair, we take into account double-, single- and non-resonant Feynman diagrams, interferences, and off-shell effects of the top quark and the $W$ gauge boson. The latter effects are included via incorporating Breit-Wigner propagators for the unstable particles. Consequently, the full off-shell calculation is free of ambiguities related to disentangling single- and double-resonant top-quark contributions. Such calculations, complete from the point of view of fixed-order perturbation theory, can be very time consuming. Therefore, different approximations are used to obtain results in a reasonable amount of time. A commonly used approximation to calculate processes with unstable particles is the narrow-width approximation. In this approach, the propagator of the unstable particle is expanded in the limit $\Gamma/m\to 0$ leading to
\begin{equation}
	\frac{1}{\left(p^2-m^2\right)^2+m^2\Gamma^2} \quad \longrightarrow \quad \frac{\pi}{m\Gamma}\delta\left(p^2-m^2\right).
\end{equation}
This forces the unstable particle to be on-shell, and suppresses the contributions from the Feynman diagrams without such a resonant propagator. The cross section factorises strictly into on-shell production and on-shell decay. This approximation reduces the complexity of the process enormously, as only double resonant diagrams are taken into account, whereas all single- and non-resonant diagrams and their interferences are systematically neglected. In our case, we treat the two top quarks and the two $W$-bosons in the NWA, which leads to the following decay chain 
\begin{equation}
pp\to t\bar{t}H\to W^+ W^- b\bar{b}\, H\to e^+\nu_e\, \mu^-\bar{\nu}_\mu \, b\bar{b}\,H.
\end{equation}
In this case, the Higgs boson can only be produced in the production of the top quark pair, because $m_t < m_W+m_H$.  With both calculations at our disposal, we can examine the range of applications of theoretical predictions in NWA at the integrated and differential fiducial level. In addition to the NWA and the full off-shell calculation, we will also present the results for a NWA, in which no QCD corrections in the decay of the top quarks are included. We label the latter $\textrm{NWA}_{\textrm{LOdec}}$.  By comparing the NWA and NWA${}_{\rm LO_{dec}}$ predictions, on the other hand,
we can examine how the inclusion of QCD corrections in the top-quark decays
affects our fixed-order NLO calculation at ${\cal O}(\alpha_s^3\alpha^5)$ \footnote{ For the NLO QCD corrections in top-quark decays, the production process is only computed at LO. Therefore, this difference cannot be
used directly as a measure of NLO QCD corrections in top-quark
decays. For this the inclusion of effects that are NNLO in nature is
required.}. In the NWA case, the top quark width as given  in Eq. (\ref{eq:width_nwa}) is used. Specifically, we employ $\Gamma^{\rm NLO}_{t,\, {\rm NWA}}$  for the NWA results and $\Gamma^{\rm LO}_{t,\, {\rm NWA}}$ for  $\textrm{NWA}_{\textrm{LOdec}}$ at NLO. Furthermore, we do not expand the top quark width in a series in the strong coupling constant in the NWA calculation, since this is not possible in the full off-shell calculation and therefore would affect our study of the size of the non-factorisable corrections for the process.  The calculations for the NWA and $\textrm{NWA}_{\textrm{LOdec}}$ are also performed within the \textsc{Helac-Nlo} framework, that has recently been extended to provide theoretical predictions in this approximation \cite{Bevilacqua:2019quz}. The Catani-Seymour dipole formalism is used, and the phase-space restriction on the dipoles is employed to check the independence of the real emission results on this parameter.
\begin{table}[t!]
	\begin{center}
	\begin{tabular}{|lccc|}
		\hline
		&$\mu_0$&$\sigma_{\textrm{LO}}$&$\sigma_{\textrm{NLO}}$\\
		&&$[$fb$]$&$[$fb$]$ \\ \hline
		&&&\\[-0.4cm]
		full off-shell&$H_T/2$&$2.2130(2)^{+30.1\%}_{-21.6\%}$&$2.728(2)^{+1.1\%}_{-4.7\%}$\\[0.2cm]
		&$\mu_{fix}$&$2.3005(2)^{+30.8\%}_{-21.9\%}$&$2.731(2)^{+0.6\%}_{-5.4\%}$\\ [0.1cm] \hline
		&&&\\[-0.4cm]
		NWA&$H_T/2$&$2.2235(2)^{+30.1\%}_{-21.6\%}$&$2.738(1)^{-3.0\%}_{-4.7\%}$\\[0.2cm]
		&$\mu_{fix}$&$2.3074(2)^{+30.7\%}_{-21.9\%}$&$2.742(1)^{-3.8\%}_{-5.3\%}$\\[0.1cm] \hline
		&&&\\[-0.4cm]
		$\textrm{NWA}_{\textrm{LOdec}}$&$H_T/2$&-&$2.862(1)^{+6.3\%}_{-9.4\%}$\\[0.2cm]
		&$\mu_{fix}$&-&$2.897(1)^{+5.1\%}_{-9.0\%}$\\[0.1cm] \hline
	\end{tabular}
	\end{center}
	\caption{\label{tab:nwa} \it
	Integrated fiducial cross section at LO and NLO QCD for the $pp\to e^+\nu_e\mu^-\bar{\nu}_{\mu}b\bar{b}\,H$ process at the LHC with $\sqrt{s}=13\textrm{ TeV}$. Results for three different approaches, full off-shell case,  NWA and $\textrm{NWA}_{\textrm{LOdec}}$,  are shown. Theoretical predictions are given for $\mu_0=H_T/2$ and $\mu_0=\mu_{fix}$. The NNPDF3.1 PDF set is employed. }
\end{table}

In Table \ref{tab:nwa}, the integrated cross section at LO and NLO QCD is shown for $\mu_0=H_T/2$ and $\mu_0=\mu_{fix}$. Results
are obtained with the default PDF set NNPDF3.1. The full off-shell calculation, NWA and $\textrm{NWA}_{\textrm{LOdec}}$ results are presented with the corresponding Monte Carlo error and a theoretical uncertainty estimate obtained  through scale variation. We note here, that in the full off-shell calculation the scale uncertainties are calculated with the $7$-point scale variation while for NWA and $\textrm{NWA}_{\textrm{LOdec}}$ the scale uncertainties are obtained with the $3$-point scale variation. The latter approach, however, leads to similar scale uncertainties as the former one, as discussed in section \ref{sec:tth-int}. At LO, the deviations between the NWA and the full off-shell calculation are about $0.5\%$ for $\mu_0=H_T/2$ and $0.3\%$ for $\mu_0=\mu_{fix}$. Thus, they are negligible compared to the large scale uncertainties of about $30\%$. At NLO, the off-shell effects are about $0.4\%$ for both scale settings and, again, significantly smaller than the scale uncertainties of about $5\%$. These differences are consistent with the expected error of the NWA that is of the order of $\Gamma_t/m_t\approx 0.8\%$ \cite{Fadin:1993kt}. We notice that the maximum of the scale variation for both scales is essentially reproduced in the NWA and that the differences between the full off-shell calculation and the NWA do not exceed $0.5\%$ for all scale choices required in the $3$-point scale variation. The integrated fiducial cross section in the $\textrm{NWA}_{\textrm{LOdec}}$ is about $5\%$ $(6\%)$ larger for $\mu_0=H_T/2$ ($\mu_0=\mu_{fix}$) than the result in the NWA. Thus, the inclusion of QCD corrections in the top-quark decays leads to a reduction of the integrated cross section. Moreover, the scale uncertainties increase to about $9\%$ when the NLO QCD corrections in the top-quark decays are omitted.
\begin{figure}[t!]
	\begin{center}
		\includegraphics[width=0.49\textwidth]{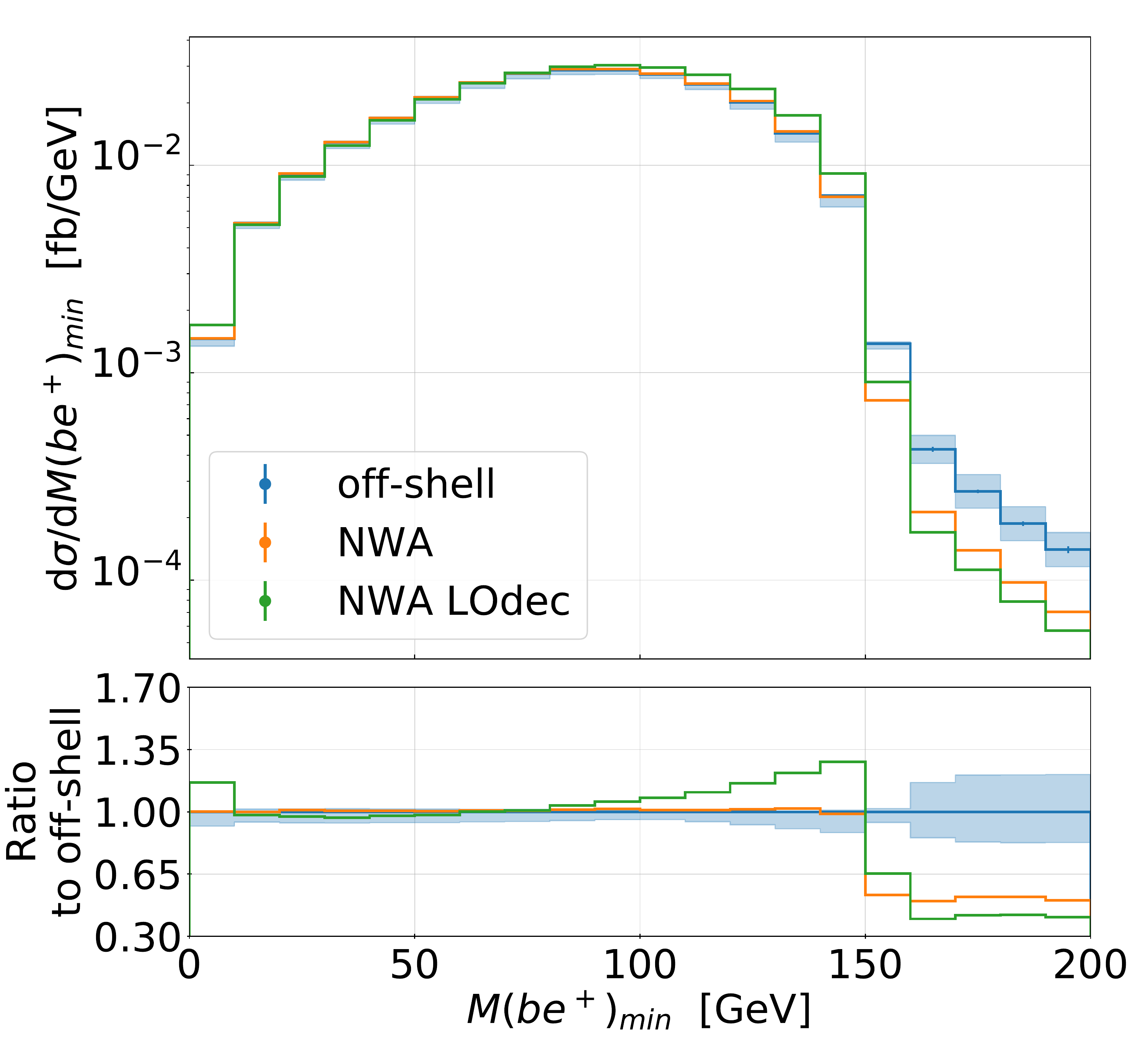}
		\includegraphics[width=0.49\textwidth]{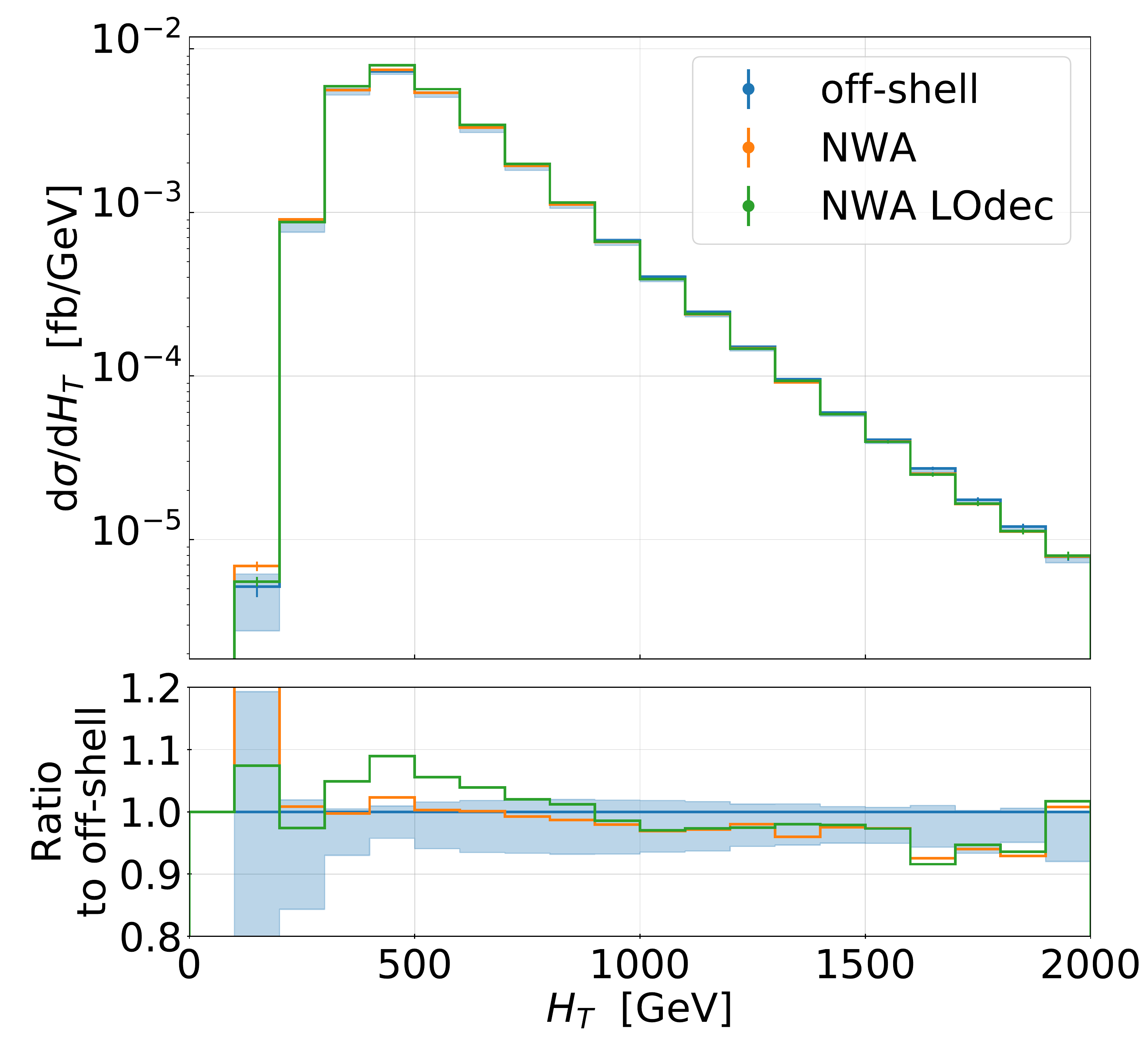}
		\includegraphics[width=0.49\textwidth]{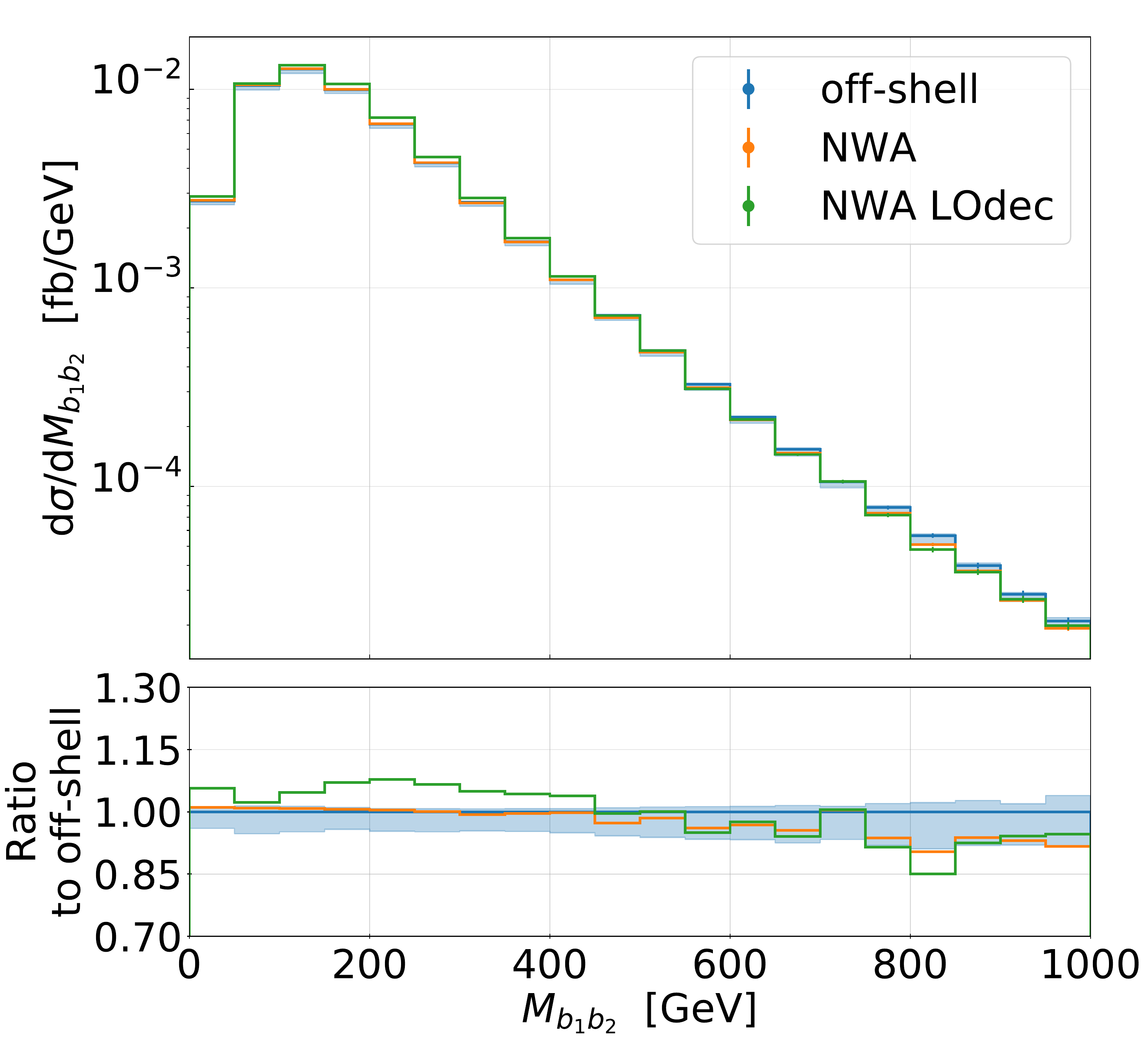}
		\includegraphics[width=0.49\textwidth]{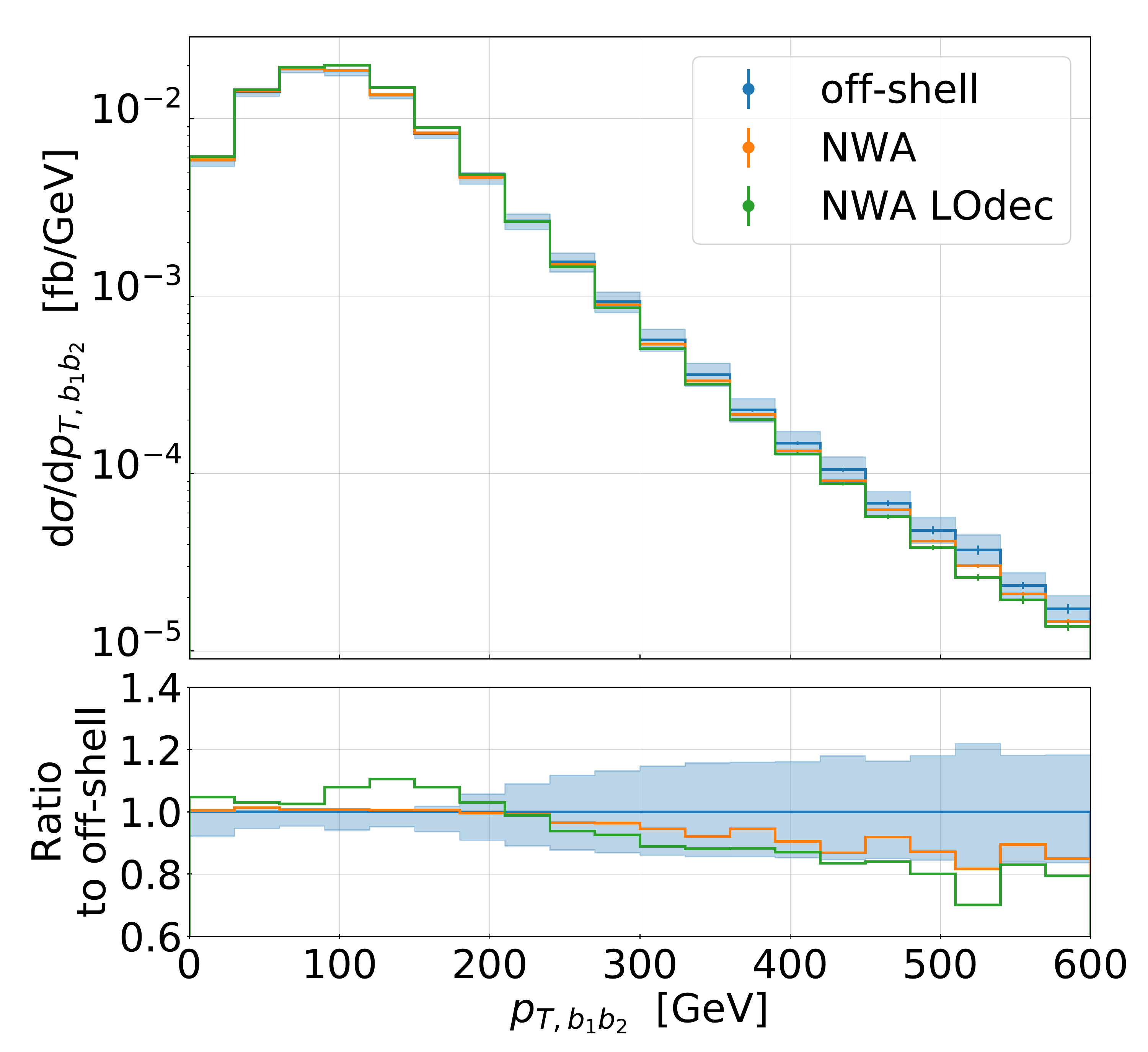}
	\end{center}
	\caption{\label{fig:diff-nwa1} \it
		Differential distributions at NLO QCD for the observables $M(be^+)_{min}$, $H_T$, $M_{b_1b_2}$ and $p_{T,\,b_1b_2}$ for the $pp\to e^+\nu_e\mu^-\bar{\nu}_{\mu}b\bar{b}\,H$ process at the LHC with $\sqrt{s}=13\textrm{ TeV}$. Results are presented for $\mu_0=H_T/2$. The NNPDF3.1 PDF set is employed. The upper panels show full off-shell, NWA and $\textrm{NWA}_{\textrm{LOdec}}$ results. We additionally provide theoretical uncertainties as obtained from the scale dependence for the full off-shell case. Also given are Monte Carlo integration errors. The lower panels display the ratios of the two NWA results to the full off-shell result. }
\end{figure}

Even though full off-shell effects of the top quarks and $W$ gauge bosons are formally suppressed by the corresponding widths and are small for the inclusive cross section, they can be strongly enhanced in exclusive observables. Specifically, for dimensionful observables at the differential level, the contribution from single- and non-resonant Feynman diagrams can be enhanced and even become larger than the contribution from double resonant diagrams in certain phase-space regions. The latter comprise high $p_T$ regions and kinematical edges, that are  related to the masses of the on-shell particles and the cuts that all final states have to pass. As an example, in Figure \ref{fig:diff-nwa1}, we show differential distribution at NLO in QCD for the observables $M(be^+)_{min}$, $H_T$, $M_{b_1b_2}$ and $p_{T,b_1b_2}$. We provide theoretical results with the full off-shell effects included as well as predictions in the NWA and $\textrm{NWA}_{\textrm{LOdec}}$. We utilize $\mu_0=H_T/2$ and the NNPDF3.1 PDF set. The lower panels display the ratios of the corresponding NWA result to the full off-shell one. We additionally provide theoretical uncertainties as obtained from the scale dependence for the full off-shell case. The observable $M(be^+)_{min}$ is defined as the minimum mass of the positron and a $b$-jet. At LO in the NWA, this observable is limited by $\sqrt{m_t^2-m_W^2} \approx 153$ GeV due to the finite mass of the top quarks and $W$-bosons. At NLO, this kinematical edge is smeared out by real radiation, but the NWA is still unable to recover the full calculation in the vicinity of $ M(be^+)_{min} \approx 153$ GeV. For  $M(be^+)_{min} \le 153$ GeV, the NWA leads to good agreement with the off-shell calculation and the differences between the two cases are significantly smaller than the scale uncertainties. The missing corrections in the decay of the top quark lead to further shape distortions in these regions of the phase space, which, this time however, are larger than the scale variation of the full result. For $M(be^+)_{min} \ge 153$ GeV, on the other hand, we obtain substantial non-factorizable corrections that are of the order of $50\%-60\%$. For $H_T$, we observe that the differences between the NWA and the off-shell calculation are much smaller than the scale uncertainties at the beginning of the spectrum. Towards the tails, these differences rise up to about $10\%$, which is roughly the size of the corresponding scale uncertainties. In the tails, NWA and $\textrm{NWA}_{\textrm{LOdec}}$ are very similar, while for $H_T \approx 500$ GeV we find additional shape distortions for the latter approach, close to $10\%$. For $M_{b_1b_2}$ and $p_{T,\,b_1b_2}$ we once more observe that the NWA and the off-shell calculation agree well for small and moderate values. However, in the tails, the differences are equal to the scale uncertainties of about $10\%$ for $M_{b_1b_2}$ and $20\%$ for $p_{T,\,b_1b_2}$. Also here the $\textrm{NWA}_{\textrm{LOdec}}$ leads to additional shape distortions of about $10\%$. 
\begin{figure}[t!]
	\begin{center}
		\includegraphics[width=0.49\textwidth]{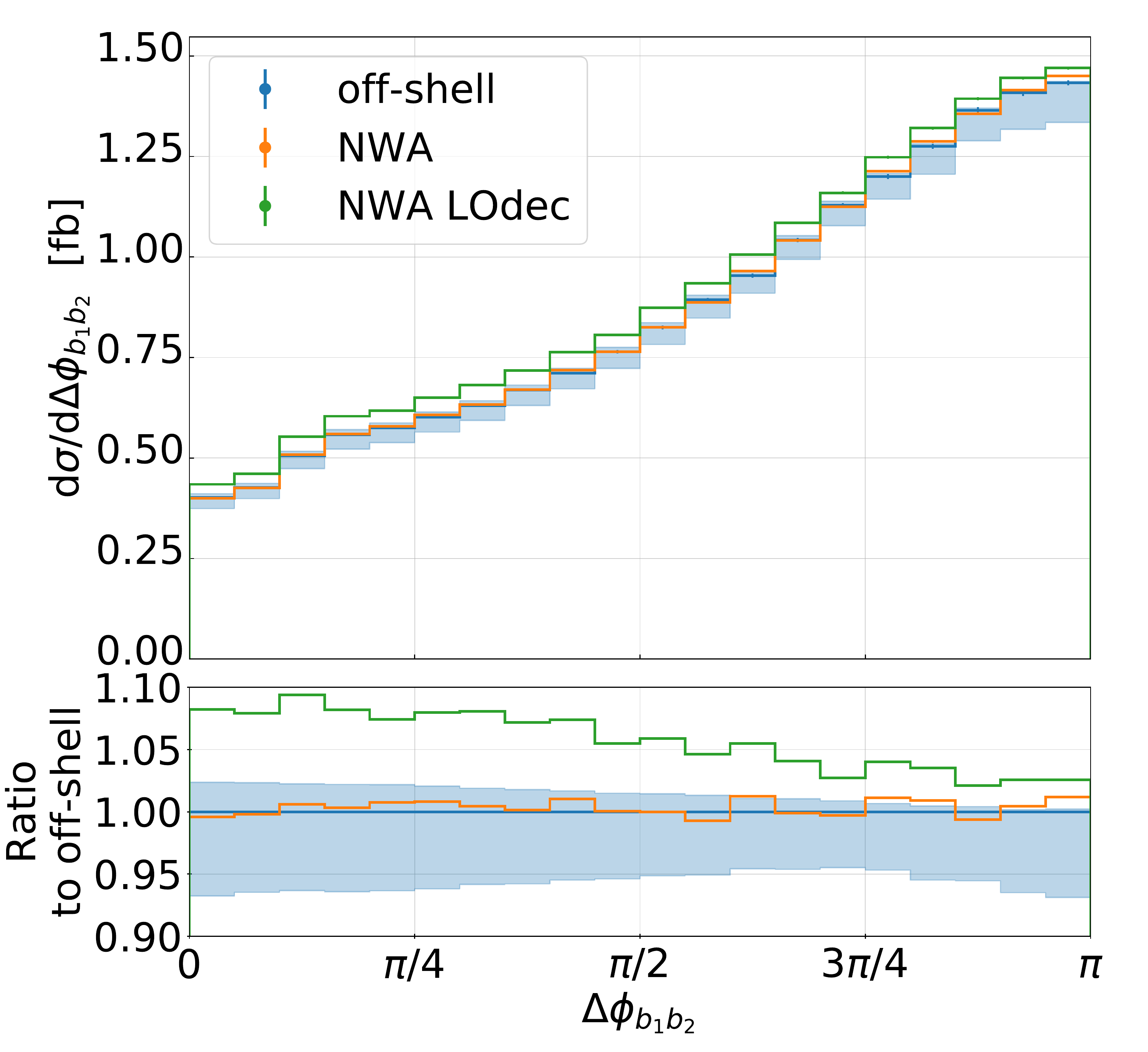}
		\includegraphics[width=0.49\textwidth]{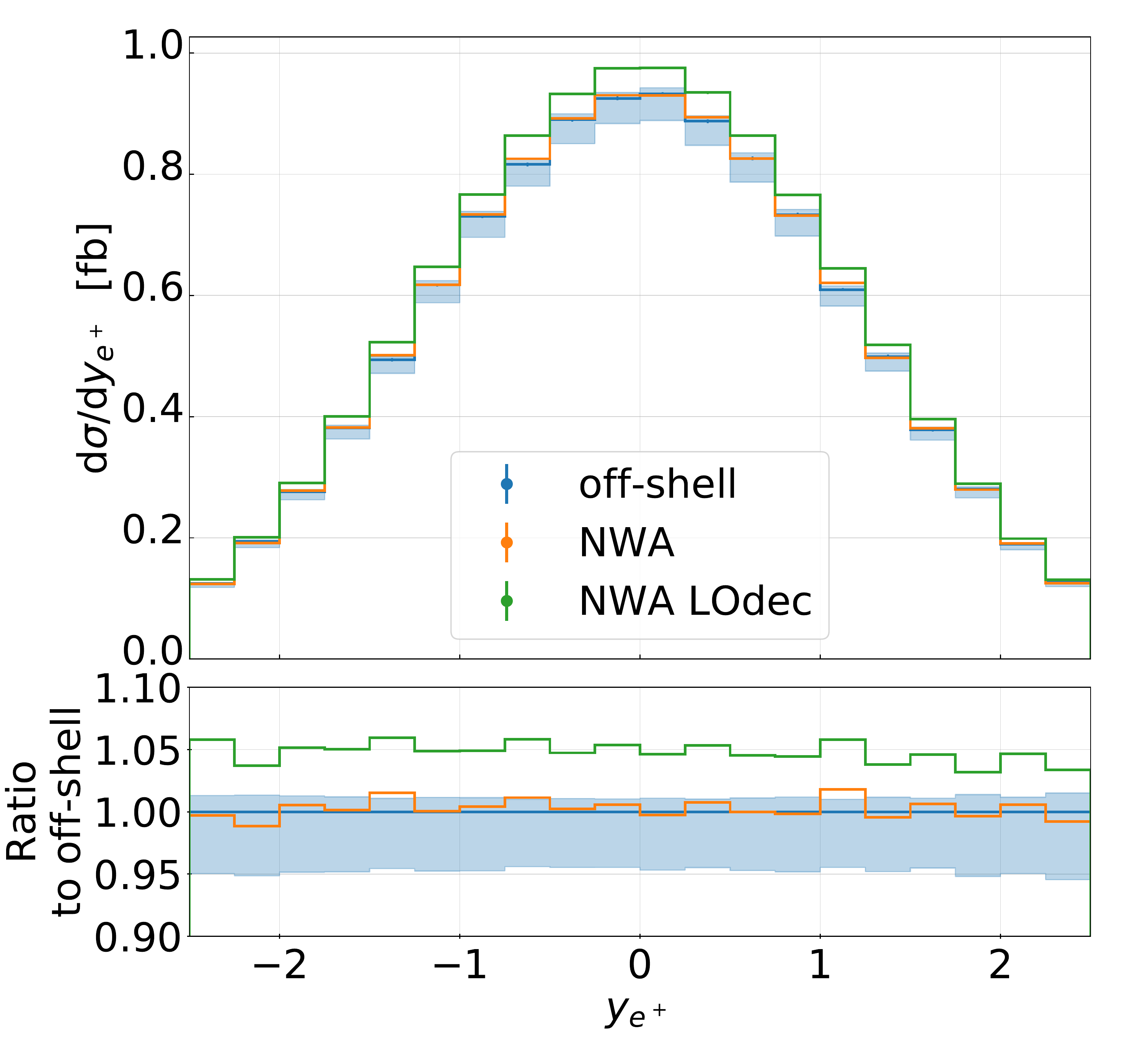}
	\end{center}
	\caption{\label{fig:diff-nwa2} \it
		Differential distributions at NLO QCD for the observables $\Delta\phi_{b_1b_2}$ and $y_{e^+}$ for the $pp\to e^+\nu_e\mu^-\bar{\nu}_{\mu}b\bar{b}\,H$ process at the LHC with $\sqrt{s}=13\textrm{ TeV}$. Results are presented  for $\mu_0=H_T/2$. The NNPDF3.1 PDF set is employed. The upper panels show full off-shell, NWA and $\textrm{NWA}_{\textrm{LOdec}}$ results. We additionally provide theoretical uncertainties as obtained from the scale dependence for the full off-shell case. Also given are Monte Carlo integration errors. The lower panels display the ratios of the two NWA results to the full off-shell result.}
\end{figure}

In Figure \ref{fig:diff-nwa2}, we exhibit differential distributions for the two dimensionless observables $\Delta\phi_{b_1b_2}$ and $y_{e^+}$. For $\Delta\phi_{b_1b_2}$, we find that the NWA is able to recover the full off-shell calculation over the entire range. Thus, full off-shell effects are negligible compared to scale uncertainties. On the other hand, $\textrm{NWA}_{\textrm{LOdec}}$ leads to shape distortions in the range of $2\%-10\%$. The latter cannot be simply corrected by a re-scaling factor. Likewise, for the rapidity of the positron, $y_{e^+}$, we find that full off-shell effects are rather insignificant. Furthermore, the ${\rm NWA}_{\rm LOdec}$ approach introduces an overall normalisation correction of the order of $+5\%$. 

Concluding this part, full off-shell effects can be safely neglected at the integrated level, while they might be substantial for precise predictions at the differential level. For dimensionless observables, the off-shell effects are negligible, but for dimensionful observables close to kinematical edges or in the high-energy tails of various dimensionful distributions, these effects become significant. They are either similar in size or well in excess of the corresponding scale uncertainties. Consequently, they cannot simply be omitted. When studying differential cross sections for the process at hand, a dedicated and careful analysis of such effects is required. We further add that the $\textrm{NWA}_{\textrm{LOdec}}$ does not provide the correct normalisation at the integrated fiducial level. At the differential level, already for some angular distributions, additional shape distortions from the missing NLO QCD corrections in the top quark decays are observed. We also find additional effects for many dimensionful observables over the entire range. These shape distortions are usually larger than the scale uncertainties of the full  off-shell calculation.

%
\section{Bottom quarks in the initial state}
\label{sec:tth-bottom}
%

The last point in the discussion of the $pp \to e^+\nu_e\, \mu^-\bar{\nu}_\mu \, bb\,H$ process with a stable Higgs boson that we would like to scrutinise concerns the contribution of initial state bottom quarks. Such contributions are often neglected in similar calculations due to the suppressed bottom quark PDFs. We investigate the size of these contributions in comparison to the other theoretical uncertainties discussed in the Sections \ref{sec:tth-int} and \ref{sec:tth-diff}. The inclusion of bottom quarks in the initial leads at NLO to additional sub-processes such as $bg\to e^+\nu_e\, \mu^-\bar{\nu}_\mu \, b\bar{b}b\,H$ and
$\bar{b}g\to e^+\nu_e\, \mu^-\bar{\nu}_\mu \, b\bar{b}\bar{b}\,H$, in which we have three bottom quarks in the final state. As we are working in the five-flavour scheme were the $b$ quarks are treated as massless, such sub-processes have an additional infrared singularity because of the gluon splitting into a bottom quark pair, which has to be resolved. This is possible if we take into account the flavour of the bottom quarks in the jet algorithm and require that two $b$-jets with opposite charge are recombined into a gluon $b\bar{b}\to g$, such that these events do not pass the event selection. Such collinear divergences do not affect the recombination of two $b$ partons with the same charge. Thus, for $bb$ and $\bar{b}\bar{b}$ we can choose which definition we would like to apply. In particular, we use two different variants as described in Ref. \cite{Bevilacqua:2021cit}. 

In the first scheme, that we label {\it charge-blind $b$-jet tagging}, the charge of the $b$-jets is neglected. We treat two $b$ partons with the same charge  in the same way as a $b\bar{b}$ pair and also recombine them into a gluon. Therefore, we have the following recombination rules for this scheme:
\begin{equation}
bg \to b\,, \quad\quad \quad \bar{b}g\to \bar{b}\,, \quad\quad \quad  b\bar{b}\to g\,, \quad\quad\quad     
bb\to g\,, \quad\quad\quad  \bar{b}\bar{b}\to g\,.
\end{equation}
Since the charge is neglected, we require that at least two $b$-jets with arbitrary charge  pass the cuts outlined in Section \ref{sec:setup}, otherwise the event is rejected. Consequently, already at LO, besides the sub-process $b \bar{b} \to e^+ \nu_e \, \mu^- \bar{\nu}_\mu \, b\bar{b} \, H$, we must also include the following two additional sub-processes
\begin{equation}
bb\to e^+\nu_e\, \mu^-\bar{\nu}_\mu \, bb\,H\qquad\qquad\qquad \textrm{and}\qquad\qquad\qquad
\bar{b}\bar{b}\to e^+\nu_e\, \mu^-\bar{\nu}_\mu \, \bar{b}\bar{b}\,H.
\end{equation}

In the second scheme, that we label {\it charge-aware $b$-jet tagging}, we no longer neglect the charge of the $b$-jet constituents and require the following recombination rules 
\begin{equation}
bg \to b\,, \quad\quad\quad \bar{b}g\to \bar{b}\,, \quad\quad\quad b\bar{b}\to g\,, \quad\quad   \quad  
bb\to b\,, \quad\quad\quad  \bar{b}\bar{b}\to \bar{b}\,.
\end{equation}
We further require that at least two $b$-jets with opposite charge pass the cuts. The two new sub-processes, which are necessary in the charge-blind scheme, do not pass this requirement and can be safely neglected. From an experimental point of view, the additional charge tagging of the $b$-jets might lead to smaller $b$-tagging efficiency and larger statistical uncertainties. 

In order to validate our results for both schemes, the computation has been performed for two extreme cases of the phase-space restriction parameter of the subtraction terms in the Nagy-Soper subtraction scheme. In this way, we check the independence of the real emission part on this parameter. For each case, a perfect agreement has been found. 
\begin{table}[t!]
	\begin{center}
		\begin{tabular}{|ccccccc|}
			\hline
			&$\mu_0$&$\sigma_\textrm{no\,b}$&$\sigma_{\textrm{aware}}$&$\sigma_{\textrm{blind}}$&$\delta_{\textrm{aware}}$&$\delta_{\textrm{blind}}$\\
			& &$[$fb$]$&$[$fb$]$&$[$fb$]$&& \\ \hline
			&&&&&&\\[-0.4cm]
			LO&$H_T/2$&$2.2130(2)^{+30.1\%}_{-21.6\%}$&$2.2169(2)^{+30.0\%}_{-21.5\%}$&$2.2170(2)^{+30.0\%}_{-21.5\%}$&$0.18\%$&$0.18\%$ \\[0.1cm] \hline
			&&&&&&\\[-0.4cm]
			NLO&$H_T/2$&$2.728(2)^{+1.1\%}_{-4.7\%}$&$2.734(2)^{+1.3\%}_{-4.8\%}$&$2.736(2)^{+1.3\%}_{-4.8\%}$&$0.22\%$&$0.29\%$ \\[0.1cm] \hline
			&&&&&&\\[-0.4cm]
			LO&$\mu_{fix}$&$2.3005(2)^{+30.8\%}_{-21.9\%}$&$2.3044(2)^{+30.7\%}_{-21.9\%}$&$2.3045(2)^{+30.7\%}_{-21.9\%}$&$0.17\%$&$0.17\%$ \\[0.1cm] \hline
			&&&&&&\\[-0.4cm]
			NLO&$\mu_{fix}$&$2.731(2)^{+0.6\%}_{-5.4\%}$&$2.738(2)^{+0.7\%}_{-5.1\%}$&$2.740(2)^{+0.7\%}_{-5.1\%}$&$0.26\%$&$0.33\%$ \\[0.1cm] \hline
		\end{tabular}
	\end{center}
	\caption{\label{tab:bot} \it
	Integrated fiducial cross section at LO and NLO QCD for the $pp\to e^+\nu_e\mu^-\bar{\nu}_{\mu}b\bar{b}\,H$ process at the LHC with $\sqrt{s}=13\textrm{ TeV}$. Results are shown for $\mu_0=H_T/2$ and $\mu_0=\mu_{fix}$. The NNPDF3.1 PDF set is employed. Predictions are given without and with initial state bottom quark contributions in the charge-aware and charge-blind scheme. Last two columns present the ratio $\delta_X=(\sigma_X-\sigma_{\textrm{no b}})/\sigma_X$.}
\end{table}

In Table \ref{tab:bot}, the integrated fiducial cross sections at LO and NLO QCD are given without $(\sigma_\textrm{no\,b})$ and with the initial state bottom quark contribution in the charge-aware and charge-blind scheme (denoted as $\sigma_{\rm aware}$ and $\sigma_{\rm blind}$ respectively). Also shown is the ratio $\delta_X=(\sigma_X-\sigma_{\textrm{no b}})/\sigma_X$, where $X={\rm aware}\,, {\rm blind}$. Furthermore, Monte Carlo errors and scale uncertainties are reported for each case. We use the two scale settings $\mu_0=H_T/2$ and $\mu_0=\mu_{fix}$, and our default PDF set NNPDF3.1. As mentioned before, our dynamical scale setting does not require top-quark reconstruction, and, therefore,  a precise estimate of the bottom-initiated contribution can be straightforwardly obtained. On the other hand, for $\mu_0 = \mu_{dyn}$, an additional procedure would have to be introduced to correctly assign the two $b$-jets to the corresponding top quarks. At LO, we find that the initial state bottom quark contribution amounts to about $0.2\%$ independently of the scale setting and the scheme employed. Indeed, the contribution of the additional $bb$ and $\bar{b}\bar{b}$ channels in the charge-blind scheme is below the integration error of about $0.01\%$. Compared to the scale uncertainties of about $30\%$, the bottom-initiated contribution can be completely neglected at LO for the process at hand.  At NLO, we find that the initial state bottom quark contribution increases due to the additional $gb/g\bar{b}$ channels with up to three $b$-jets in the final state. However, the overall impact is still very small. Specifically, in the charge-blind scheme  the bottom-initiated contribution accounts to $0.3\%$ for both $\mu_0=H_T/2$ and $\mu_0=\mu_{fix}$. In the charge-aware scheme we have $0.2\%$ for $\mu_0=H_T/2$ and $0.3\%$ for $\mu_0=\mu_{fix}$. The bottom-initiated contribution that is more than fifteen times smaller than the corresponding scale uncertainties, has no phenomenological impact and can be safely omitted. As this is a negligible fraction of the integrated fiducial cross section, the choice of a particular scheme is irrelevant. On the other hand, when only sub-processes with initial state bottom quarks are considered, theoretical predictions for the charge-blind scheme are about $25\%-30\%$ larger than the corresponding ones for the charge-aware scheme.
\begin{figure}[t!]
	\begin{center}
		\includegraphics[width=0.49\textwidth]{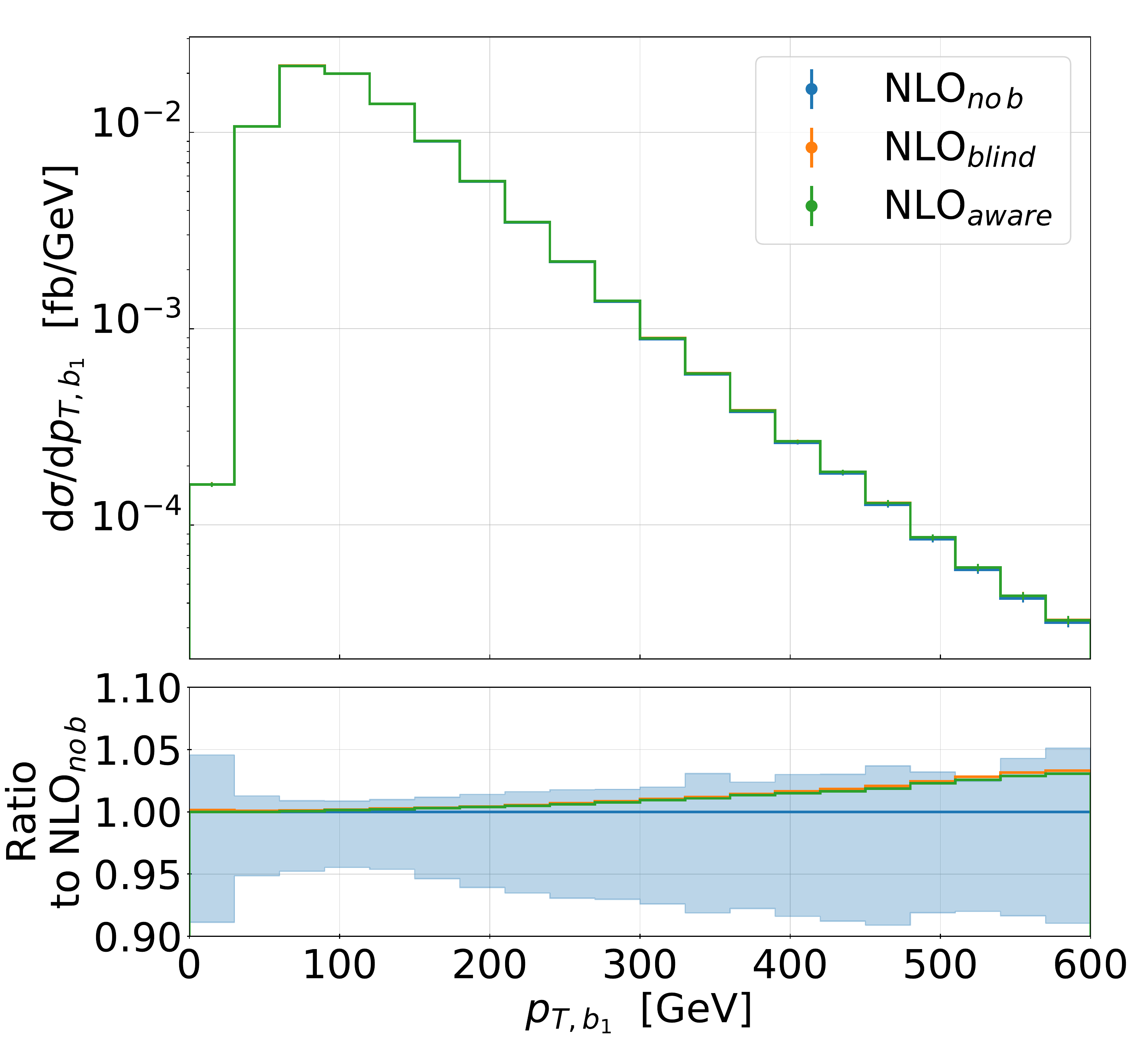}
		\includegraphics[width=0.49\textwidth]{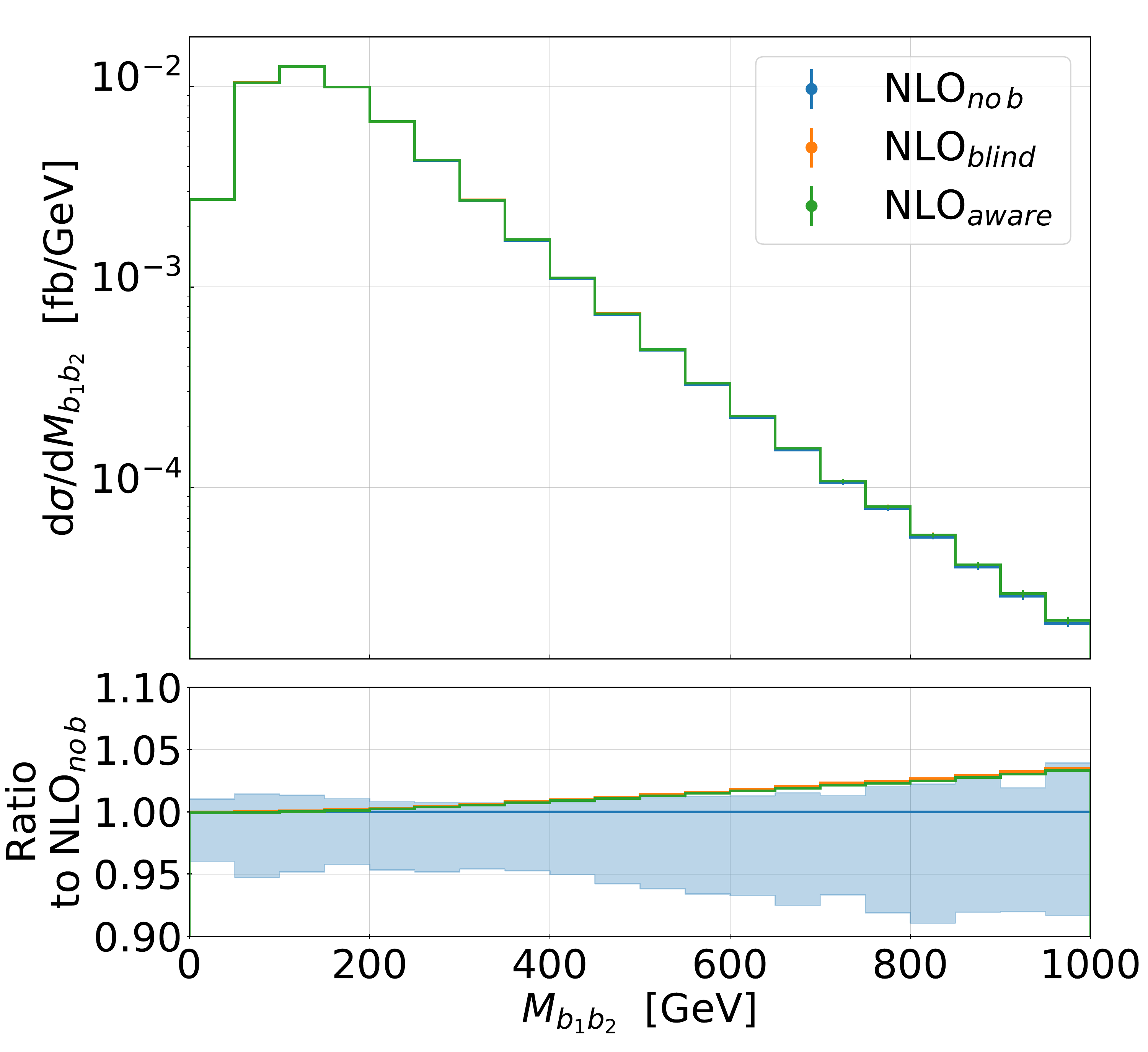}
		\includegraphics[width=0.49\textwidth]{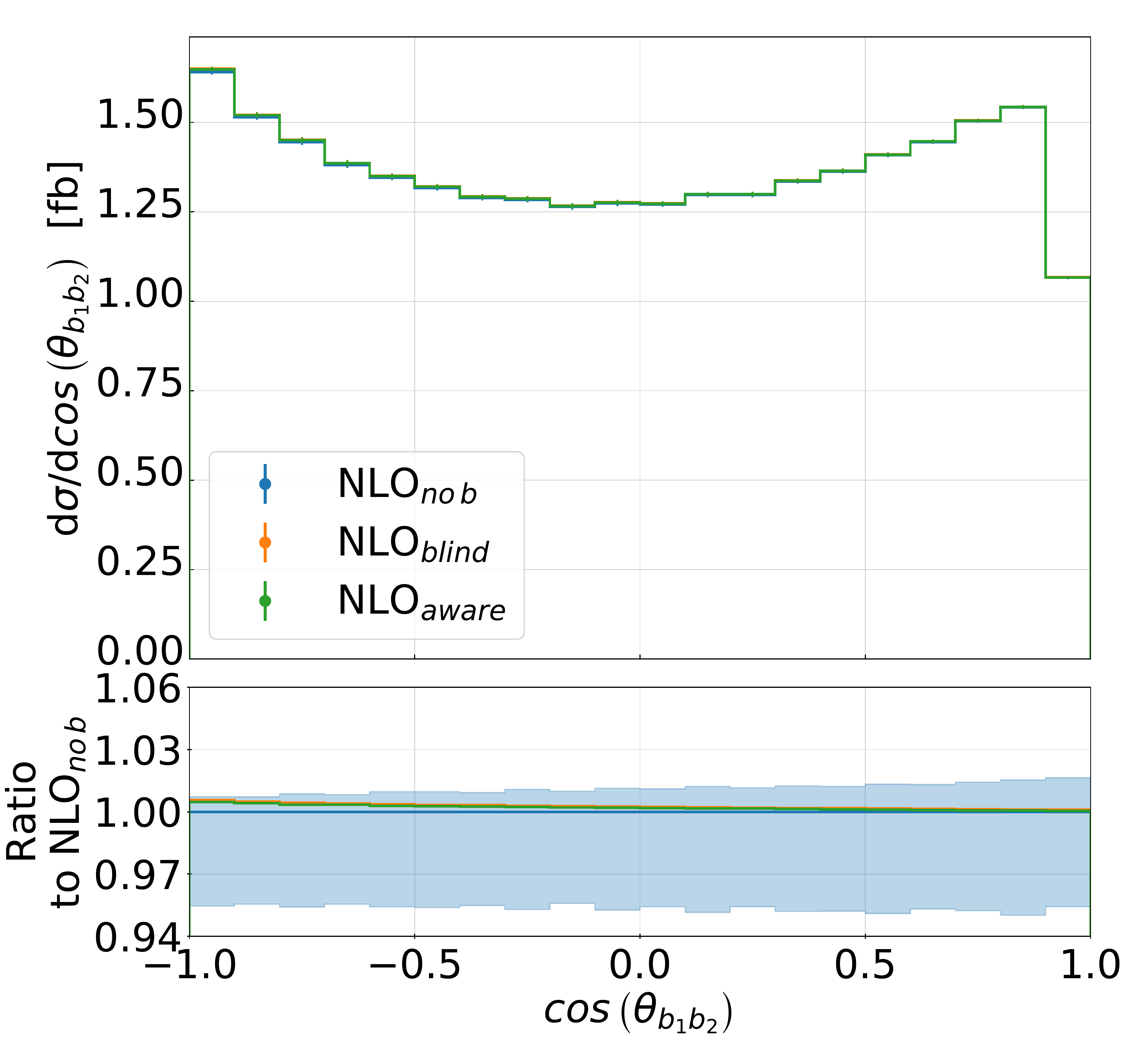}
		\includegraphics[width=0.49\textwidth]{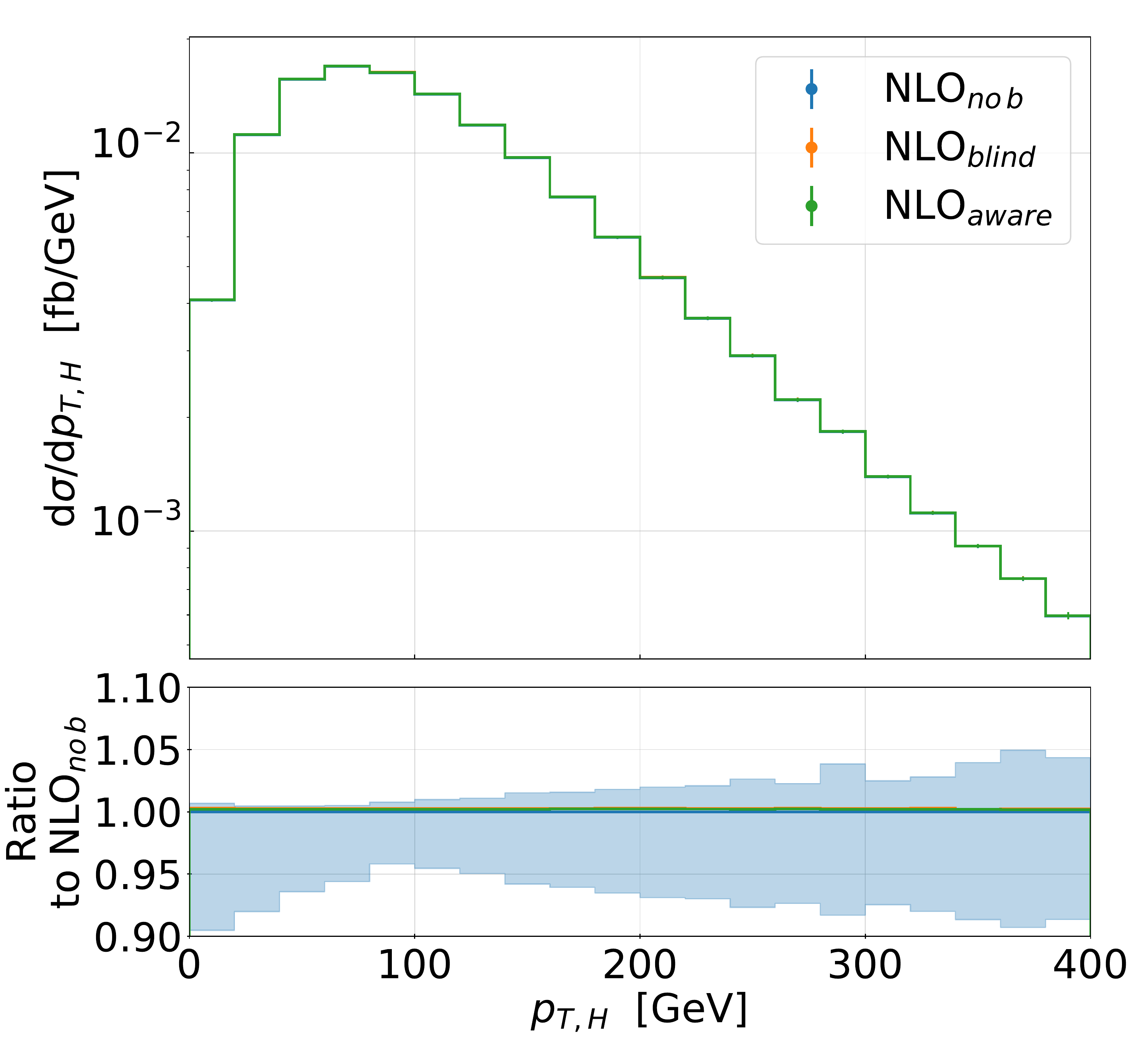}
	\end{center}
	\caption{\label{fig:bot} \it
		Differential distributions at NLO QCD for the observables $p_{T,b_1}$, $M_{b_1b_2}$, $\cos\left(\theta_{b_1b_2}\right)$ and $p_{T,H}$ for the $pp\to e^+\nu_e\mu^-\bar{\nu}_{\mu}b\bar{b}\,H$ process at the LHC with $\sqrt{s}=13\textrm{ TeV}$. Results are given for $\mu_0=H_T/2$ and  NNPDF3.1 PDF set. Theoretical predictions are shown without and with initial state bottom quark contributions in the charge-aware and charge-blind scheme. Also given are Monte Carlo integration errors. The lower panels display ratio to the case without the bottom-initiated contribution together with its scale uncertainties.}
\end{figure}

We also conduct a similar comparison at the differential level to check whether the bottom-initiated contribution is enhanced in some particular regions of the phase space. We expect that the largest impact, if any, should be found in hadronic observables due to the presence of the additional $b$-jet in the final state. To this end, we show in Figure \ref{fig:bot} the following three observables $p_{T,b_1}$, $M_{b_1b_2}$ and $\cos\left(\theta_{b_1b_2}\right)$. The transverse momentum of the Higgs boson is also given in Figure \ref{fig:bot} because we would like to study the impact of the bottom-initiated sub-processes on this important observable. All results are obtained  for $\mu_0=H_T/2$ and the NNPDF3.1 PDF set. Indeed, we observe that for $p_{T,b_1}$ and $M_{b_1b_2}$ the initial state bottom quark contribution increases towards the tails relative to the calculation without this contribution. Specifically, we find effects up to about $3\%$. Compared to the asymmetric scale uncertainties the initial state bottom quark contribution is similar in size, although,  still smaller than the conservative scale uncertainty estimate of about $10\%$. The latter can be obtained by selecting the maximum of the two results in the bin that represents the theoretical error for the observable under consideration. Moreover, we find no significant differences between the two tagging schemes. 

For hadronic angular distributions such as $\cos\left(\theta_{b_1b_2}\right)$, the differences between the calculations do not exceed $0.6\%$ at large angles and decrease even further towards small angles. For large angles, we again find that the initial state bottom quark contribution is similar in size to the asymmetric scale uncertainties but significantly smaller than the conservative scale uncertainties of about $5\%$. Finally, for non-hadronic observables such as the transverse momentum of the Higgs  boson we find no significant effects. 

To sum up this part of the paper, the bottom-initiated contribution is negligible at the integrated and differential level when comparing to theoretical uncertainties as estimated by varying the renormalization and factorization scales in $\alpha_s$ and PDFs. However, due to the fact that in the tails of some dimensionful hadronic observables these contributions can increase up to $3\%$, they can be similar in size or even larger than the corresponding  internal PDF uncertainties. For dimensionless hadronic and non-hadronic observables the initial state bottom quark contribution does not play an important role.

%
\section{Theoretical predictions with Higgs boson decays}
\label{sec:tth-decay}
%

In the last part of the paper we discuss the cross section at the integrated and differential level but this time various decay channels of the Higgs boson are included as well. As it is well known, the Higgs boson can only be detected via its decay products. Thus, it is important to obtain precise theoretical results for fully realistic final states. In our framework, this can be done without further time consuming calculations, because we simply reuse the LHEFs generated for the $pp\to e^+\nu_e\mu^-\bar{\nu}_{\mu}b\bar{b}\,H$  process and model the decay of the Higgs boson in the NWA afterwards.
\begin{table}[t!]
	\begin{center}
	\begin{tabular}{|lccc|}
		\hline
		&$\sigma_{\textrm{LO}}$&$\sigma_{\textrm{NLO}}$&$\mathcal{K}$\\
		&$[$fb$]$&$[$fb$]$& \\ \hline
		&&&\\[-0.4cm]
		Stable Higgs&$2.2130(2)^{+30.1\%}_{-21.6\%}$&$2.728(2)^{+1.1\%}_{-4.7\%}$&1.23\\[0.1cm]
		\hline
		&&&\\[-0.4cm]
		$H \to b\bar{b}$&$0.8304(2)^{+44.4\%}_{-28.7\%}$&$0.9456(8)^{+2.5\%}_{-9.5\%}$&1.14\\[0.1cm]
		\hline
		&&&\\[-0.4cm]
		$H \to \tau^+\tau^-$&$0.11426(2)^{+30.0\%}_{-21.6\%}$&$0.1418(1)^{+1.2\%}_{-4.8\%}$&1.24\\[0.1cm]
		\hline
		&&&\\[-0.4cm]
		$H \to \gamma\gamma$&$0.0037754(8)^{+30.0\%}_{-21.6\%}$&$0.004552(4)^{+0.9\%}_{-4.1\%}$&1.21\\[0.1cm]
		\hline
		&&&\\[-0.4cm]
		$H \to e^+e^-e^+e^-$&$1.0083(7)\cdot 10^{-5}{}^{+30.2\%}_{-21.6\%}$&$1.313(4)\cdot 10^{-5}{}^{+1.8\%}_{-6.2\%}$&1.30\\[0.1cm] \hline
	\end{tabular}
    \end{center}
	\caption{\label{tab:decay} \it
		Integrated fiducial cross section at LO and NLO QCD for the $pp\to e^+\nu_e\mu^-\bar{\nu}_{\mu}b\bar{b}\,H$ process at the LHC with $\sqrt{s}=13\textrm{ TeV}$. Also listed are predictions with the following Higgs boson decays: $b\bar{b}\,, \tau^+\tau^-\,, \gamma\gamma$ and $e^+e^-e^+e^-$.  Results are given for  $\mu_0=H_T/2$ and $\mu_{R,\,dec}=m_H$. The default NNPDF3.1 PDF set is employed.  The last column shows the ${\cal K}$-factor defined as ${\cal K}=\sigma_{\rm NLO}/\sigma_{\rm LO}$.}
\end{table}

In Table \ref{tab:decay}, integrated cross sections are provided for the $pp\to e^+\nu_e\mu^-\bar{\nu}_{\mu}b\bar{b}\,H$ process with  Higgs boson decays into $b\bar{b}$,
$\tau^+\tau^-$, $\gamma\gamma$ and $e^+e^-e^+e^-$. For comparison, predictions with a stable Higgs boson have also been included. All results are obtained for $\mu_0=H_T/2$ with our default NNPDF3.1 PDF set. For the decay into a bottom quark pair, in which NLO QCD corrections are taken into account, we also use the following renormalisation scale setting $\mu_{R, \,dec}=m_H$ in the decay. In addition, Monte Carlo errors and scale uncertainties are presented. Also given in the last column is the $\mathcal{K}$ factor. For all cases, except for the decay into a bottom quark pair, the $7$-point scale variation is used. For $H\to b\bar{b}$ a $21$-scale variation is used instead, where $\mu_{R,\,dec}$ is varied independently of the renormalisation and factorisation scales. The cross sections of the different decay channels are smaller than the naive multiplication of the production cross section with the branching ratio. This is of course due to the additional cuts that are applied on the Higgs boson decay products. We can estimate the number of events in the LHEFs that pass the event selection by the ratio of these two values. For all $1\to 2$ decays approximately  $70\%-80\%$ of events pass the event selection, while for the $1\to 4$ decay chain only about $15\%$ of events  remain. Thus, the cuts on the final state of the last decay mode have a greater effect on the kinematics of this process and also lead to larger statistical uncertainties due to the extremely limited available phase space. For Higgs decays into $\tau^+\tau^-$ and $\gamma\gamma$, we find that the integrated cross section behaves very similarly to the case with a stable Higgs boson. The scale uncertainties at LO and NLO and the $\mathcal{K}$-factor are very similar,  since both decay channels differ only slightly in the overall kinematics due to the cuts. In contrast, for the Higgs boson decay into $e^+e^-e^+e^-$, we observe that the theoretical uncertainties increase from about $5\%$ to $6\%$ and that the $\mathcal{K}$-factor rises from $1.23$ to $1.30$. The differences between these three decay channels, however, are rather small compared to the Higgs boson decay into a $b\bar{b}$ pair. For the latter case, we find already at LO that the scale uncertainties increase from $30\%$ to $44\%$. The growth is caused by the running of the Yukawa coupling in the $\overline{\textrm{MS}}$ scheme that is taken into account in the scale variation. At NLO, the scale uncertainties drop from $44\%$ to about $10\%$, but are still a factor of $2$ larger than the corresponding scale uncertainties as estimated for the case of a stable Higgs boson. In addition, for the $H\to b\bar{b}$ decay, the ${\cal K}$-factor decreases from $1.23$ to $1.14$ leaving us with smaller QCD corrections. 

To investigate the size of NLO QCD corrections in the Higgs boson decay, we again present the integrated fiducial cross section at NLO for the $pp\to e^+\nu_e\mu^-\bar{\nu}_{\mu}b\bar{b}\,H$ process, but this time  the $H\to b\bar{b}$ decay is included only at LO. Our NLO QCD result, as obtained with the NLO Yukawa coupling, is 
\begin{equation}
	\notag \sigma_{\textrm{NLO}_{\textrm{LOdec}_H}}=0.8956(8)^{+13.8\%}_{-14.2\%}\textrm{ fb}.
\end{equation}
We note that, when NLO QCD corrections in the Higgs boson decay are neglected, the scale uncertainties increase from about $10\%$ to $14\%$. Furthermore, we find that $\sigma_{{\rm NLO}_{{\rm LOdec}_H}}$ is about $5\%$ smaller than $\sigma_{\rm NLO}$, where NLO QCD corrections are included both at the production and decay stage. The ${\cal K}$-factor, defined this time as ${\cal K}=\sigma_{{\rm NLO}_{{\rm LOdec}_H}}/\sigma_{\rm LO}$, is equal to ${\cal K}=1.08$. Had we used the same LO value for the Yukawa coupling for $\sigma_{\rm LO}$ and $\sigma_{{\rm NLO}_{{\rm LOdec}_H}}$, we would rather find  ${\cal K}=1.23$. The latter result is in perfect agreement with the ${\cal K}$-factor obtained in the case of a stable Higgs boson. Thus, taking the NLO value of the Yukawa coupling results in smaller NLO QCD corrections for  both cases, $\sigma_{{\rm NLO}_{{\rm LOdec}_H}}$ and $\sigma_{\rm NLO}$. 

At the differential level, we first discuss the $pp\to e^+\nu_e\mu^-\bar{\nu}_{\mu}b\bar{b}\,H$ process with a subsequent $H\to b\bar{b}$ decay. Since the production and decay of the Higgs boson are disentangled, the QCD corrections in the Higgs boson decay do not affect substantially non-hadronic observables except for the overall normalisation. On the other hand, we expect additional shape distortions for hadronic observables that can be constructed from the four $b$-jets. Indeed, the latter can now originate from the production or from the decay of the Higgs boson. For example, in the case of $b$-jets originating from top quark decays, these shape distortions can be caused by the possible recombination of the bottom quark from $t \to b W^+$ or $\bar{t}\to \bar{b}W^-$ decay with a gluon emitted in $H\to b\bar{b}$. Furthermore, an additional gluon in the Higgs boson decay, if resolved as an extra light jet by the jet algorithm and independently of whether it is observed or not, can affect the reconstruction of the Higgs boson system. Such effects can impact various $b$-jet distributions mainly in the soft and collinear regions of the phase space. Thus, we might expect some modifications both in dimensionful and dimensionless observables. 
\begin{figure}[t!]
	\begin{center}
		\includegraphics[width=0.49\textwidth]{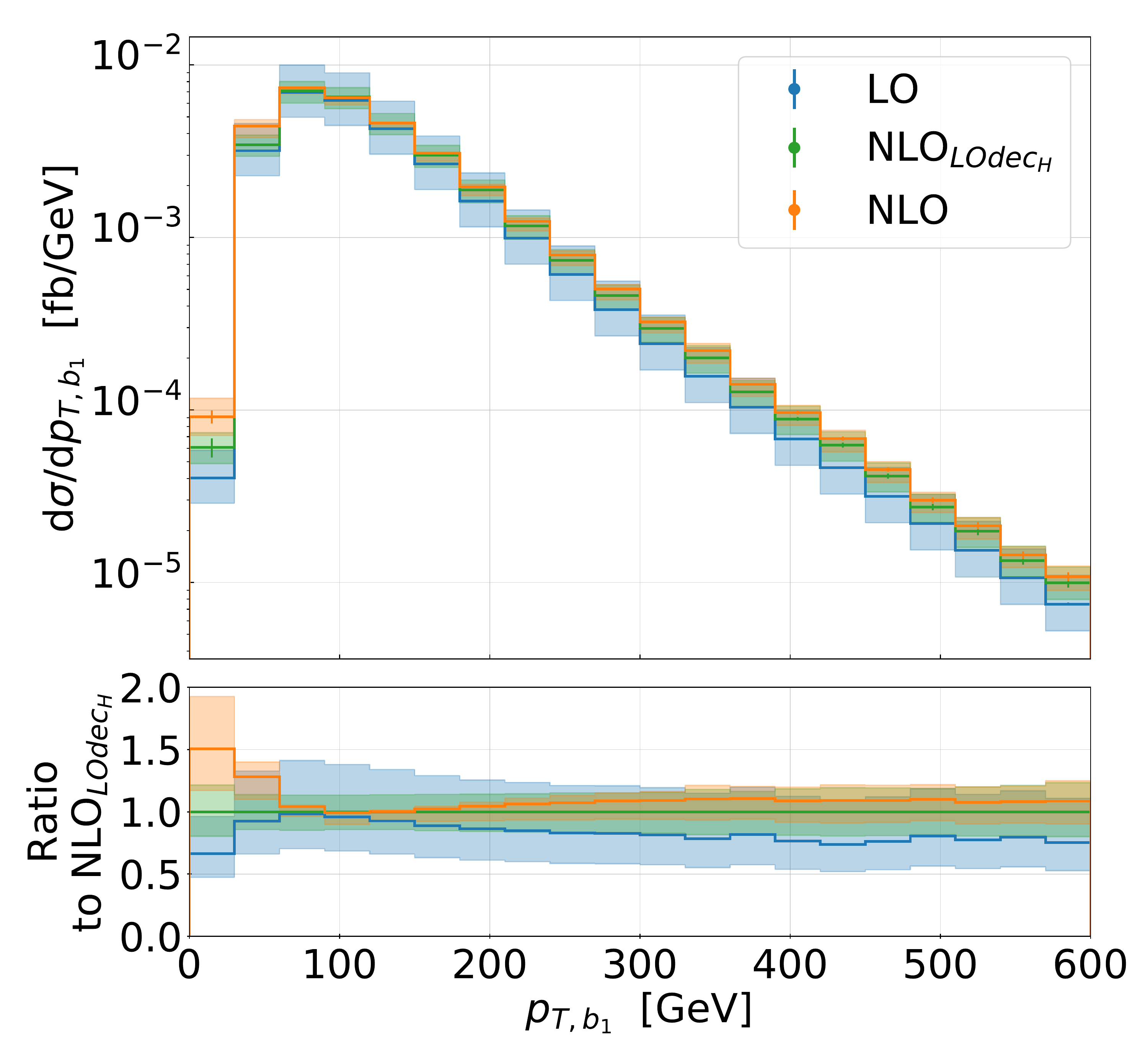}
		\includegraphics[width=0.49\textwidth]{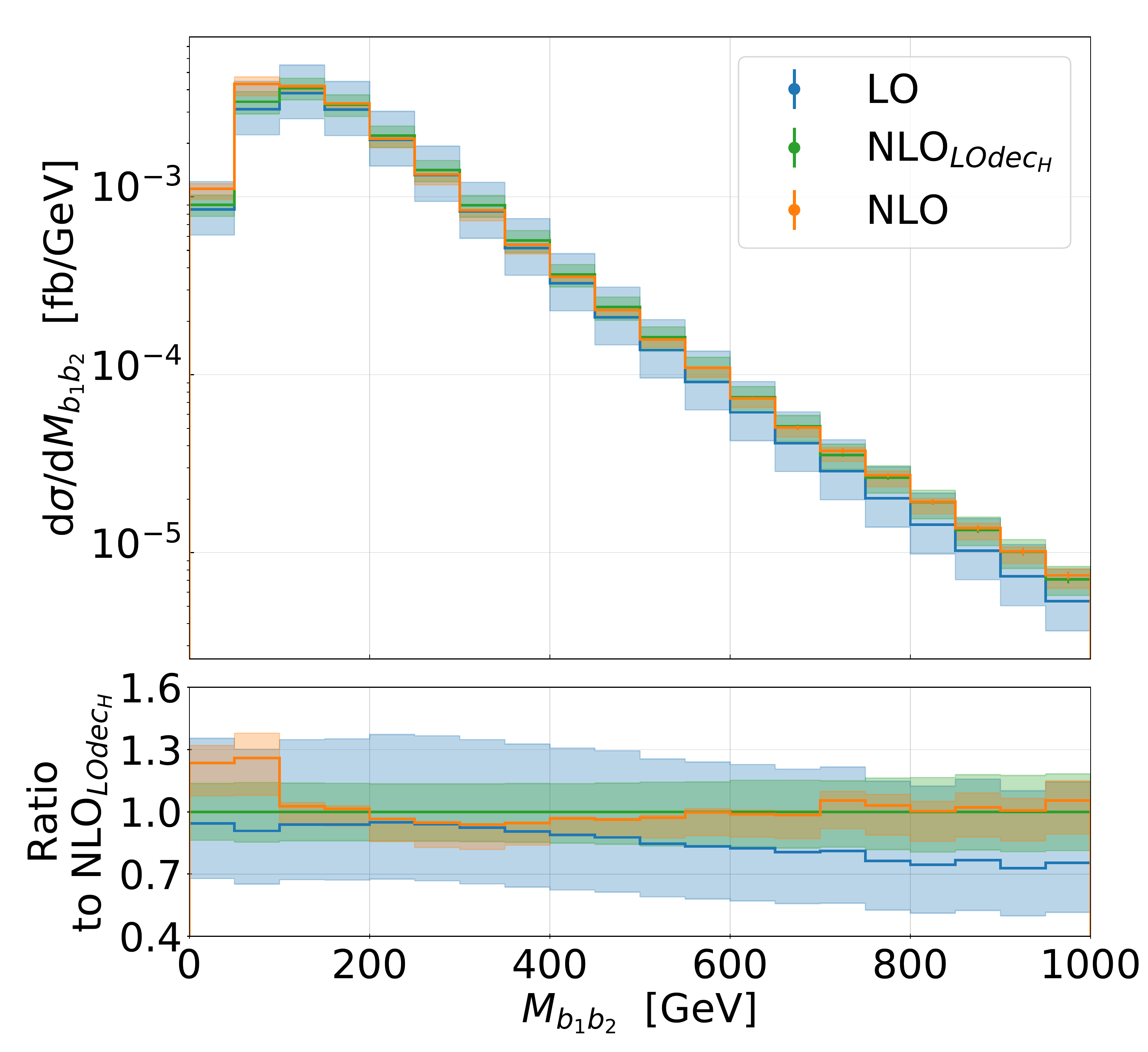}
		\includegraphics[width=0.49\textwidth]{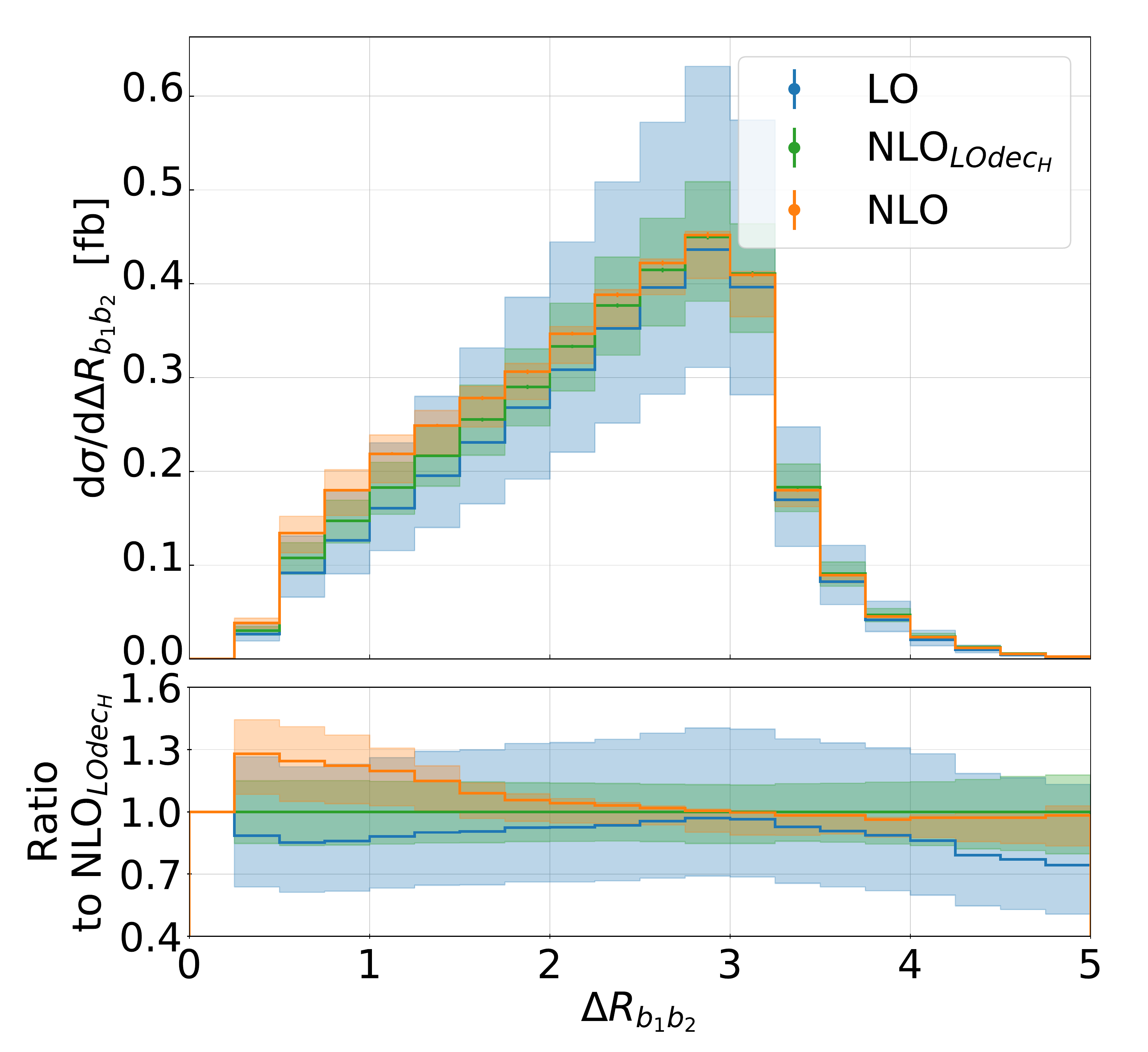}
		\includegraphics[width=0.49\textwidth]{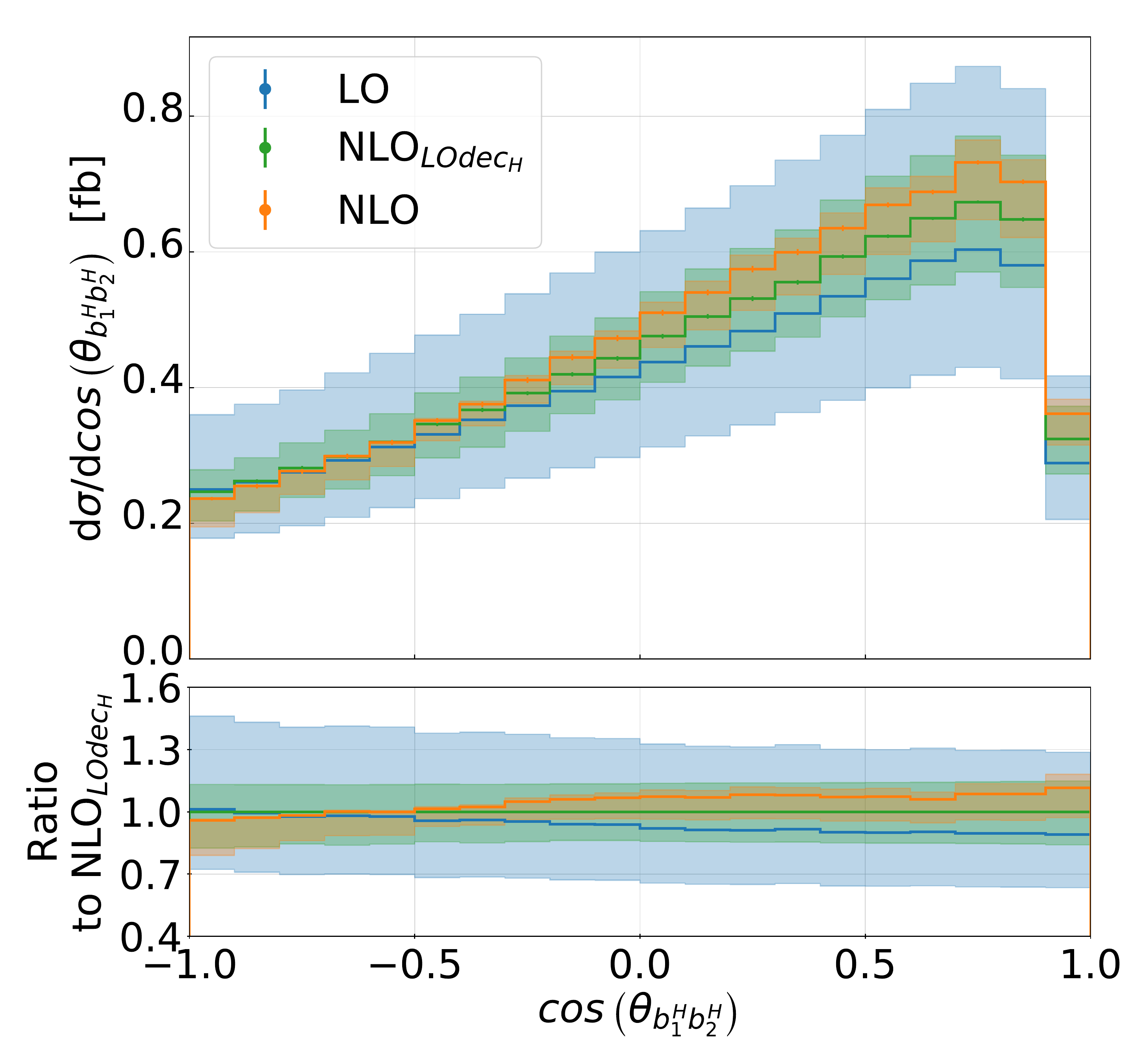}
	\end{center}
	\caption{\label{fig:hbb_lodec} \it
		Differential distributions for the observables $p_{T,\,b_1}$, $M_{b_1b_2}$, $\Delta R_{b_1b_2}$ and $\cos(\theta_{b_1^H b_2^H})$ for the $pp\to e^+\nu_e\mu^-\bar{\nu}_{\mu}b\bar{b}\,H\, (H\to b\bar{b})$ process at the LHC with $\sqrt{s}=13\textrm{ TeV}$. Results are given for $\mu_0=H_T/2$ and $\mu_{R,\,dec}=m_H$. The NNPDF3.1 PDF set is employed. We show predictions at LO, NLO${}_{{\rm LOdec}_H}$ and NLO. Also given are Monte Carlo integration errors. The lower panels display the ratio to the NLO${}_{{\rm LOdec}_H}$ result.}
\end{figure}

In Figure \ref{fig:hbb_lodec}, we show differential distributions for the following observables $p_{T,\,b_1}$, $M_{b_1b_2}$, $\Delta R_{b_1b_2}$ and $\cos(\theta_{b_1^H b_2^H})$ for the $pp\to e^+\nu_e\mu^-\bar{\nu}_{\mu}b\bar{b}\,H\, (H\to b\bar{b})$ process at the LHC with $\sqrt{s}=13$ TeV. We use our default NNPDF3.1 PDF set and employ the following scale settings $\mu_0=H_T/2$ and $\mu_{R,\,dec}=m_H$. Results are again presented at LO, NLO and for the  ${\rm NLO}_{{\rm LOdec}_H}$ case. The lower panels display the ratio to the latter case. We distinguish between $b$-jets coming from the top quarks $(b_1, b_2)$ and the Higgs boson $(b_{1}^{H}, b_2^H)$ via the superscript $H$. The Higgs boson reconstruction is performed as described in Section \ref{sec:setup}. For the transverse momentum of the hardest $b$-jet coming from top quark decays, denoted as $p_{T,b_1}$, we find that the scale uncertainties in the tails are substantially reduced when NLO QCD corrections are included. Indeed, they decrease from about $50\%$ at LO to $20\%$ at ${\rm NLO}_{{\rm LOdec}_H}$. The inclusion of higher-order corrections in the decays of the Higgs boson additionally influences these theoretical uncertainties, reducing them even further to about $15\%$. All three results are in good agreement within the corresponding uncertainty bands. For small values of $p_{T,b_1}$ an enhancement in the range of  $30\%-50\%$ can be noticed. The latter is caused by gluon emission in the Higgs boson decay that affects the Higgs boson reconstruction. On the other hand, in the tails of this distribution, we find only small corrections of the order of $10\%$. For the invariant mass of two $b$-jets originating from the $t\bar{t}$ pair, we observe a very similar reduction in scale uncertainties when going from LO to NLO. Again, all results agree well within the corresponding scale uncertainties. Moreover, at the beginning of the spectrum, the inclusion of the NLO QCD corrections in the Higgs boson decay increases the cross section by about $25\%$, while we observe no  significant corrections in the tails. Looking at the angular separation between the two $b$-jets, denoted as $\Delta R_{b_1b_2}$, we find that, close to $\Delta R_{b_1b_2} \approx 3$, the scale uncertainties decrease from $45\%$ at LO to $15\%$ at $\textrm{NLO}_{\textrm{LOdec}_H}$ and $10\%$ at NLO. This phase-space region corresponds to the characteristic back-to-back production of the two $b$-jets from the top-quark pair. On the other hand, for small values of $\Delta R_{b_1b_2}$, we find an enhancement up to even $30\%$, which clearly shows that these corrections are caused by the mismatch in the Higgs boson reconstruction.  We plan to investigate this issue in more detail in the near  future. To this end, different reconstruction techniques, that mimic  experimental analyses as closely as possible, will be employed  and
their impact on various observables will be investigated. Finally, in the case of the cosine of the two $b$-jets originating from the Higgs boson decays, denoted as $\cos\theta_{b_1^H b_2^H}$, the scale uncertainties decrease from about $45\%$ at LO to $15\%-20\%$ at ${\rm NLO}_{{\rm LOdec}_H}$. The additional NLO QCD corrections in the Higgs boson decay lead to a further reduction of theoretical uncertainties down to about $10\%$. Furthermore,  these corrections are below $5\%$ at large angles and of the order of $10\%$ for small values of $\theta_{b_1^H b_2^H}$. 
\begin{figure}[t!]
	\begin{center}
		\includegraphics[width=0.49\textwidth]{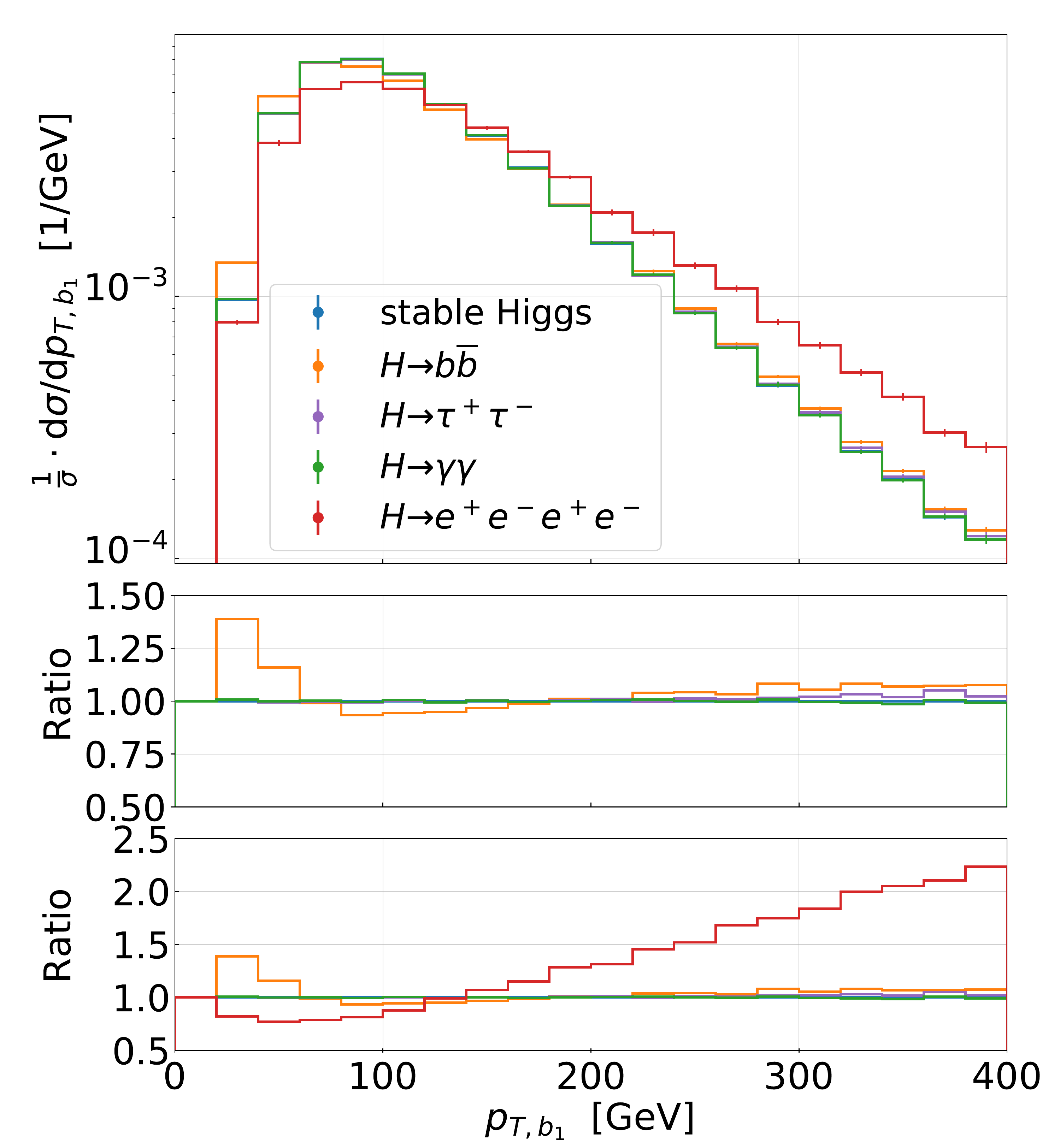}
		\includegraphics[width=0.49\textwidth]{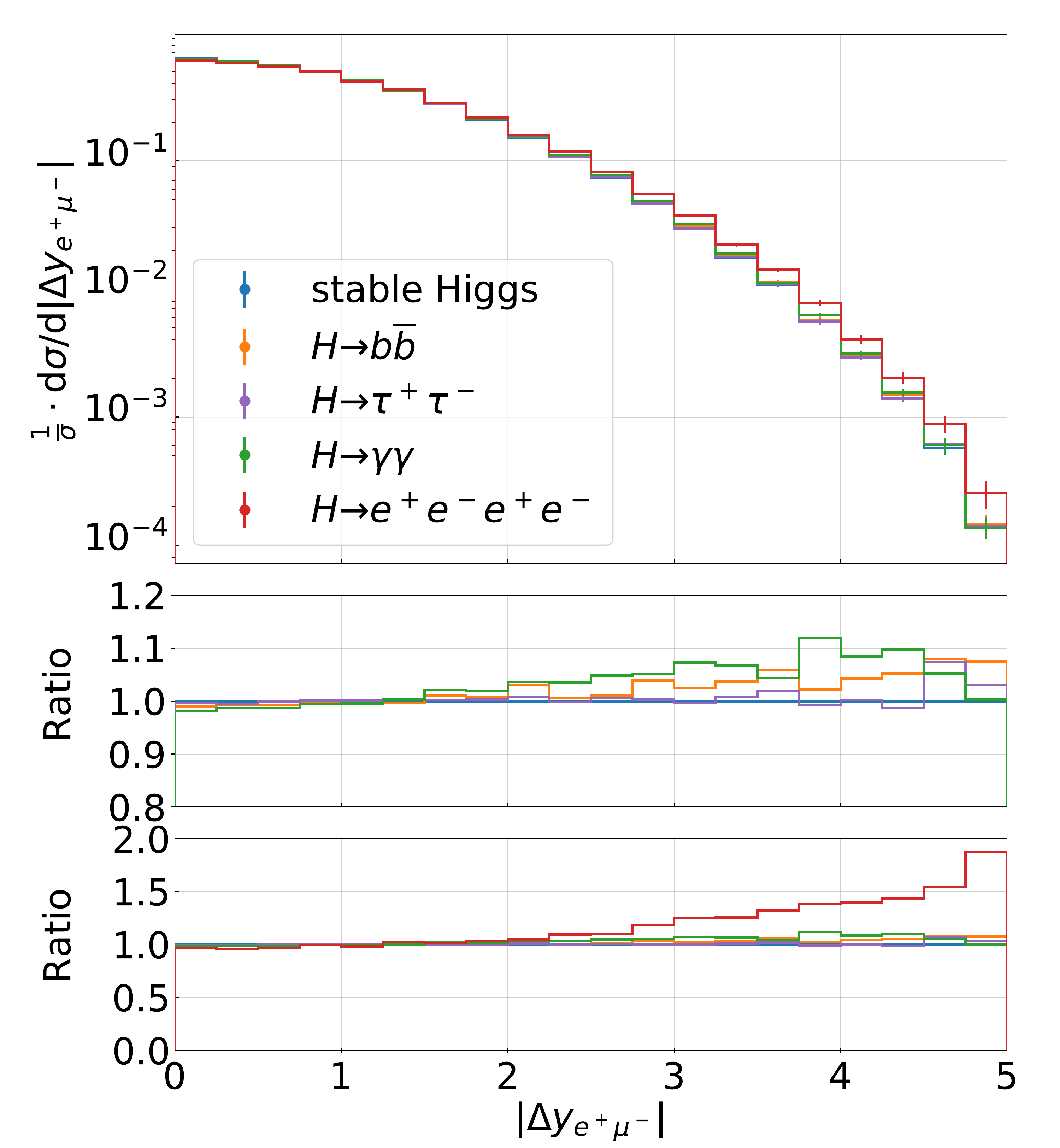}
		\includegraphics[width=0.49\textwidth]{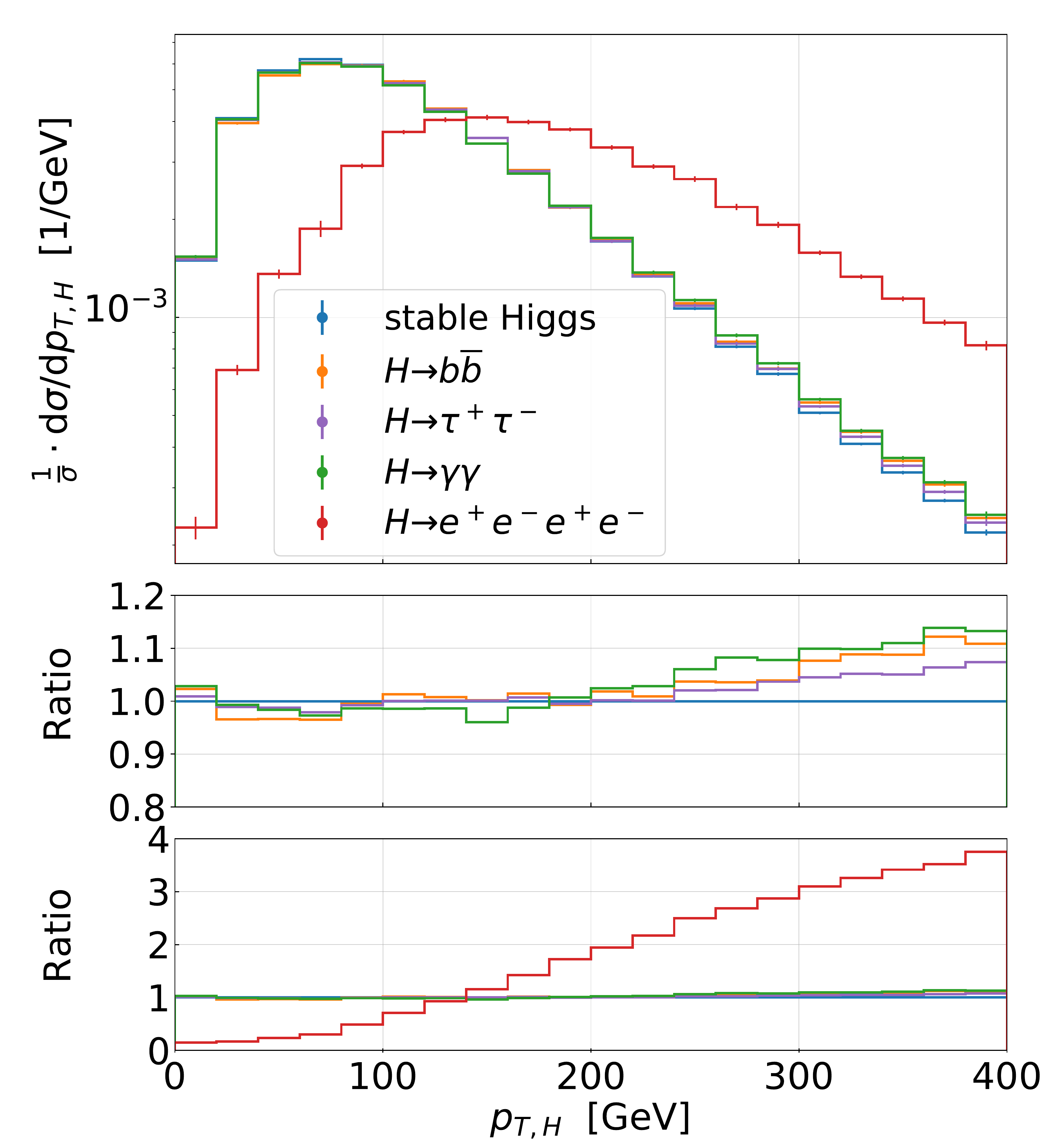}
		\includegraphics[width=0.49\textwidth]{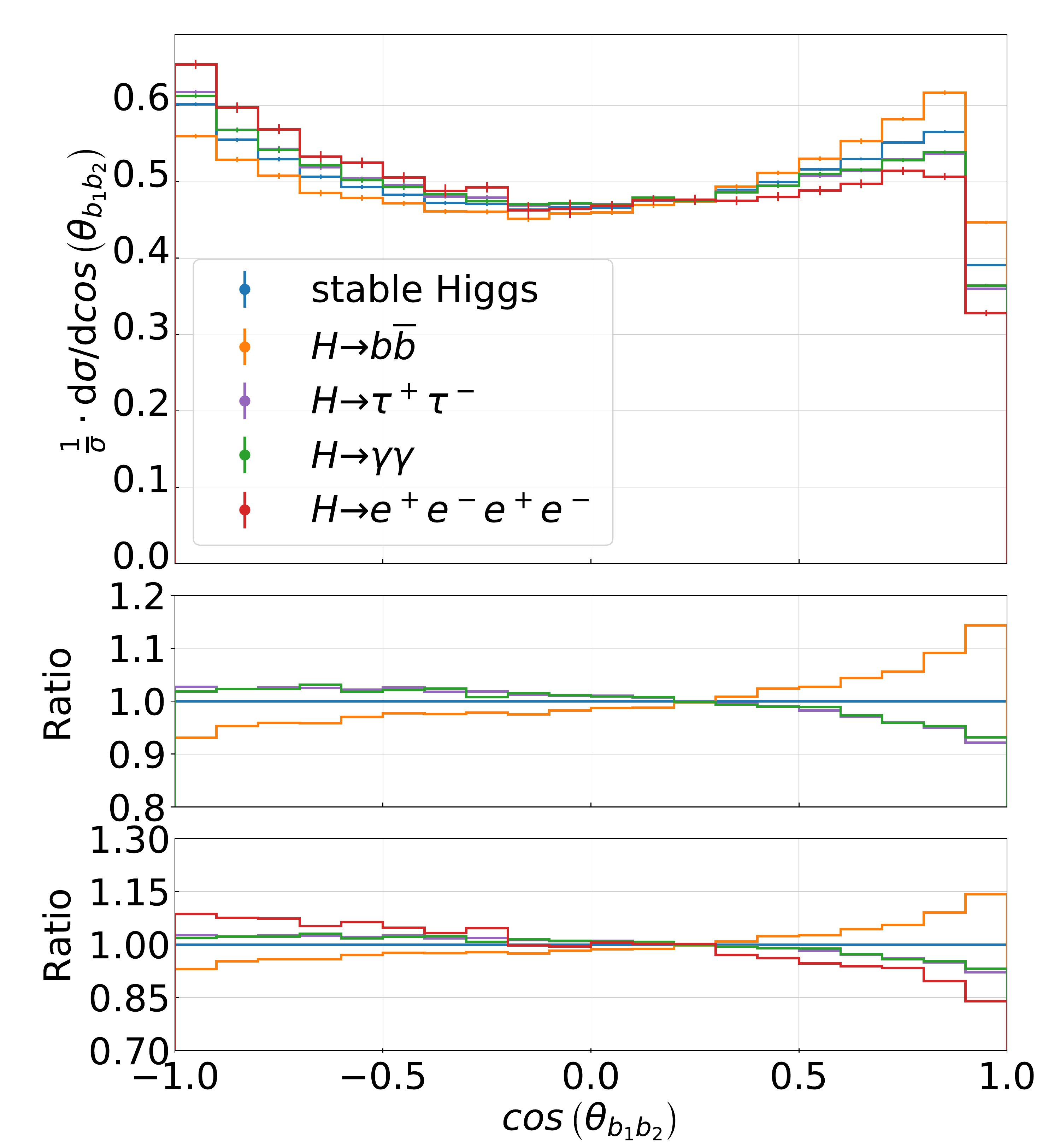}
	\end{center}
	\caption{\label{fig:hdecays} \it
		Normalised differential distributions for the observables $p_{T,\,b_1}$, $| \Delta y_{e^+\mu^-} |$, $p_{T,H}$ and $\cos\left(\theta_{b_1 b_2}\right)$ for the $pp\to
		e^+\nu_e\mu^-\bar{\nu}_{\mu}b\bar{b}\,H$ process at the LHC with $\sqrt{s}=13$ TeV. Results are given for $\mu_0=H_T/2$ and $\mu_{R,\,dec}=m_H$. The NNPDF3.1 PDF set is employed. The case of the stable Higgs is displayed as well as predictions with the Higgs boson decays into $b\bar{b}$, $\tau^+\tau^-$, $\gamma\gamma$ and $e^+e^-e^+e^-$. Also given are Monte Carlo integration errors. The lower panels display the ratio to the case with the stable Higgs boson.}
\end{figure}

In a next step, we discuss the differences between the various Higgs boson decay channels at the differential level. For this purpose, we examine the normalised differential cross-section
distributions to assess whether only the overall normalisation is different, or whether we can
find further shape distortions. In Figure \ref{fig:hdecays} we present results for the $pp\to
e^+\nu_e\mu^-\bar{\nu}_{\mu}b\bar{b}\,H$ process with a subsequent Higgs boson decay into
$b\bar{b}$, $\tau^+\tau^-$, $\gamma\gamma$ or $e^+e^-e^+e^-$. For comparison, also the case with
a stable Higgs boson is displayed. We present the following observables $p_{T,\,b_1}$, $| \Delta
y_{e^+\mu^-} |$, $p_{T,H}$ and $\cos\left(\theta_{b_1 b_2}\right)$. The lower panels display the
ratio to the case with a stable Higgs boson. We use the same scale settings and PDF set as for
the integrated fiducial cross section.  As in the previous discussion, we distinguish the decay products from the top quark and Higgs decays by using the Higgs boson reconstruction procedure outlined in Section \ref{sec:setup}. We use the Higgs boson reconstruction for all decay channels, but for the decay channels without bottom quarks the MC truth and the Higgs boson reconstruction is exactly the same due to the absence of additional electroweak radiation.  For the transverse momentum of the hardest $b$-jet, we find that the decay into four electrons leads to a completely different shape. This is not surprising, of course, since we are talking about the decay pattern $1 \to 4$. In particular, the $p_{T,b_1}$ spectrum is significantly harder due to the strict event selection. This observation essentially holds for all dimensionful observables. By comparison, our results for all $1\to 2$ decay modes are quite similar. The NLO QCD corrections in the $H\to b\bar{b}$ decay lead to an enhancement at the beginning of the spectrum up to $40\%$. However, in the tails, the decay channel $b\bar{b}$ differs only by about $10\%$ from the case with a stable Higgs boson. For the remaining two decay channels, $\tau^+\tau^-$ and $\gamma\gamma$, differences up to $2\%$ are only observed. For the  rapidity separation of the two charged leptons $|\Delta y_{e^+\mu^-}|$, we find that all results are very similar for small values. If the two leptons are well-separated in rapidity, the differences increase, especially for the Higgs boson decay into four electrons. For the $1\to 2$ decay channels, discrepancies are below $10\%$. Indeed, in the case of $H \to b\bar{b}$ and $H \to \gamma\gamma$ we find differences up to $8\%$, which are reduced to $3\%$ for the $H\to \tau^+\tau^-$ case. For the transverse momentum of the Higgs boson, we find differences up to about $10\%$ for $\gamma\gamma$, $b\bar{b}$ and for $\tau^+\tau^-$. Thus, the inclusion of QCD corrections in the Higgs boson decay into $b\bar{b}$ does not lead to significant enhancements in any specific phase-space region. However, these differences are comparable in size to the scale uncertainties and thus can not be simply ignored. We note, of course, that for the $H \to e^+e^-e^+e^-$ decay chain, the $p_{T,H}$ spectrum has a completely different shape. Last but not least, for $\cos\left(\theta_{b_1 b_2}\right)$ we observe for $b\bar{b}$ once again that the region of small angles is enhanced by up to $15\%$ due to the mismatch in the reconstruction of the Higgs boson. Similar effects in absolute value can be observed for the $H \to e^+e^-e^+e^-$ decay chain. We find that the other two decay channels lead to similar results, which still differ by about $7\%-8\%$ from the prediction for a stable Higgs boson.  Also in this case, these shape distortions are similar in size to the scale uncertainties. For large values of $\theta_{b_1 b_2}$, smaller discrepancies well below $10\%$, can be noticed for $H \to b\bar{b}$ and $H \to e^+e^-e^+e^-$. They are again similar in absolute magnitude but opposite in sign. 

%
\section{Summary}
\label{sec:sum}
%

In this paper, we discussed the production and decay of the Higgs boson in association with a $t\bar{t}$
pair. In particular, we calculated NLO QCD corrections to $pp\to e^+ \nu_e \mu^- \bar{\nu}_{\mu} b\bar{b}\, H\, (H\to X)$ at the LHC with $\sqrt{s}=13\textrm{ TeV}$ including full off-shell effects of the top quarks and $W$ gauge bosons. In the computation these unstable particles have been described by Breit-Wigner distributions. Furthermore double-, single- and non-resonant top quark/$W$ gauge boson contributions as well as all interference effects have been consistently incorporated already at the matrix element level. The calculation of the production process was performed within the \textsc{Helac-Nlo} framework. The decays of the Higgs boson were modelled in the NWA with the help of the LHEFs. 

For Higgs production, we first compared our results with the findings from Ref.
\cite{Denner:2015yca} both at the integrated and differential level. For the two scale settings,
$\mu_0=\mu_{dyn}$ and $\mu_0=\mu_{fix}$, that were employed there, we found an overall good
agreement between the two calculations. In the next step, we introduced an additional scale choice, $\mu_0=H_T/2$, and performed a comparison between the three scales. At the integrated fiducial level, we found that, for all scale settings, the NLO QCD corrections are of the order of $20\%$. The scale uncertainties decreased from about $30\%$ at LO to $5\%$ at NLO. All results are in good agreement within the corresponding scale uncertainties. Furthermore, we reported the PDF uncertainties for the following PDF sets NNPDF3.1, CT18 and MSHT20. These PDF uncertainties
are up to only $1\%-2\%$. Thus, the theoretical uncertainties due to the scale dependence are the
dominant source of the theoretical systematics.

A similar comparison was performed at the differential level. We found that the fixed scale setting led to perturbative instabilities for dimensionful observables in high-energy tails. Even at the level of differential cross sections, the fixed scale choice resulted in larger scale uncertainties. The differences between the other two dynamical scales are only minor. For $\mu_0=H_T/2$ we obtained NLO QCD corrections up to $20\%-30\%$ and scale uncertainties of the order of $5\%-10\%$. In addition, the PDF uncertainties can increase in the tails of dimensionful observables. Specifically, for the CT18 PDF set we observed PDF uncertainties of about $5\%-7\%$. The latter are similar in size to the corresponding scale uncertainties. However, for our default NNPDF3.1 PDF set, the PDF  uncertainties decreased to $2\%-3\%$ only. Therefore, for our default setup, also at the differential level, the theoretical error is dominated by the theoretical uncertainties resulting from the scale variation.

In the next step, we discussed the off-shell effects of the top quarks by comparing the full off-shell calculation with the results in the NWA and $\textrm{NWA}_{\textrm{LOdec}}$ at the integrated and differential level. At the integrated level, the off-shell effects are of the order of $0.3\%-0.5\%$. The missing NLO QCD corrections in the decays of the top quarks increase the integrated fiducial cross section by about $5\%$ and the scale uncertainties to $9\%$. At the differential level, the full off-shell top quark effects are significant close to kinematical edges and in high energy regions of the phase space. In the tails of dimensionful observables the full off-shell effects are in the range of $10\%-20\%$ and thus similar in size to  the scale uncertainties. In the case of special observables, like for example $M(be^+)_{min}$, they can be as high as $50\%-60\%$. For angular distribution, the off-shell effects do not play a crucial role. In addition, the $\textrm{NWA}_{\textrm{LOdec}}$ leads to a wrong normalisation and further shape distortions in dimensionful observables and even angular distributions. 

Subsequently, we studied the bottom-induced contributions to the process at hand. The latter is often neglected in similar calculations due to the suppressed bottom quark PDF. We used two different schemes for $b$-jet tagging, the so-called charge-blind and charge-aware tagging of $b$-jets, to obtain IR-safe results. At the integrated level the contribution is negligible of the order of $0.2\%-0.3\%$ only. At the differential level the contribution from the initial state bottom quarks can be enhanced in the tails of dimensionful hadronic observables, reaching up to $3\%$. The two different schemes lead essentially to the same phenomenological results.

Last but not least, we combined $t\bar{t}H$ production with Higgs boson decays. We considered $H\to b\bar{b}$, $H\to  \tau^+\tau^-$, $H\to \gamma\gamma$ and $H\to e^+e^-e^+e^-$. At
the integrated fiducial level, we observed that the scale dependence and the $\mathcal{K}$-factor are very similar to the case of a stable Higgs boson for all decays except for $H \to b\bar{b}$. For this decay channel, we included NLO QCD corrections in the decay and performed the renormalisation of the bottom Yukawa coupling in the $\overline{\rm MS}$-scheme, which led to a scale-dependent Yukawa coupling. Consequently, the scale uncertainties are about $45\%$ at LO and of the order of $10\%$ at NLO. The $\mathcal{K}$-factor is $1.14$ and, therefore, smaller than the one for the case of a stable Higgs boson. This reduction is caused by the smaller Yukawa coupling of the bottom quark at NLO. Without corrections in the decay the scale uncertainties increased to $14\%$ and the cross section decreased by about $5\%$.  At the differential level the scale dependence is reduced by about $5\%$ due to the NLO QCD corrections in the decay. We obtain large shape distortions of about $25\%$ for $b$-jet observables at lower values of dimensionful observables and for small opening angles due to the Higgs reconstruction. On the other hand, the Higgs boson decay into four electrons leads to completely different normalised distributions. The latter effect can be especially visible for dimensionful observables, where significantly harder spectra are observed due to the strict cuts on the four charged leptons. For the Higgs boson decays into $\tau^+\tau^-$ and $\gamma\gamma$ in normalised differential distributions shape distortions up to $14\%$ have been found, compared to the case with a stable Higgs boson. 

To recapitulate, non-factorisable NLO QCD corrections as well as higher-order QCD effects in top-quark decays impacted significantly the $pp\to e^+ \nu_e \mu^- \bar{\nu}_{\mu} b \bar{b} \,H$ cross section in various
phase-space regions. Furthermore, realistic predictions for different Higgs boson decay modes showed significant distortions compared to the case with a stable Higgs boson. Therefore, to obtain accurate predictions, all these effects should be taken into account in future comparisons between theoretical results and experimental data.

\acknowledgments{
The work was supported by the Deutsche Forschungsgemeinschaft (DFG) under grant 396021762 - TRR 257: {\it P3H - Particle Physics Phenomenology after the Higgs Discovery}. Support by a grant of the Bundesministerium f\"ur Bildung und Forschung (BMBF) is additionally acknowledged.

Simulations were performed with computing resources granted by RWTH
Aachen University under project {\tt rwth0414}.
}




\providecommand{\href}[2]{#2}\begingroup\raggedright\endgroup

\end{document}